\documentclass[acmsmall,nonacm]{acmart}

\input{preamble}

\begin{document}

\title{The Hypervolume Indicator: Problems and Algorithms}

\author{Andreia P. Guerreiro}
\orcid{0000-0003-0757-2709}
\affiliation{
    \institution{INESC-ID}
    \streetaddress{Rua Alves Redol, 9}
    \postcode{1000-029}
    \city{Lisbon}
    \country{Portugal}
}
\affiliation{
    \department{CISUC, Department of Informatics Engineering}
    \institution{University of Coimbra}
    \streetaddress{Polo II}
    \postcode{3030-290}
    \city{Coimbra}
    \country{Portugal}
}
\email{andreia.guerreiro@tecnico.ulisboa.pt}
\author{Carlos M. Fonseca}
\orcid{0000-0001-5162-2457}
\affiliation{
    \department{CISUC, Department of Informatics Engineering}
    \institution{University of Coimbra}
    \streetaddress{Polo II}
    \postcode{3030-290}
    \city{Coimbra}
    \country{Portugal}
}
\email{cmfonsec@dei.uc.pt}
\author{Lu\'is Paquete}
\orcid{0000-0001-7525-8901}
\affiliation{
    \department{CISUC, Department of Informatics Engineering}
    \institution{University of Coimbra}
    \streetaddress{Polo II}
    \postcode{3030-290}
    \city{Coimbra}
    \country{Portugal}
}
\email{paquete@dei.uc.pt}

\acmJournal{CSUR}

\begin{abstract}

The hypervolume indicator is one of the most used set-quality indicators
for the assessment of stochastic multiobjective optimizers, as well
as for selection in evolutionary multiobjective optimization algorithms.
Its theoretical properties justify its wide acceptance, particularly
the strict monotonicity with respect to set dominance which is still
unique of hypervolume-based indicators.
This paper discusses the computation of hypervolume-related problems,
highlighting the relations between them,
providing an overview of the paradigms and techniques used, a description
of the main algorithms for each problem,
and a rundown of the fastest algorithms regarding asymptotic complexity and
runtime.
By providing a complete overview of the computational problems associated to the hypervolume indicator,
this paper serves as the starting point for the development of new algorithms, and 
supports users in the identification of the most appropriate implementations available for each problem.
\end{abstract}



\maketitle

\section{Introduction}
In multiobjective optimization, a single optimal solution seldomly exists due to the (usually)
conficting nature of objectives. Instead, there is typically a set of Pareto-optimal solutions.
The set of all Pareto-optimal solutions (in decision space) is known as the Pareto-optimal set,
and the corresponding images in objective space is known as the Pareto front. In such case,
the best solution of a given problem depends on the (subjective) preferences of the Decision Maker (DM).
As the Pareto front may be very large, and even infinite, the aim of optimizers
under a no-preference information scenario is to find, and present to the DM,
a set of solutions whose corresponding point set
in objective space is a representative and finite subset of the Pareto front.
In the search for such subset, the task of comparing the point sets into which solution sets map to
becomes unavoidable, and therefore, also the need to define preferences over point sets.
It is commonly accepted that the quality of a point set should be evaluated based on its closeness to the
Pareto front (the closer the better), on the diversity in the set
(the more evenly distributed they are, the better), and its spread~\citep{ZitzlerDebThieleEC00}.

Set-quality indicators facilitate the evaluation process of Pareto-front approximations
by reconciling, in a single real value, characteristics such as proximity to the Pareto front,
and diversity. Even though different set-quality indicators possibily
value different characteristics, the knowledge of an indicators' properties is essential
to the understanding of its inner preferences, and consequently, allow for a
more conscious choice of the most appropriate quality indicator.
Given its easy interpretation and its good properties, the hypervolume
indicator rapidly became, and still is,
one of the most widely used quality indicators among the many existing indicators~\cite{LiYaoSurvey2019}.

The hypervolume set-quality indicator maps a point set in $\Rd$ to the
measure of the region dominated by that set and (assuming minimization) bounded above by a given
reference point, also in $\Rd$, where $d$ is the number of objectives.
It was first referred to
as the ``size of the space covered"~\citep{Zitzler98,zt1999a}, and as
``size of the dominated space"~\citep{Zitz1erPhD99}.
Alternative designations have also been used,
such as $\Sc$-metric~\citep{zt1999a,nopHOY} and ``Lebesgue measure"~\citep{Fleischer03}.
Different definitions of this indicator have been proposed.
For example, it has been defined
based on the union of polytopes~\citep{zt1999a} and, more generally
based on the (integration of the) attainment function~\citep{HVIRevisted07,GF2012}.
The problem of computing the hypervolume indicator is known to be a special case of
Klee's Measure Problem (KMP)~\citep{HOY}, which is
the problem of measuring the region resulting from the union of axis-parallel boxes.
The hypervolume indicator is, in fact, a special case of KMP on unit cubes, and of KMP on
grounded boxes~\citep{YildizSuri2012}. 
See~\cite{OrderKMP} for a review on KMP's special cases and their relation
to one another.

The hypervolume indicator was first proposed as a method for assessing multiobjective
optimization algorithms~\citep{zt1999a}. 
It evaluates the optimizer outcome by simultaneously taking into account
the proximity of the points to the Pareto front, diversity, and spread.
The indicator's unique properties quickly led to its integration in
Evolutionary Multiobjective Optimization Algorihms (EMOAs),
as a bounding method for archives~\citep{Knowles2003}, an environmental selection
method~\citep{EmmBeuNauj2005}, a ranking method~\citep{CMA-MO}, and as a fitness
assignment method~\citep{IBEA,Bader2010,ZTB2010}.
The integration of preferences in the indicator~\citep{HVIRevisted07,HSSP2DWHypE2009,BaderWHypE}
has also been the subject of discussion, and so has the integration of diversity in the decision
space~\citep{ubz2010a}.
Currently, the hypervolume indicator is one of the indicators used in the
Black-Box Optimization Benchmarking (BBOB) tool~\citep{BBOBGecco15}
to continuously evaluate the external archive containing all
nondominated solutions EMOAs generate during their execution.

The merits of the hypervolume indicator are well recognized,
however, its main drawback lies in its computational cost.
This is particularly relevant as hypervolume-based
EMOAs and benchmarking tools such as BBOB depend heavily on its computation.
This imposes strong limitations on the number
of objectives considered and/or on EMOA parameters such as the number of generations and number
of offspring.
In order to overcome such a limitation, approximation algorithms~\citep{Bader2010} have been proposed,
as well as objective reduction methods~\citep{ObjRedHV07}.

The main goal of this paper is, firstly, to instigate the development
of new algorithms for hypervolume-based problems by providing a broad overview
of the current computational approaches to solving hypervolume-based problems, and
by highlighting the intrinsic relation between the problems which can be exploited.
Secondly, this review is meant to promote best practices by
providing summaries of the currently fastest algorithms for each problem,
both regarding asymptotical and runtime performance,
and by providing links to available implementations.
This promotes the use of the most adequate algorithm either for application
purposes, \ie, to make a more efficient use of hypervolume-based problems,
and for benchmarking purposes, for example, to have fairer/adequate comparison tests with
hypervolume-based EMOAs.

In the following sections, the theoretical advantages and the computational aspects of
hypervolume indicator are discussed in more detail,
mostly in the context of EMOAs.
In Section~\ref{ch:hvstoa:probs}, the hypervolume indicator and some related problems
are formally defined, and its properties are reviewed.
A review of the state-of-the-art algorithms for hypervolume-related problems is provided
in Sections~\ref{s:hvstoa:algs} to~\ref{ch:hvstoa:algs:hssp}.
Concluding remarks are drawn in Section~\ref{stoa:conc}.

\section{Hypervolume-related Problems}\label{ch:hvstoa:probs}

\subsection{Notation}\label{s:hvstoa:probs:not}

\begin{description}
    \setlength\itemsep{0.0cm}
    \item[Spaces]
    $(x,y)$-, $(x,y,z)$- and $(x,y,z,w)$-spaces will be referred to as,
    2-, 3- and 4-dimensional spaces or, for brevity, as 2D, 3D and 4D spaces, respectively.
    \item[Problem size] The lower-case letter $n$ is used for the problem size, which is typically the size of the input set.
    \item[Number of dimensions] The lower-case letter $d$ is used to represent the number of dimensions considered.
    \item[Points and sets] Points are represented by lower-case letters (in italics) and sets by Roman capital letters. For example, $p,q\in\Rd$
and $\Xr,\Sr\subset\Rd$.
    \item[Coordinates (for $d\leq 4$)] Letters $x$, $y$, $z$ and $w$ in subscript denote the coordinates
        of a point in an $(x,y,z,w)$-space. This notation is used only for spaces up to 4 dimensions.
        For example, if $p\in\Rt$ then $p=(\cx{p},\cy{p}, \cz{p})$.
    \item[Coordinates (general case $d\geq 2$)] In general $d$-dimensional spaces, an index in subscript is used
        to identify the coordinate.
        For example, if $p\in\Rd$ then $p=(\cd{p}{1},\cd{p}{2}, \ldots, \cd{p}{d})$, where $\cd{p}{i}$ denotes
        the $i^{th}$ coordinate of $p$, $i\in\{1,\ldots,d\}$. 
    \item[Enumeration] Numbers in superscript are used to enumerate points or sets, \eg, $p^1,p^2,p^3\in\Rd$ and
$\Sr^1,\Sr^2\subset\Rd$.
    \item[Projections] Projection onto $(d-1)$-space by omission of the last coordinate is denoted by an asterisk. 
For example, given the point set $\Xr=\{p,q\}\subset\Rt$, $p^*$ and $\Xr^*$ denote the projection of the point
$p$ and of the point set $\Xr$ on the $(x,y)$-plane, respectively, \ie, $p^*=(\cx{p},\cy{p})$
and $\Xr^*=\{(\cx{p},\cy{p}), (\cx{q},\cy{q})\}$.
    \item[Dominance] A point $p\in\Rd$ is said to \emph{weakly dominate} a point $q\in\Rd$ if
$p_i\leq q_i$  for all $1\leq i\leq d$. This is represented as $p\leq q$. 
If, in addition $p\not\leq q$, then $p$ is said to \emph{(strictly) dominate} $q$,
which is represented here as $p < q$.
If $p_i < q_i$ for all $1\leq i\leq d$, then $p$ is said to \emph{strongly
dominate} $q$, and this is represented as $p \ll q$.

\end{description}

\subsection{Definitions}\label{bg:def}

\begin{figure*}[t!]
  \center
  \hspace{-0.2cm}
  \subfigure[$H(\{p^1,\ldots,p^4\})$]{\label{fig:defs:2D:hv}\includegraphics[width=3.5cm,height=3.5cm]{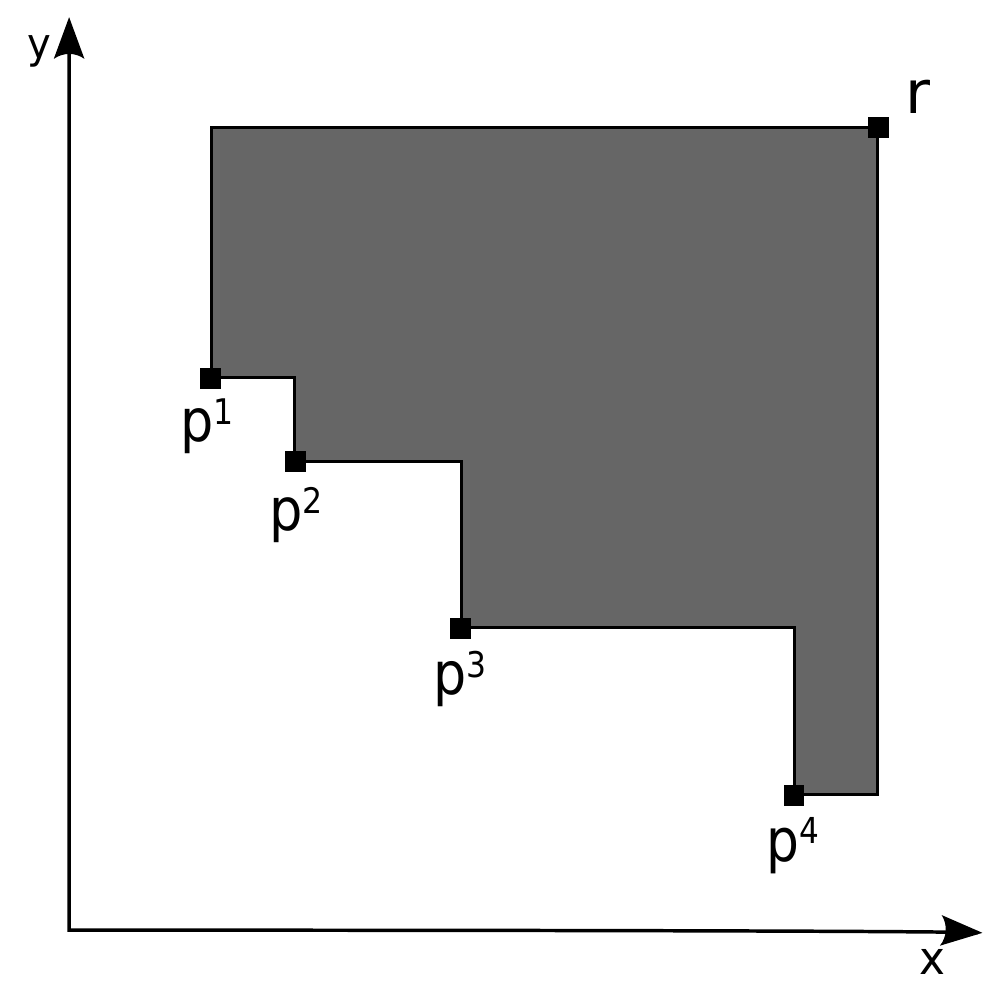}}
  \hspace{-0.3cm}
  \subfigure[$H(\{p^2,p^3\},\{p^1,p^4\})$]{\label{fig:defs:2D:hvcs}\includegraphics[width=3.5cm,height=3.5cm]{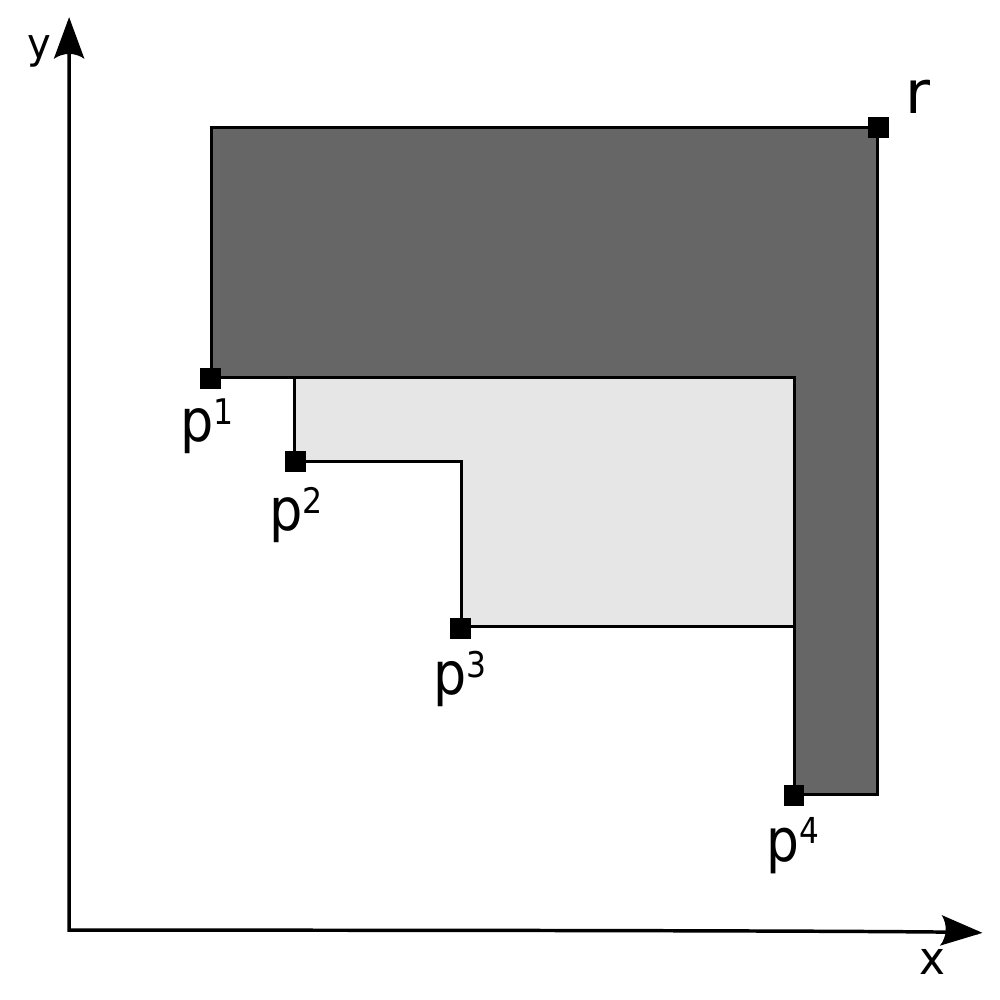}}
  \hspace{-0.3cm}
  \subfigure[$H(p^3,\{p^1,p^2,p^4\})$]{\label{fig:defs:2D:hvc}\includegraphics[width=3.5cm,height=3.5cm]{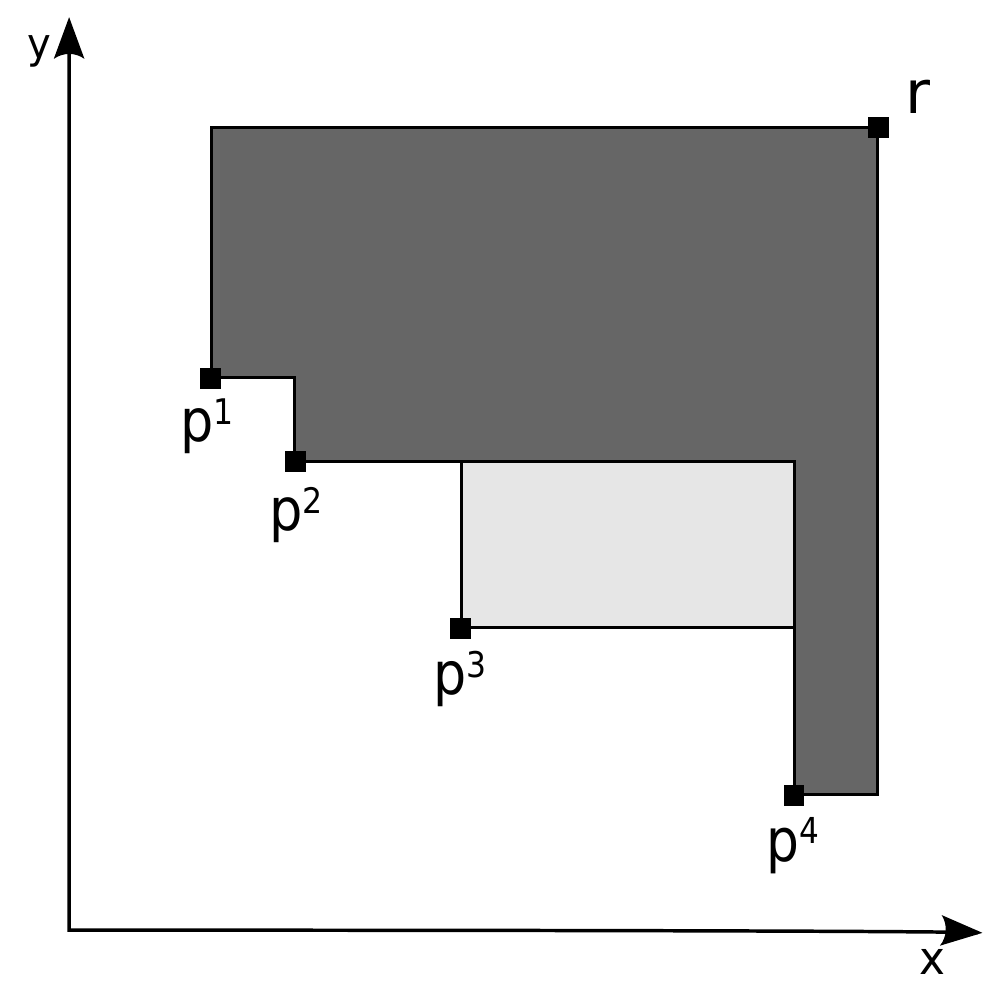}}
  \hspace{-0.3cm}
  \subfigure[$H(p^2,p^3,\{p^1,p^4\})$]{\label{fig:defs:2D:jhvc}\includegraphics[width=3.5cm,height=3.5cm]{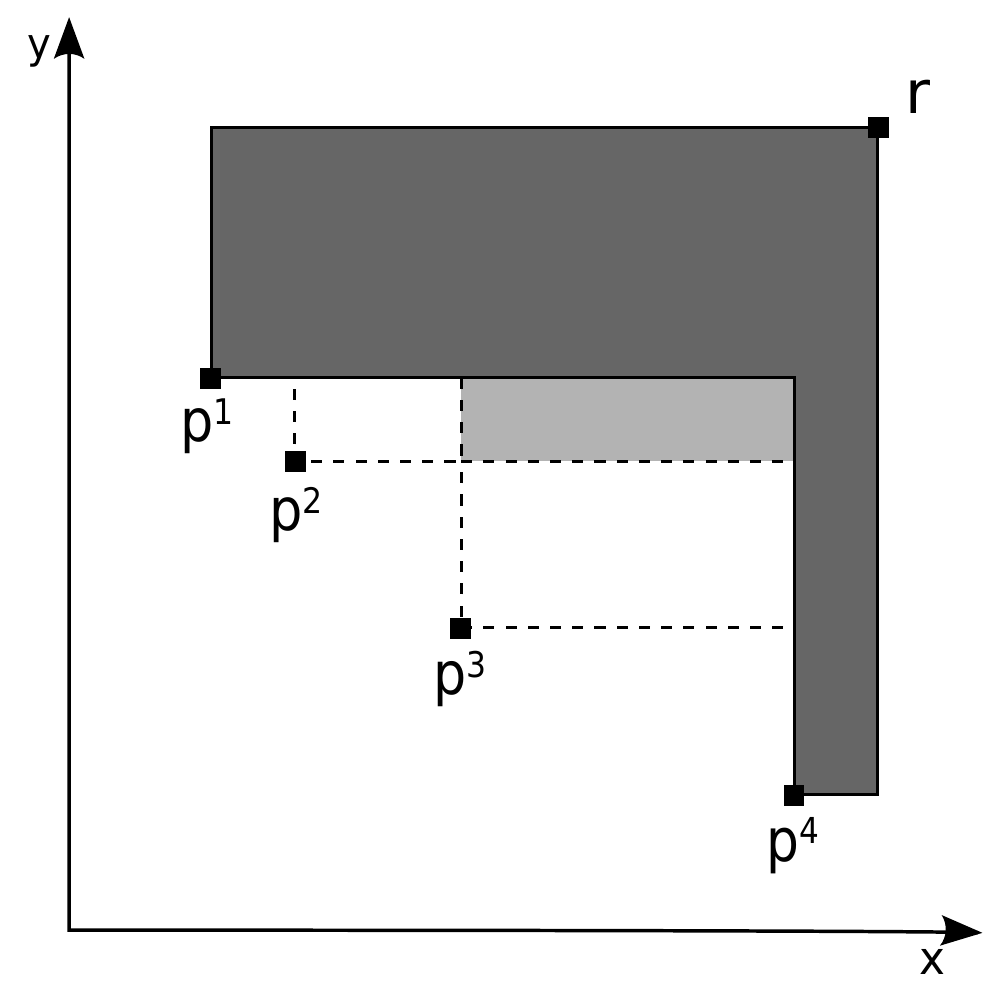}}
  \caption[Hypervolume-related definitions: 2D examples]
  {Examples in two-dimensions:
    (a) hypervolume indicator (dark gray region), 
    (b) hypervolume contribution of a point set (light gray region), 
    (c) hypervolume contribution of a point (light gray region), 
    (d) joint hypervolume contribution  (mid gray region).}
  \label{fig:defs:2D}
\end{figure*}

\begin{figure*}[t]
  \center
\begin{tabular}{lll}
  \subfigure[$H(\{p^1,\ldots,p^4\})$]{\label{fig:defs:3D:hv}
  \begin{overpic}[width=3.5cm,height=3.5cm]{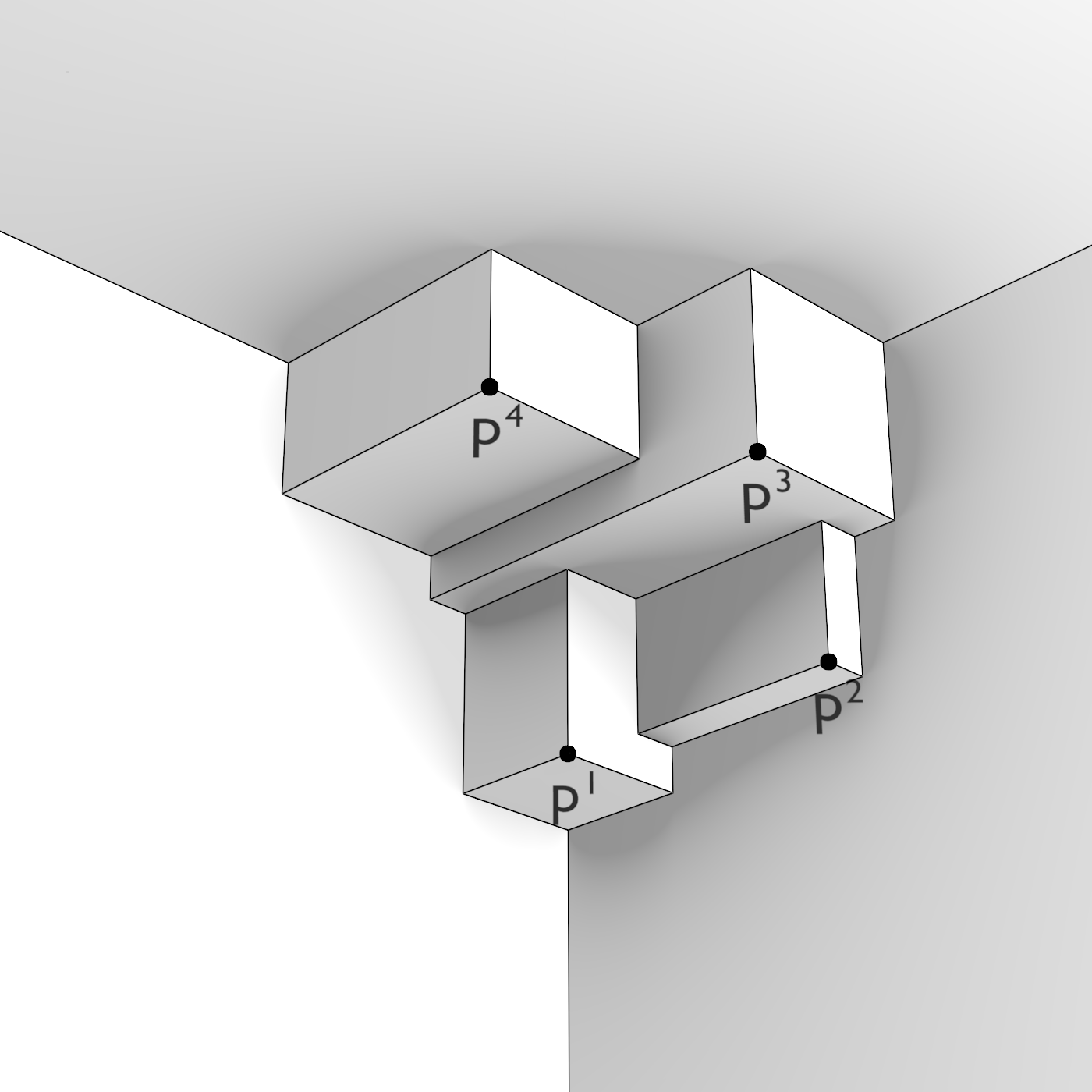}
     \put(3,3){\includegraphics[width=0.85cm,height=0.8cm]{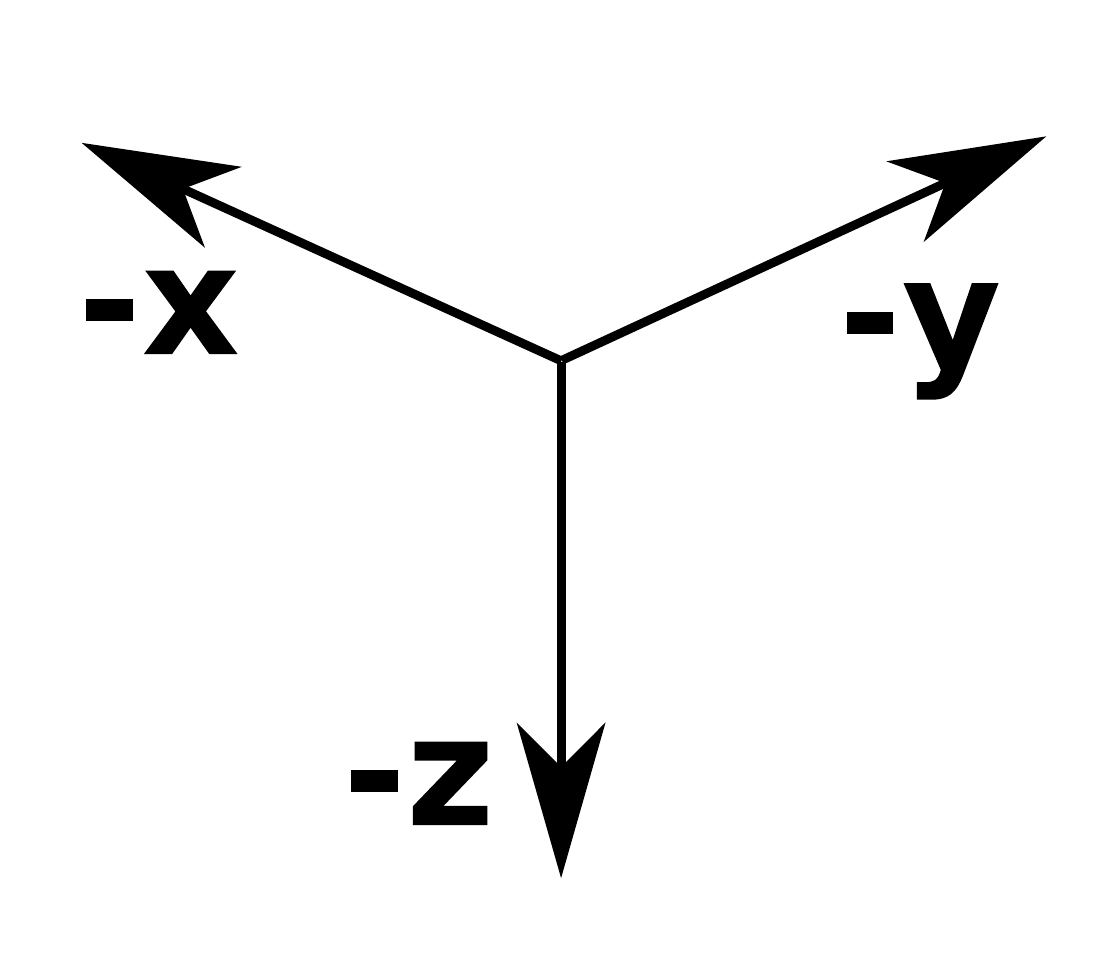}}  
  \end{overpic}
  }
  &\subfigure[$H(p^3,\{p^1,p^2,p^4\})$]{\label{fig:defs:3D:hvc}\includegraphics[width=3.5cm,height=3.5cm]{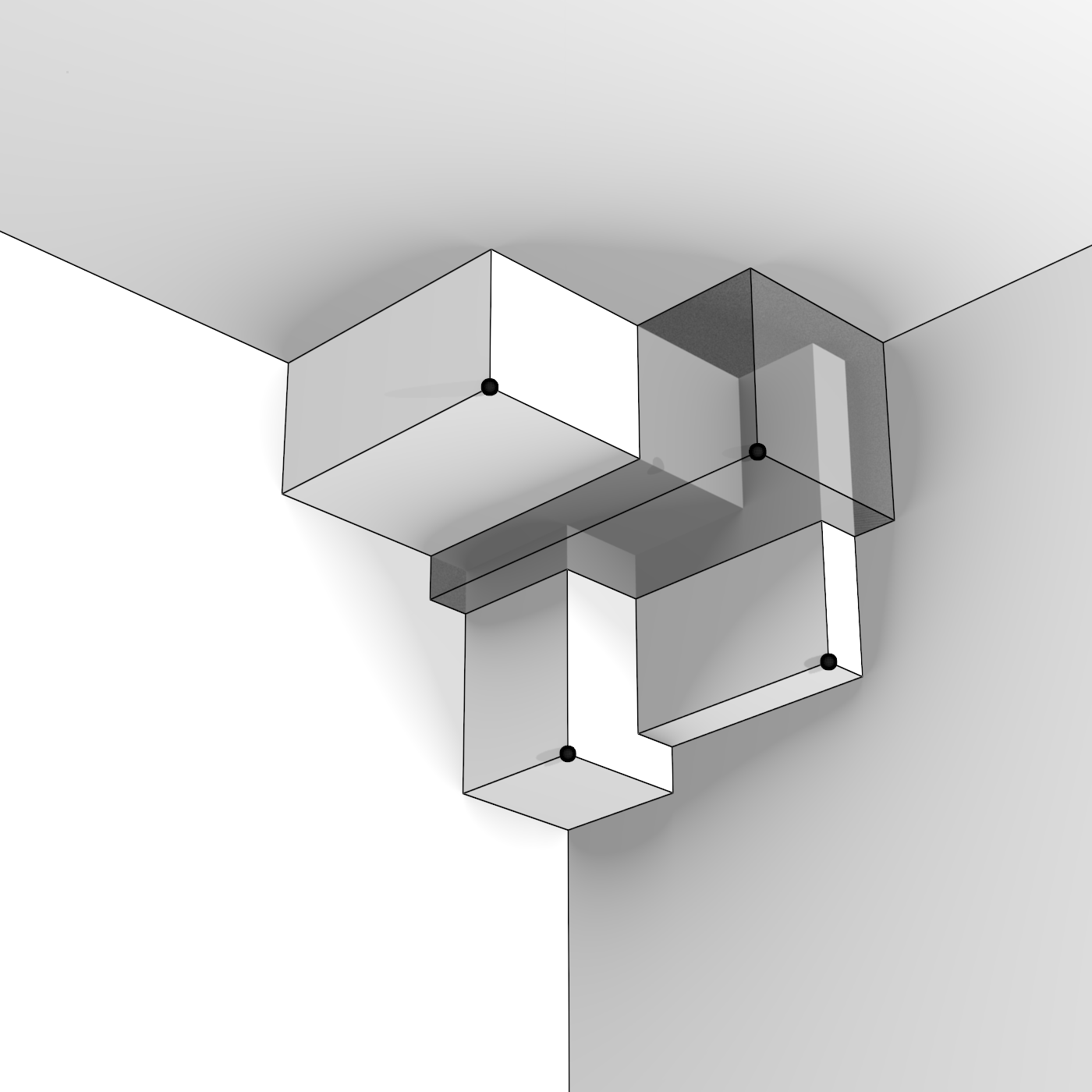}} 
  &  \subfigure[$H(p^3,\{p^1,p^2\})$,\newline $H(p^4,\{p^1,p^2\})$]{\label{fig:defs:3D:hvc2}\includegraphics[width=3.5cm,height=3.5cm]{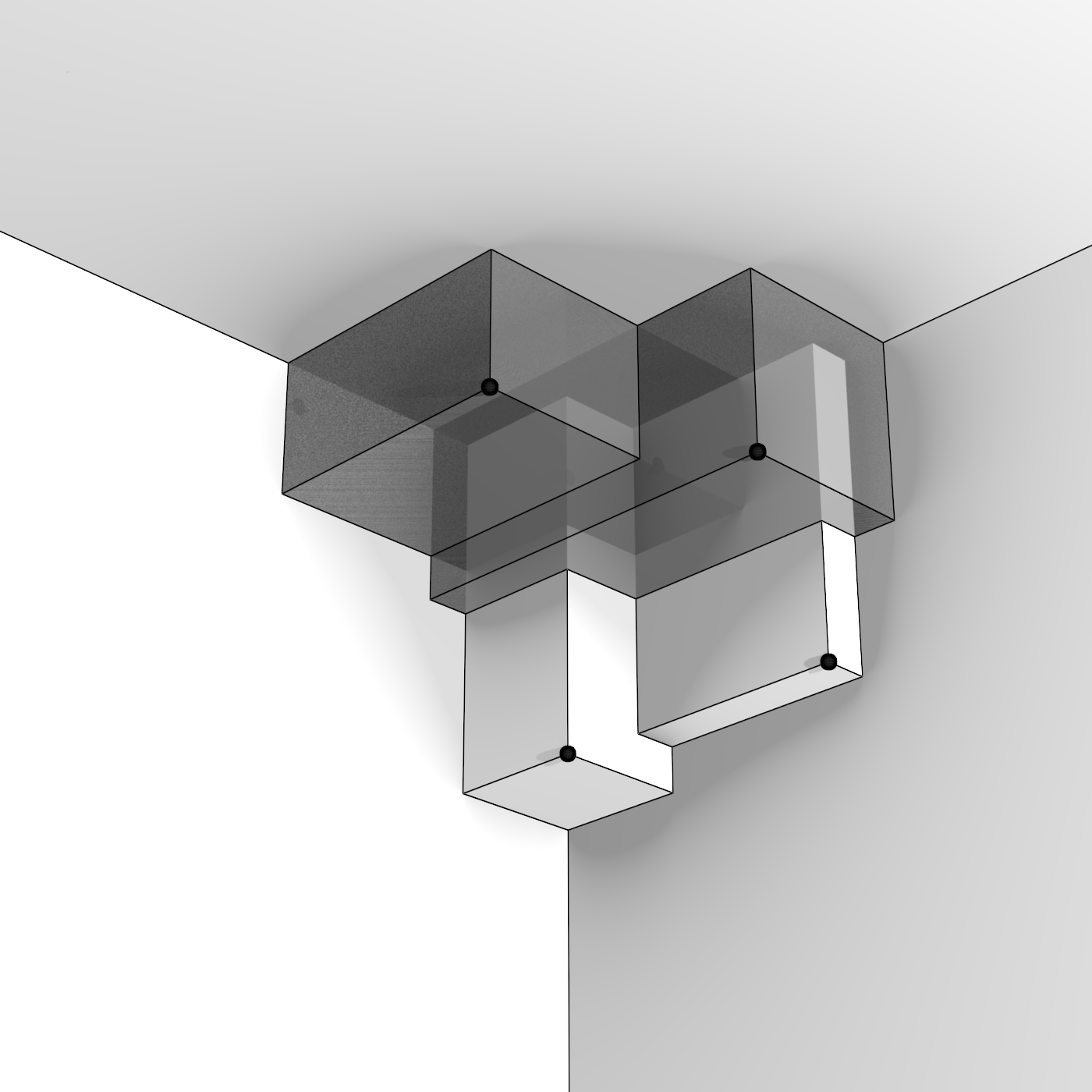}}
\\
  \hspace{-0.2cm}
  \subfigure[$H(p^3,\{p^1,p^2\})$]{\label{fig:defs:3D:jhvcq}\includegraphics[width=3.5cm,height=3.5cm]{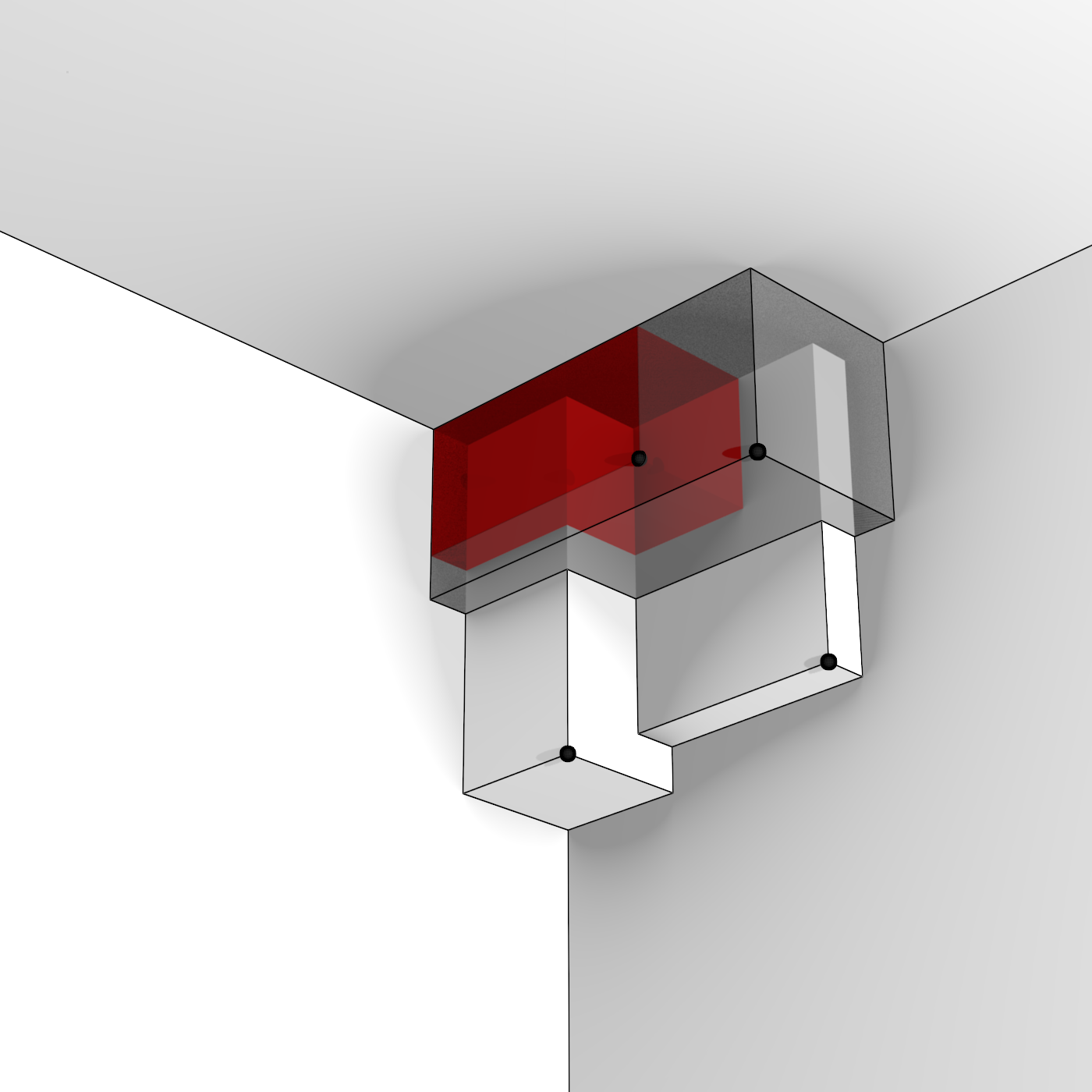}}
  \hspace{-0.3cm}
  &\subfigure[$H(p^4,\{p^1,p^2\})$]{\label{fig:defs:3D:jhvcp}\includegraphics[width=3.5cm,height=3.5cm]{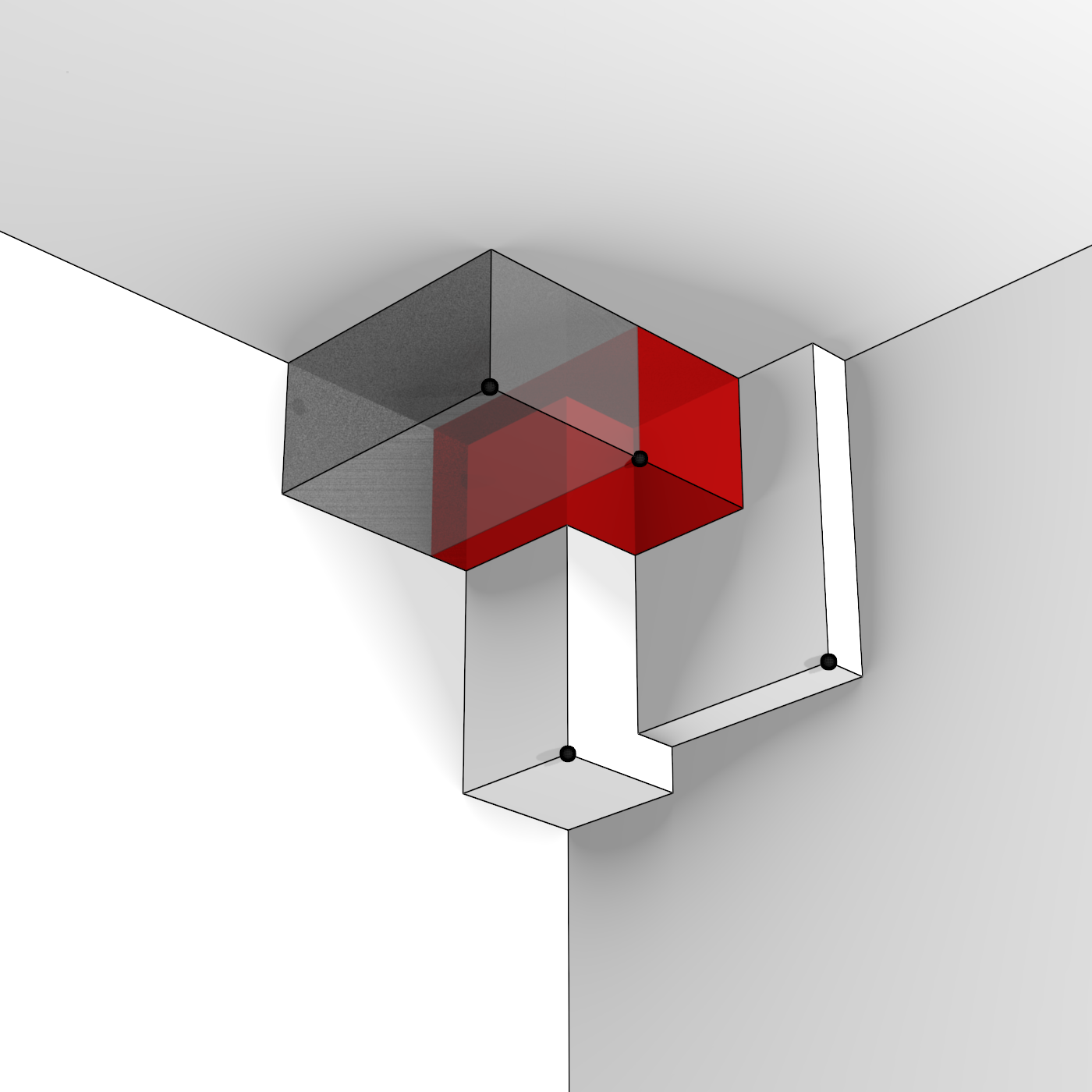}} 
  \hspace{-0.3cm}
  &\subfigure[$H(p^3,p^4, \{p^1,p^2\})$]{\label{fig:defs:3D:jhvc}\includegraphics[width=3.5cm,height=3.5cm]{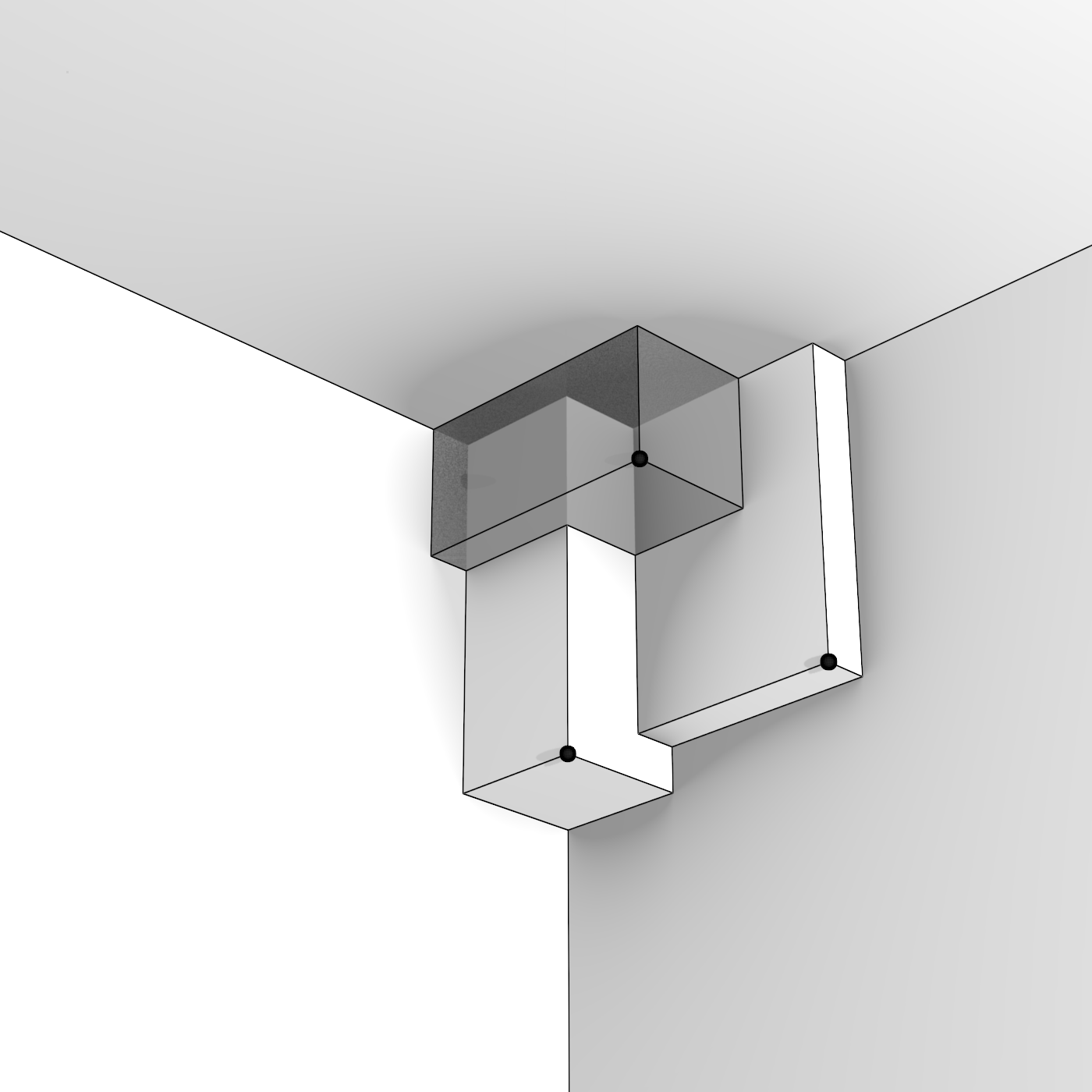}}
\end{tabular}\caption[Joint Contribution: 3D example]
  {Three-dimensional examples: (a) hypervolume indicator (opaque volume), (b-c) hypervolume contribution (transparent volume),
  (d-f) joint contribution of $p^3$ and $p^4$ to $\Sr=\{p^1,p^2\}$.
  Transparent red highlights the part of a contribution also dominated by the omitted point.
  }
  \label{fig:defs:3D}
  \label{fig:defs:3D2}
    \vspace{-15pt}
\end{figure*}

The hypervolume indicator~\citep{Knowles2003, Zitzler98} is formally defined as follows:
\begin{definition}[Hypervolume Indicator]
Given a point set $\Sr\subset\Rd$ and a reference point $r\in\Rd$, the hypervolume indicator
of $\Sr$ is the measure of the region weakly dominated by $\Sr$ and bounded above by $r$, \ie:
\begin{equation*}
H(\Sr) = \Lambda(\{ q \in \mathbb{R}^d \mid \exists p \in \Sr : p \leq q \andl q \leq r\})
\end{equation*}%
where $\Lambda(\cdot)$ denotes the Lebesgue measure.
Alternatively, it is interpreted as the measure of the union of boxes:
\begin{equation*}
    H(\Sr) = \Lambda\left(\bigcup_{p\in \Sr \atop p\leq r} [p, r]\right)
\end{equation*}
where $[p,r]=\{ q \in \mathbb{R}^d \mid p \leq q \andl q \leq r\}$
denotes the box delimited below by $p\in\Sr$ and above by $r$.
\label{def:hv}%
\end{definition}%
Since a fixed reference point, $r$, is assumed throughout this paper, it is omitted
as an argument of $H(\cdot)$ function.
Figure~\ref{fig:defs:2D:hv} shows a two-dimensional example of the hypervolume (an area)
and Figure~\ref{fig:defs:3D:hv} shows a three-dimensional example (a volume).

The hypervolume contribution of a point set to some reference point set~\citep{ArchAlgsBF2014, PhDBrockhoff2009}
is formally defined based on the definition of hypervolume indicator:
\begin{definition}[Hypervolume Contribution of a Point Set]\label{def:hvcs}
Given two point sets $\Xr,\Sr\subset\Rd$, and a reference point $r\in\Rd$, the (hypervolume)
contribution of $\Xr$ to $\Sr$ is:
\begin{equation*}
    H(\Xr, \Sr) = H(\Xr\cup\Sr) - H(\Sr\setminus\Xr)
\end{equation*}
\end{definition}%
Note that if $\Xr\cap\Sr=\emptyset$ then the contribution of $\Xr$ to $\Sr$ is simply
$H(\Xr, \Sr) = H(\Xr\cup\Sr) - H(\Sr)$.
See Figure~\ref{fig:defs:2D:hvcs} for a two-dimensional example.
The particular case of $|\Xr|=1$ is used more frequently and is defined as in~\cite{ArchAlgsBF2014}:
\begin{definition}[Hypervolume Contribution]\label{def:hvc}
The hypervolume contribution of a point $p\in\Rd$ to a set $\Sr\subset\Rd$ is:
\begin{equation*}
    H(p, \Sr) = H(\Sr\cup\{p\}) - H(\Sr\setminus\{p\})
\end{equation*}%
\end{definition}%
The hypervolume \textit{contribution} of a point is sometimes referred in the literature
as the \textit{incremental} hypervolume or the \textit{exclusive} hypervolume~\citep{IWFG}.
Moreover, the contribution of a point $p$ to the empty set is sometimes called the \textit{inclusive}
hypervolume~\citep{IWFG}.
See Figures~\ref{fig:defs:2D:hvc} and~\ref{fig:defs:3D:hvc} for two-
and three-dimensional examples of a hypervolume contribution, respectively.

As pointed out in~\cite{ArchAlgsBF2014}, the above Definition~\ref{def:hvc}
is consistent with the case where
$p\in\Sr$, and the contribution is the hypervolume lost when $p$ is removed from $\Sr$,
as well as with the case where $p\notin\Sr$, and the contribution of $p$ is the hypervolume
gained when adding $p$ to $\Sr$. While this is certainly convenient,
it does not reflect the fact that the hypervolume gained by ``adding'' a point $p$
to a set already including it is zero. However, this last situation can be
handled easily as a special case by checking whether $\Sr$ includes $p$ before
applying the definition.

In some cases, such as when determining the decrease in the contribution of a given point $p\in\Rd$ to a
set $\Sr\subset\Rd$ due to the addition of another point $q\in\Rd$ to $\Sr$,
it is also useful to consider the contribution dominated simultaneously and exclusively
by two points~\cite{gHSSECJ2016}.
%
%
\begin{definition}[Joint Hypervolume Contribution]\label{def:jhvc}
The joint hypervolume contribution of $p,q\in\Rd$ to $\Sr\subset\Rd$ is:

\begin{equation*}
    H(p,q, \Sr) = H((\Sr\setminus\{p, q\})\cup\{p\vee q\}) - H(\Sr\setminus\{p, q\})
\end{equation*}
where $\vee$ denotes the \emph{join}, or component-wise maximum between two points.
A general definition of the joint contribution to $\Sr$ of $t$ points can also be
found in the literature~\cite{HSSPsmallk}.

\label{def:jcont}%
\end{definition}%
%
Figures~\ref{fig:defs:2D:jhvc} shows an example for the two-dimensional case
and Figures~\ref{fig:defs:3D:hvc2}-\subref{fig:defs:3D:jhvc} for the three-dimensional case.
In the $d=3$ example, Figure~\ref{fig:defs:3D:hvc2} shows the individual contribution of $p^3$ and $p^4$
to $\Sr=\{p^1,p^2\}$, which partially overlap. This partially overlapped volume is the joint contribution
(represented in transparent gray in Figure~\ref{fig:defs:3D:jhvc}).
This joint contribution can also be interpreted as the region of the contribution of $p^3$ to $\Sr$
that is also dominated by $p^4$ (the transparent red region in Figure~\ref{fig:defs:3D:jhvcq}).
Analogously, it can be interpreted as the region of the contribution of $p^4$ to $\Sr$
that is also dominated by $p^3$ (the transparent red region in Figure~\ref{fig:defs:3D:jhvcp}).

\begin{figure}[t]
\center
  \hspace*{\fill}%
  \subfigure[Delimiters of $p$ and elements of $\Jr$]{\label{fig:defs:2D:del}\includegraphics[width=3.5cm,height=3.5cm]
  {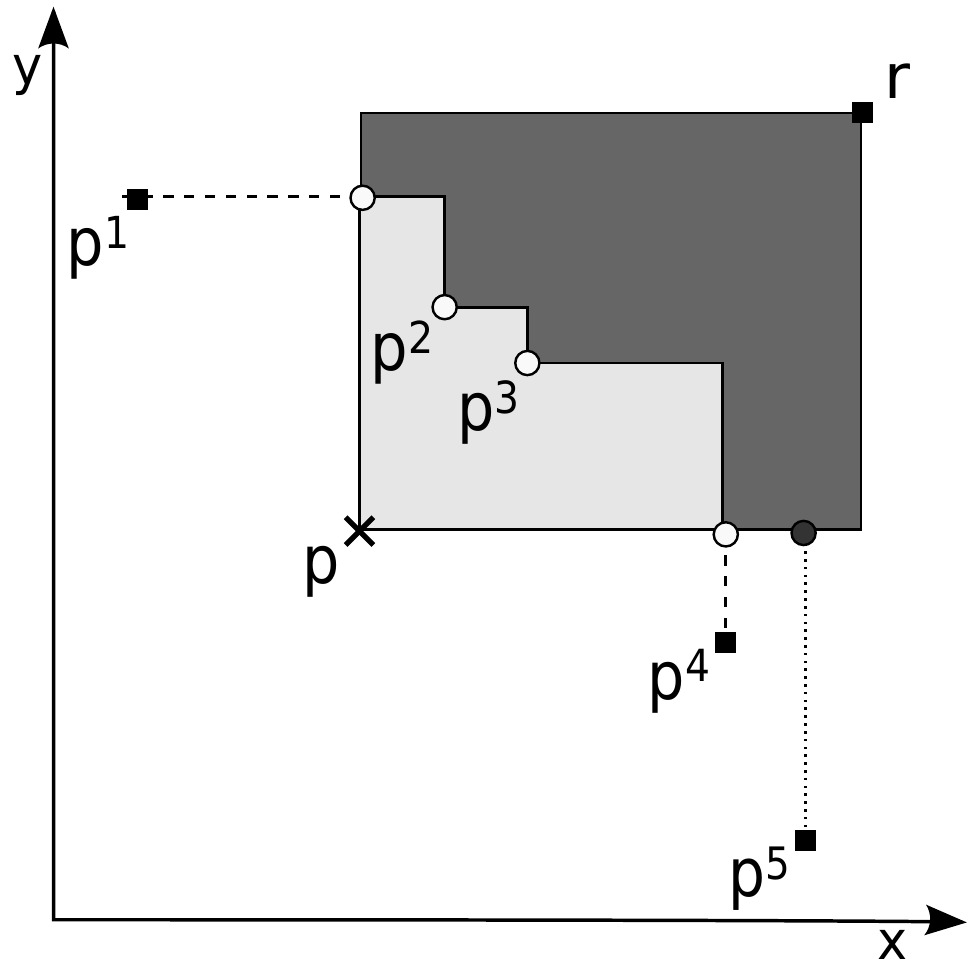}
  }
  \hspace*{\fill}%
  \subfigure[Strong and non-strong delimiters of $p$]{\label{fig:bg:delms:ex1}
                \includegraphics[width=3.5cm,height=3.5cm]
                    {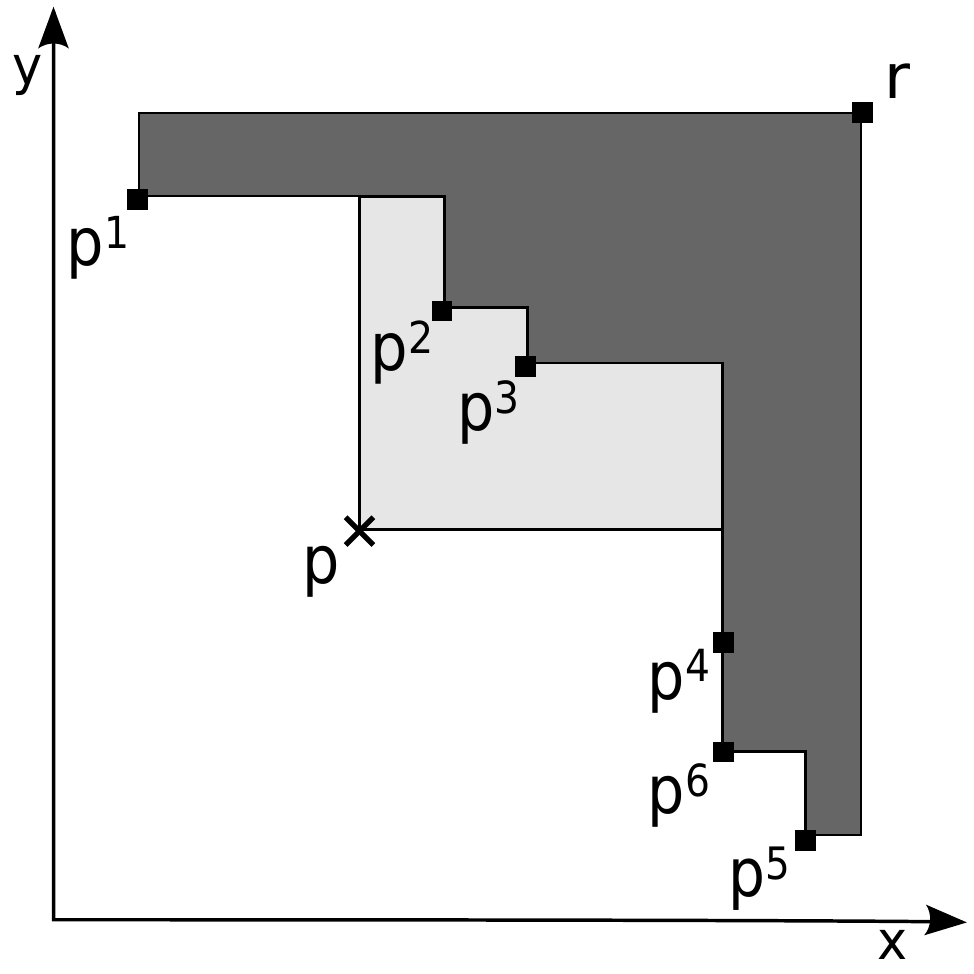}
                }
  \hspace*{\fill}%
  \subfigure[Proper and non-proper delimiters of $p$]{\label{fig:bg:delms:ex2}
                \includegraphics[width=3.5cm,height=3.5cm]
                {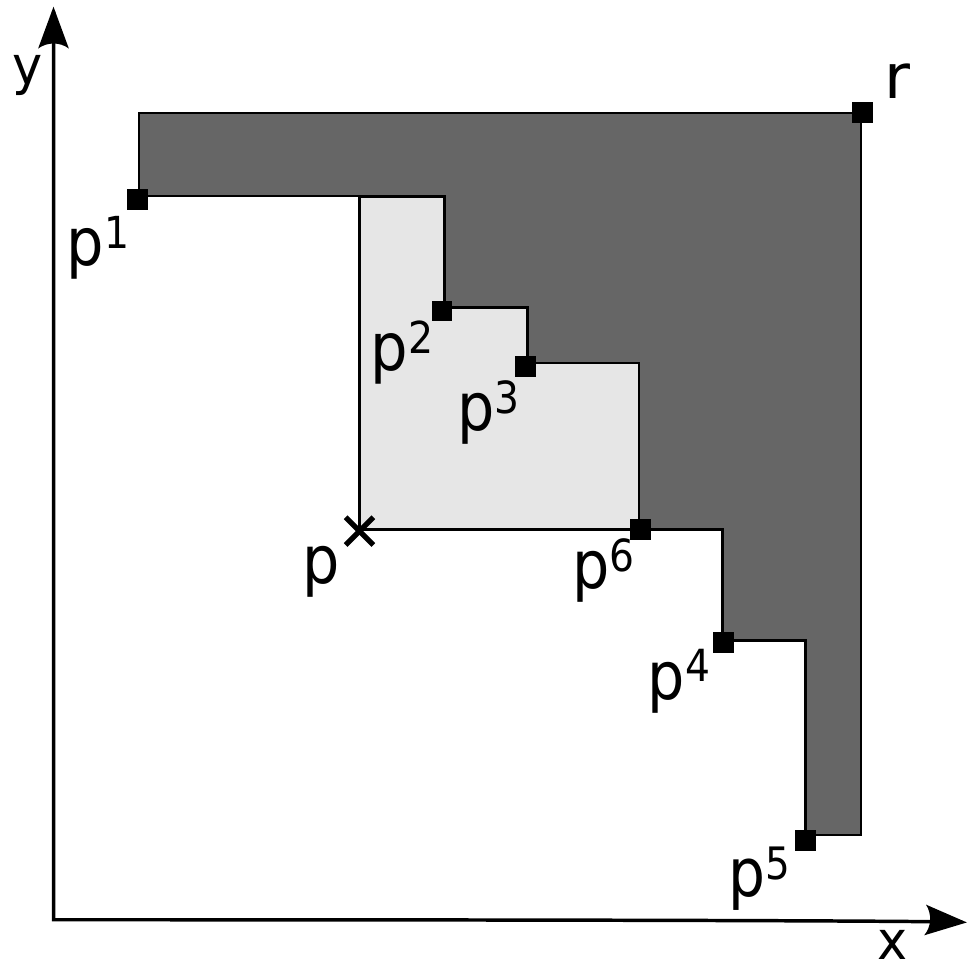}
                }
  \hspace*{\fill}%
  \vspace{-0.3cm}
  \caption[Examples of delimiters]
  {Examples of the delimiters of the contribution of $p$ to a point set.
  (a) delimiters of $p$ ($p^1,\ldots,p^4$), as well as dominated point $(p \vee p^5)$, represented
   by a filled circle, and nondominated points $(p \vee p^1)$, $(p \vee p^2)$, $(p \vee p^3)$, $(p \vee p^4)$, elements of $\Jr$, 
   represented by hollow circles (see text for more details), (b) strong ($p^1,\ldots,p^3$) and non-strong ($p^4,p^6$) delimiters,
   (c) proper ($p^1,p^2,p^3,p^6$) and non-proper ($p^4$) delimiters.}

\label{fig:bg:delms}
    \vspace{-7pt}
\end{figure}

Moreover, the contribution of a point $p$ to a set $\Sr$ is bounded above by certain points $q\in\Sr$
that shall be referred to as \emph{delimiters}, and are defined as follows~\cite{HVCTEC2017}:
\begin{definition}[Delimiter]\label{def:dlmtr}
Given a point set $\Sr\subset\Rd$ and a point $p\in\Rd$, let $\Jr = \ndd(\{(p\vee q)\mid q\in \Sr\setminus\{p\}\})$.
Then, $q\in\Sr$ is called a \textit{(weak) delimiter} of the contribution of $p$ to $\Sr$
iff $(p \vee q)\in \Jr$.
If, in addition, $H(p,q,\Sr)>0$, then $q$ is also a \textit{strong delimiter} of the contribution of $p$ to $\Sr$.
\vspace{0.2cm}
\end{definition}
Where $\ndd(\Xr)=\{s\in \Xr\mid \forall_{t\in \Xr}, t\leq s \Rightarrow s\leq t \}$ denotes the set of
nondominated points in $\Xr$.
%
%
Note that $\Jr$ is the smallest set of points weakly dominated by $p$ that delimits its contribution
to $\Sr$, that is, $H(p,\Sr)=H(p,\Jr)$.
Consequently, all $q\in\Jr$ are such that $H(p,q, \Jr)>0$, and $\Jr$ is
such that $H(p,\Sr)=H(\{p\})-H(\Jr)$.
Figure~\ref{fig:defs:2D:del} shows an example where the
contribution of $p$ is delimited only by $p^1$, $p^2$, $p^3$ and $p^4$, where all of them are
strong delimiters.
Non-strong delimiters can only exist when $\Sr$ contains points with repeated coordinates.
In the example of Figure~\ref{fig:bg:delms:ex1}, $p^1,p^2,p^3$ are
strong delimiters while $p^4$ and $p^6$ are not strong but only weak delimiters.
This means that, in practice, only one point in such a group of delimiters is needed to bound the
contribution of $p$.
If one of them is deleted, then the contribution of $p$ remains unchanged,
whereas it increases if all are deleted.
The following extension to the notion of delimiter will also be needed in this work~\cite{HVCTEC2017}:
\begin{definition}[Outer Delimiter] 
Given a point set $\Sr\subset\Rd$ and a point $p\in\Rd$, $q\in\Sr$ is called an
\emph{outer delimiter} of the contribution of $p$ to $\Sr$ if it is a delimiter
of the contribution of $p$ to $\{s\in\Sr\mid p\not\leq s\}$. A delimiter,
$q$, of the contribution of $p$ to $\Sr$ is called an \emph{inner delimiter} if it
is not an outer delimiter, i.e., if $p\leq q$.
\label{def:outer}
\end{definition}

In general, outer delimiters may not be actual, or
\emph{proper}, delimiters in the sense of Definition~\ref{def:dlmtr},
in particular, when $p$ shares a coordinate with a point in $\Sr$.
In the example of Figure~\ref{fig:bg:delms:ex2},
points $p^1,p^2,p^3,p^6$ are the (proper) delimiters of the contribution of $p$ to
$\Sr$, of which $p^2$, $p^3$ and $p^6$ are inner delimiters.
There are two outer delimiters, $p^1$ and $p^4$. Point $p^4$ is not a proper
delimiter of the contribution of $p$ to $\Sr$ because $(p\vee p^6)=p^6 < (p\vee p^4)$.

\subsection{Problems}\label{ss:hvrp:prob}

Many computational problems related to the hypervolume indicator can be found in the literature.
The following problems extend the list of problems given in~\cite{EF2011,HVCTEC2017}.
Recall that the reference point is considered to be a constant.
\begin{problem}[\sc{\hvp}]\label{prob:hvp}
Given an $n$-point set $\Sr\subset\Rd$ and a reference point $r\in\Rd$,
compute the hypervolume indicator of $\Sr$, \ie, $H(\Sr)$.
\end{problem}

\begin{problem}[\sc{\onecp}]\label{prob:onecp}
Given an $n$-point set $\Sr\subset\Rd$, a reference point $r\in\Rd$ and a point $p\in\Rd$,
compute the hypervolume contribution of $p$ to $\Sr$, \ie, $H(p,\Sr)$.
\end{problem}

\begin{problem}[\sc{\allcp}]\label{prob:allcp}
Given an $n$-point set $\Sr\subset\Rd$ and a reference point $r\in\Rd$, compute the hypervolume contributions
$H(p, \Sr)$
of all points $p\in\Sr$ to $\Sr$.
\end{problem}
%
%

\begin{problem}[\sc{\allcps}]\label{prob:allcps}
Given an $n$-point set $\Sr\subset\Rd$, an $m$-point set $\Rr\subset\Rd$ such that
$\Sr\cap\Rr = \emptyset$
and a reference point $r\in\Rd$,
compute the hypervolume contributions $H(p,\Rr)$ of all points $p\in\Sr$ to $\Rr$. 
\end{problem}

\begin{figure}[t]
\center
%
\hspace*{\fill}%
  \subfigure
    [$\allcp$ \newline $H(p^i,\Sr)$]
    {\label{fig:prob:allcp}\includegraphics[width=3.5cm,height=3.5cm]{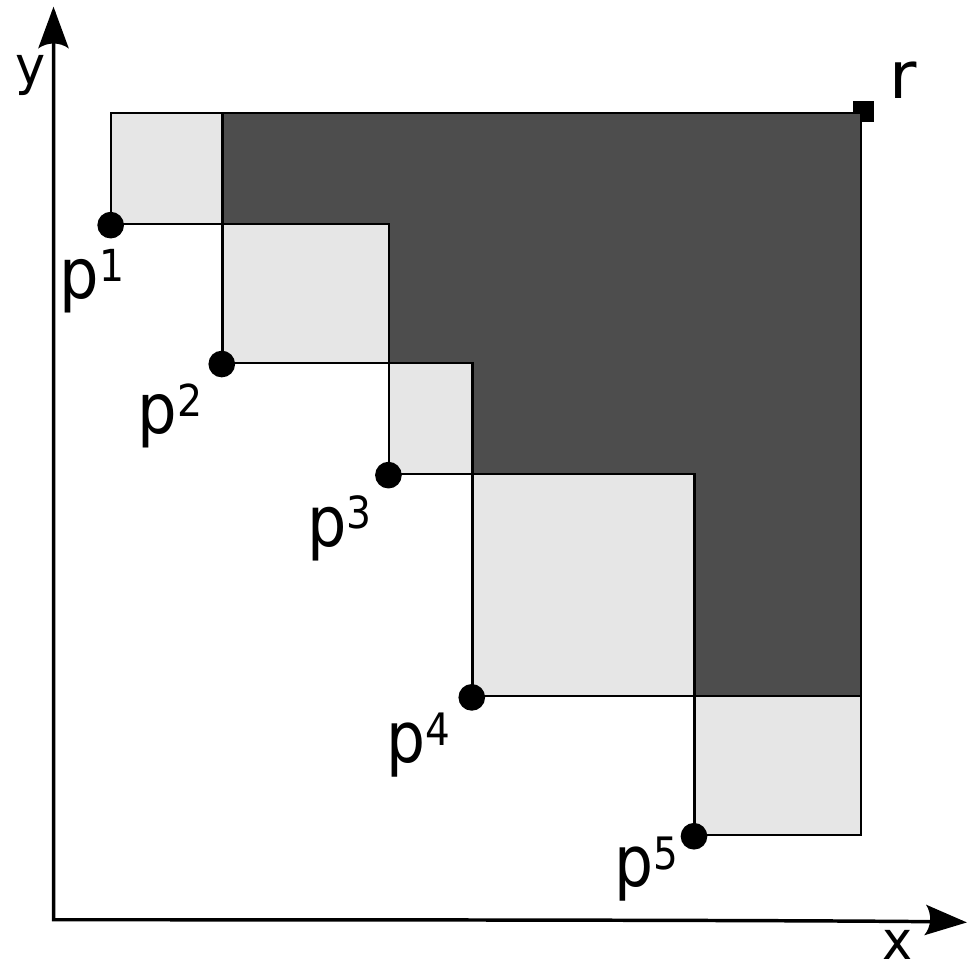}}
  \hspace*{\fill}%
  \subfigure
  [$\allcps$\newline  $H(p^i,\Rr)$]
  {\label{fig:prob:allcps}
                \includegraphics[width=3.5cm,height=3.5cm]{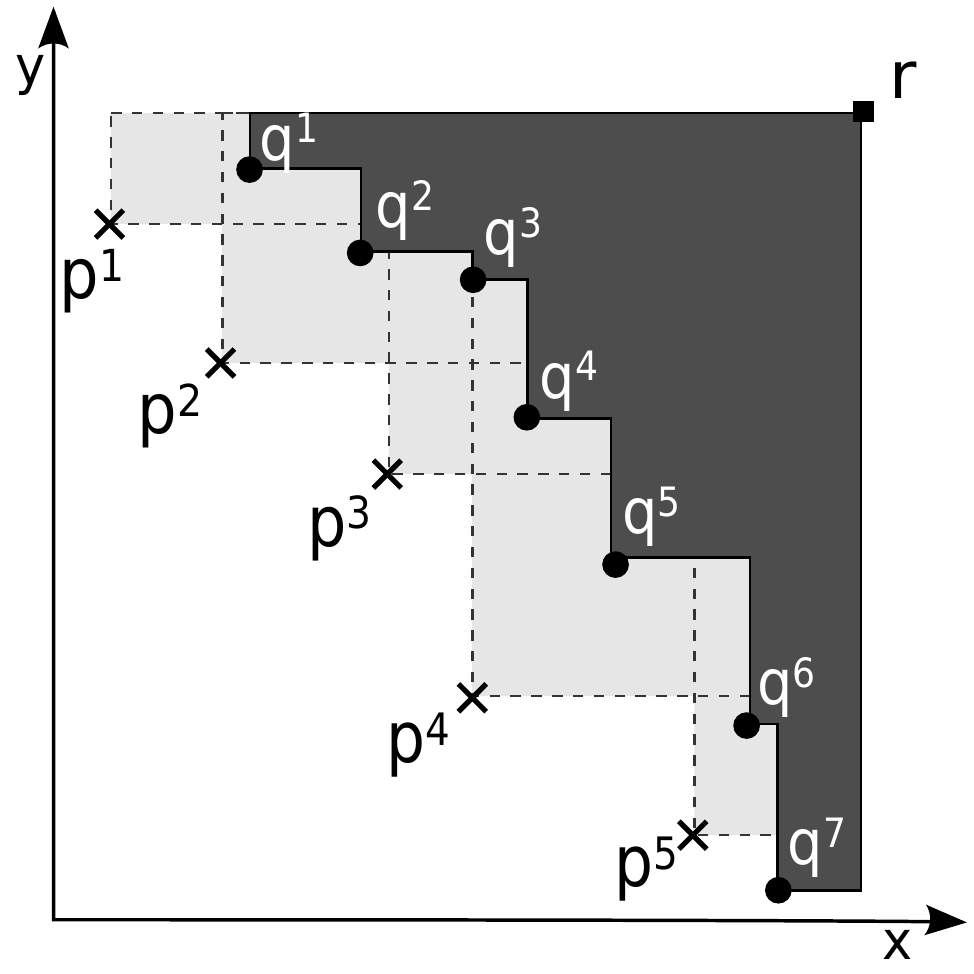}}
  \hspace*{\fill}%
  \subfigure
  [$\hssp$/$\hsspc$ $k=2$, $H(\{p^2,p^4\})$]
  {\label{fig:prob:hssp}
  \includegraphics[width=3.5cm,height=3.5cm]{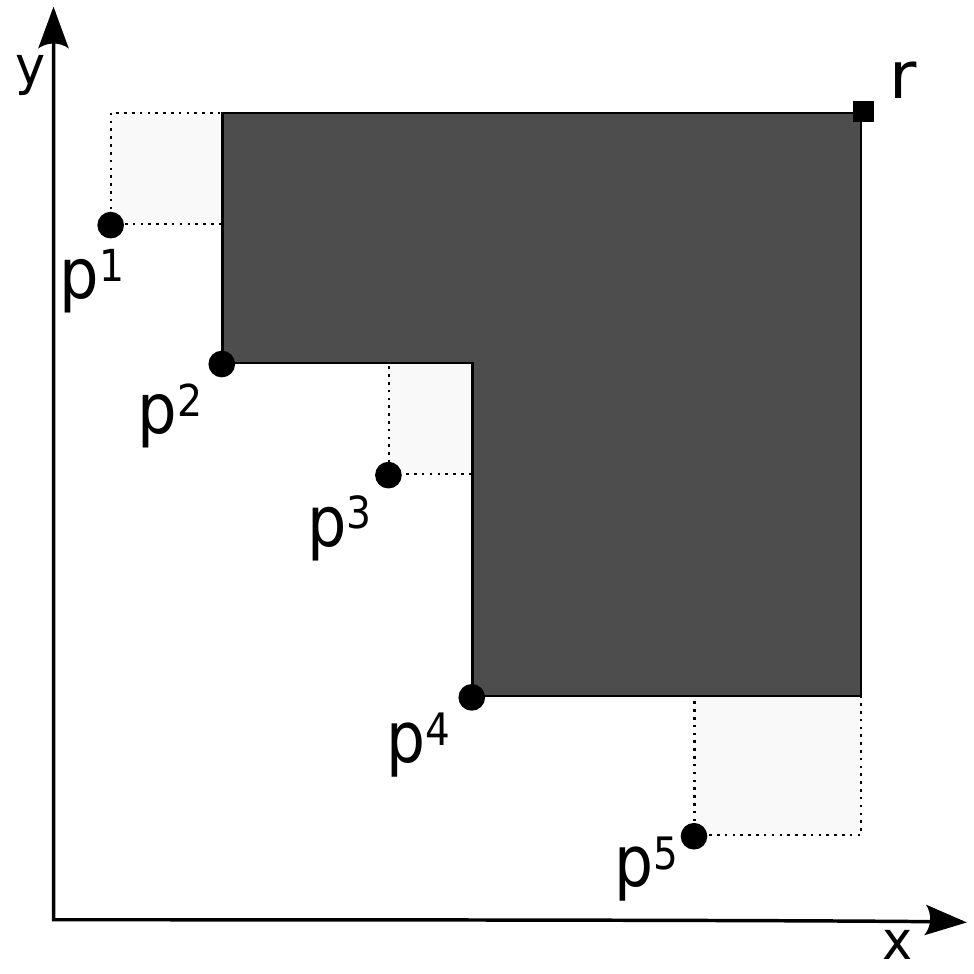}}
  \hspace*{\fill}%
  \vspace{-0.2cm}
  \caption[Examples of hypervolume-related problems]
  {Examples of hypervolume-related problems, given $\Sr=\{p^1,\ldots,p^5\}$, $\Rr=\{q^1,\ldots,q^7\}$, and $i=1,\ldots,5$.}
  \label{fig:prob}
\end{figure}

Note that, by definition, the contributions of two points $p,q\in\Sr$ to
$\Sr$ never overlap in Problem~\ref{prob:allcp} while in Problem~\ref{prob:allcps},
the contributions of $p,q\in\Sr$ to a point set $\Rr$ may overlap.
For example, in Figure~\ref{fig:prob:allcps}, given $\Sr=\{p^1,\ldots,p^5\}$ and $\Rr=\{q^1,\ldots,q^7\}$,
the contribution of the point $p^2\in\Sr$ to $\Rr$ (Problem~\ref{prob:allcps}) includes all the
lighter-gray regions dominated by $p^2$, including regions partially dominated by $p^1$, $p^3$, and $p^4$.
However, in Figure~\ref{fig:prob:allcp},
the contribution of $p^2$ to $\Sr$ (Problem~\ref{prob:allcp}) corresponds solely to
the respective exclusively dominated (lighter-gray) region.

\begin{problem}[\sc{\mincrp}]\label{prob:mincrp}
Given an $n$-point set $\Sr\subset\Rd$ and a reference point $r\in\Rd$, find a point $p\in\Sr$ with minimal
hypervolume contribution to $\Sr$. 
\end{problem}

Sometimes, the above problems are computed for a sequence of sets that
differ in a single point from the previous one,
either by adding a point to (Incremental case) or by removing a point from (Decremental case)
the previous set.
 
\begin{problem}[\sc{\hvupdp}]\label{prob:hvupdp}
Given an $n$-point set $\Sr\subset\Rd$, the reference point $r\in\Rd$, the value of $H(\Sr)$,
and a point $p\in\Rd$, compute:
    \begin{itemize}
        \item\textbf{Incremental:} $H(\Sr\cup\{p\})=H(\Sr)+H(p,\Sr)$, where $p\notin\Sr$.
        \item\textbf{Decremental:} $H(\Sr\setminus\{p\})=H(\Sr)-H(p,\Sr)$, where $p\in\Sr$.
    \end{itemize}
\end{problem}

\begin{problem}[\sc{\allcupdp}]\label{prob:allcupdp}
Given an $n$-point set $\Sr\subset\Rd$, a reference point $r\in\Rd$, the value of $H(q,\Sr)$ for every
$q\in\Sr$, and a point $p\in\Rd$:
    \begin{itemize}
        \item\textbf{Incremental:} Compute $H(q, \Sr\cup\{p\})=H(q, \Sr)-H(p,q, \Sr)$ for all $q\in\Sr$, and also $H(p,\Sr)$, where $p\notin\Sr$.
        \item\textbf{Decremental:} Compute \mbox{$H(q, \Sr\setminus\{p\})=H(q, \Sr)+H(p,q, \Sr)$}
        for all $q\in\Sr\setminus\{p\}$, where $p\in\Sr$.
    \end{itemize}
\end{problem}

%
\begin{problem}[\sc{\allcupdps}]\label{prob:allcupdps}
Given an $n$-point set $\Sr\subset\Rd$, an 
$m$-point set $\Rr\subset\Rd$,
a reference point $r\in\Rd$, the value of $H(q,\Rr)$ for every $q\in\Sr$,
and a point $p\in\Rd$:
    \begin{itemize}
        \item \textbf{Incremental:} Compute $H(q, \Rr\cup\{p\})=H(q, \Rr)-H(p,q, \Rr)$ for all $q\in\Sr$, where $p\notin\Rr \cup \Sr$. 
        \item \textbf{Decremental:} Compute $H(q, \Rr\setminus\{p\})=H(q, \Rr)+H(p,q, \Rr)$ for all $q\in\Sr$, where $p\in\Rr$ and $p\notin\Sr$.
    \end{itemize}
\end{problem}

Finally, based on the definition by~\cite{Bader2010}, 
the Hypervolume Subset Selection Problem (HSSP)
\footnote{Note that, the HSSP has also been defined in~\cite{HSSP2DWHypE2009} as 
the subset selection problem with respect to the Weighted Hypervolume Indicator
of which the hypervolume indicator is a special case.}
is formally defined here as:
\begin{problem}[\sc{\hssp}]\label{prob:hssp}
Given a $n$-point set $\Sr\subset\Rd$ and an integer $k\in\{0,1,\ldots,n\}$,
find a subset
$\Ar\subseteq\Sr$ such that $|\Ar| \leq k$ and:
\begin{equation*}
H(\Ar) = \max_{\substack{\Br\subseteq\Sr\\|\Br| \leq k}} H(\Br)
\end{equation*}%
\end{problem}%
The complement problem of the HSSP is defined as:
\begin{problem}[\sc{\hsspc}]\label{prob:hsspc}
Given a $n$-point set $\Sr\subset\Rd$ and an integer $k\in\{0,1,\ldots,n\}$,
find a subset $\Cr\subseteq\Sr$ such that $|\Cr| \geq (n-k)$ and:
\begin{equation*}
H(\Cr,\Sr) = \min_{\substack{\Br\subset\Sr\\|\Br| \geq (n-k)}} H(\Br,\Sr)
\end{equation*}%
\end{problem}%
If $\Ar\subseteq\Sr$ is a solution to the $\hssp$ given $k$ and $\Sr$, then $\Sr\setminus\Ar$
is a solution to the $\hsspc$, and vice-versa.
For example, in Figure~\ref{fig:prob:hssp}, given $\Sr=\{p^1,\ldots,p^5\}$ and $k=2$,
the optimal solution to the $\hssp$ is $\{p^2,p^4\}$, and the optimal solution
to the $\hsspc$ is $\{p^1,p^3, p^5\}$.

Note that, in the above problems, $\Sr$
is usually a nondominated point set, even though this is not mandatory. Any dominated
point $q\in\Sr$
has a null contribution to $\Sr$.
However, if $q$ is dominated by a single point $p\in\Sr$, then
the contribution of $p$ to $\Sr$ will be lower than what it would be if $q\notin\Sr$.
Moreover, the incremental scenarios of Problems~\ref{prob:hvupdp} to~\ref{prob:allcupdps}
explicitly require that $p\notin\Sr$ because the adopted definition of hypervolume
contribution does not handle adding a point to a set in which it is already included,
as discussed before, nor does it consider the multiset that would result from such an operation.
If such cases become relevant, the hypervolume contribution of repeated
points in a multiset should be considered to be zero.

\subsubsection{Relation between problems}\label{ss:stoa:algs:rmrks}

Most of the problems listed above are not expected to be efficiently solved for an arbitrary number
of dimensions $d$. 
For example, the $\hvp$~\citep{Bringmann2008} and $\onecp$~\citep{Bringmann2009approx} problems
are known to be \textbf{\#P}-hard.
Even deciding if a point is the least contributor is \textbf{\#P}-hard~\citep{Bringmann2009approx}.
Moreover, HSSP was recently shown to be \textbf{NP}-hard~\citep{Bringmann2017} for $d\geq3$.
Although these are not encouraging results for an arbitrary $d$,
this does not mean that efficient algorithms to compute the hypervolume-related problems exactly,
or to approximate the HSSP cannot be developed for a fixed and small $d$.
To develop such efficient algorithms, it is important to understand how the various hypervolume
problems relate to each other or if they arise as subproblems.

All problems above (Problems~\ref{prob:hvp} to~\ref{prob:hsspc}) are intrinsically related.
For most of them, it is possible to solve each one by solving one or more instances of the others.
Consequently, state-of-the-art algorithms frequently exploit these relations.

It is clear from Definition~\ref{def:hvc} that any algorithm that
computes $\hvp$ (Problem~\ref{prob:hvp})
can also be used to compute $\onecp$ (Problem~\ref{prob:onecp}).
In fact, by Definition \ref{def:dlmtr}, it can be computed considering only $p$ and its delimiters.
Moreover, the $\hvupdp$ problem (Problem~\ref{prob:hvupdp}) 
can be solved by computing either $\hvp$ given $\Sr\cup\{p\}$ or $\onecp$
given $\Sr$ and $p$.
On the other hand, the $\hvp$ problem can be computed by
solving a sequence of $\hvupdp$ problems as new points are added to a set.
For example, consider $\Sr=\{p^1,p^2,p^3\}$, the hypervolume $H(\Sr)$
can be computed as the sum $H(p^1,\{\})+H(p^2,\{p^1\})+H(p^3,\{p^1,p^2\})$,
more generally:
\begin{equation*}
H(\Sr) = \sum_{i=1}^{n} H(p^i, \{p^1,\ldots,p^{i-1}\}) = H(p^1,\{\})+H(p^2,\{p^1\})+\ldots+H(p^n,\{p^1,\ldots,p^{n-1}\})
\end{equation*}
where $\Sr=\{p^1,\ldots,p^n\}$.
In fact, when such points are sorted according to dimension $d$, then the sequence of
subproblems become $(d-1)$-dimensional problems, as exploited by the dimension-sweep approach
(see Section~\ref{algs:par:ds}).
Dedicated algorithms to solve $\hvupdp$ (and the other update problems)
can take advantage of previous calculations and
data structures to avoid redundant (re)computations and consequently, to save time.

It should also be clear that any algorithm that computes $\onecp$ can be used
to compute $\allcp$ (Problem~\ref{prob:allcp}), and vice-versa. 
Algorithms to solve $\allcp$ can also be used to solve $\mincrp$
(Problem~\ref{prob:mincrp}), by computing all contributions and then
selecting a point with minimal contribution, 
although it is not strictly required to know all contributions to find the least
contributor (see IHSO~\ref{HVC:IHSO} algorithm in Section~\ref{ch:hvstoa:algs:hvc}).
In the absence of dedicated algorithms, $\allcupdp$ (Problem~\ref{prob:allcupdp})
can be solved by recomputing all contributions ($\allcp$).

Analogously to Problem~\ref{prob:allcp}, algorithms to compute $\onecp$ can also be used to compute
$\allcps$ (Problem~\ref{prob:allcps}) problem. Moreover,
$\allcupdps$ (Problem~\ref{prob:allcupdps}) can also be solved by recomputing the
contributions ($\allcps$).
Despite the similarities, $\allcp$ and $\allcps$ are distinct to the point that one cannot 
be directly used to solve the other.
The same observation applies to the corresponding update problems ($\allcupdp$ and $\allcupdps$).

The definition of $\hssp$ suggests that it can be solved by enumerating 
all subsets of size $k$ and computing the hypervolume indicator of each one.
Similarly, $\hsspc$ can be computed in an analogous way.
However, this is obviously not practical as the time required would quickly be unacceptable
unless $n$ is sufficiently small and $k$ is either very small or close to $n$.
Moreover, recall that an optimal solution to the $\hssp$ can be obtained from an optimal solution
to $\hsspc$ and vice-versa.
For the particular case of $k=n-1$, the $\mincrp$ problem provides the solution
to the $\hssp$ by definition.

Finally, it is important to have in mind that the way problems are solved
has effect on the precision of the calculations.
As pointed out in~\cite{exQHV}, computing the contribution of a point
as the subtraction of two large hypervolumes raises precision problems.
It is thus recommended to avoid subtracting hypervolumes as much as possible and, ideally,
performing subtractions only between coordinates.
For example, regarding numerical stability,
using a specific algorithm to compute $\onecp$ may be preferable to
using an algorithm for $\hvp$ to compute the contribution based on a subtraction.

Section~\ref{s:hvstoa:algs} gives more details on the relation between 
hypervolume-based
problems by explaining the existing algorithms and the techniques used by them.

\subsection{Properties}\label{ch:hvstoa:props}

As an indicator that imposes a total order among point sets, the hypervolume indicator is
biased towards some type of distribution of the points on a front~\citep{HVIRevisted07}.
Understanding that bias by studying its properties
allows to better understand the underlying assumptions on DM preferences.
Among the most important properties are the monotonicity properties, sensitivity to objective
rescaling and parameter setting, and optimal $\mu$-distributions
(see~\cite{MNTNCT08} for an overview of properties of quality indicators).
In particular,
monotonic properties reflect the formal agreement between a binary relation on point sets and
the ranking imposed by a unary set-quality indicator. Given a binary relation
$R$ on point sets, 
a set-quality indicator $I$ (to be maximized)
is said to be weakly $R$-monotonic if
$\Ar R \Br$ implies $I(\Ar) \geq I(\Br)$ for any $\Ar,\Br\subset\Rd$,
and it is strictly $R$-monotonic
if $\Ar R \Br$ implies $I(\Ar) > I(\Br)$.
The study of optimal $\mu$-distributions describe how points in an indicator-optimal
subset of maximum size $\mu$ are distributed in the objective space given a known Pareto Front.

The hypervolume indicator is well acknowledged by its properties.
It is scaling independent
and it is strictly $\Strictdom$-monotonic~\citep{Zitz1erPhD99,Knowles2002PhD},
where $\Ar\Strictdom\Br$
($\Ar$ strictly dominates $\Br$) if and only if for every point $b\in\Br$
there is a point $a\in\Ar$ that weakly dominates it, but not the other
way around.

This implies that this indicator is maximal for the Pareto front~\citep{Guerreiro:PhD,Fleischer03}
and that scaling objectives
does not affect the order imposed on point sets.
Moreover, hypervolume-based selection methods provide
desirable (convergence) properties to EMOAs,
see~\cite{SeqOnlArch2011,ArchAlgsBF2014} for more details.

Concerning optimal $\mu$-distributions in two dimensions, 
the exact location of the points in an optimal subset of
a given size $\mu$ is known only for continuous linear fronts~\citep{Auger09HV}.
In this case, there is a unique optimal subset where all points are on the
Pareto front and uniformly spaced between two outer points, the position of
which depends both on the reference point and on the two extreme points of the
Pareto front~\citep{Dimo2010HV}. General fronts have also been studied, but only
in terms of point density on the Pareto front when the number of points $\mu$
tends to infinity~\citep{Auger09HV}. 
Unfortunately, the results available for the two-objective case do not
generalize easily to three objectives, and not much is known about optimal
$\mu$-distributions in this case.
The main results concern whether there exists a setting of the reference point that guarantees
the inclusion of a front's extreme points in the optimal
$\mu$-distribution~\citep{Auger10HV,HVmuDistr2014} and the derivation of the
optimal $\mu$-distributions for the special case of Pareto fronts consisting of
a line segment embedded in a three-objective space~\citep{HVmuDistr2014}.

Ulrich and Thiele~\cite{TT2012} showed that the hypervolume indicator is a
submodular function. Given a decision space $\X$ and a function $z:2^\X\rightarrow \R$,
$z$ is submodular if:
\begin{equation}
\forall\Ar,\Br\subseteq\X,\:\:  z(\Ar)+z(\Br) \geq z(\Ar\cup\Br) + z(\Ar\cap\Br)
\end{equation}
Submodularity is an important property as it relates to convexity in
combinatorial optimization~\citep{SMFConvex83}.
Additionally, a submodular function $z$ is non-decreasing (or monotone) if:
\begin{equation}
\forall\Ar\subseteq\Br\subseteq\X, \:\: z(\Ar) \leq z(\Br)
\end{equation}
The hypervolume indicator is a non-decreasing submodular function~\citep{TT2012}.
See~\cite{Nemhauser78} for alternative equivalent definitions of (non-decreasing) submodular functions
and examples.
Because the Hypervolume Subset Selection Problem (HSSP) consists of maximizing a submodular function
subject to a cardinality constraint~\citep{Friedrich2014},
the approximation of HSSP by means of a (incremental) greedy approach has an
approximation guarantee~\citep{Nemhauser78}.
Hence, the subset 
obtained by selecting $k$ points from $\Sr$ one at a time so as to maximize the hypervolume gained
at each step
is a $(1-1/e)$-approxima\-tion to the hypervolume of an optimal subset, \ie,
the ratio between the greedy solution and the optimal solution is greater than or equal
to $(1-1/e)\simeq 0.63$.
A tighter approximation bound for $k>\frac{n}{2}$ is known~\citep{Laitila2017},
which takes into account that in such case a greedy and an optimal solution must agree in $m$ points
where $m$ is at least $2k-n$.
The new bound relies both in $n$ and $k$ while the previous $(1-1/e)$ bound did not.
A weak but simple form of the new bound is: $1-\left(1-\frac{m}{k}\right)\left(1-\frac{1}{k}\right)^{k-m}$.
Moreover, an approximation guarantee of $\frac{k}{n}$ for the approximation of HSSP by means
of a decremental greedy approach was independently derived specifically for the HSSP~\citep{Guerreiro:PhD} and more generally for the maximization of monotone submodular functions
subject to a cardinality constraint~\citep{GEapprox2018}.
In this case the subset is obtained by discarding $n-k$ points from $\Xr$ one at a time so as to
minimize the hypervolume lost at each step.
Such approximation ratios do not extend to the $\hsspc$ counterpart
since the approximation ratio with either approaches can be arbitrarily large~\citep{Bringmann2010,HVCTEC2017}.

\section{Paradigms and Techniques}\label{s:hvstoa:algs}

This section is the first of four sections overviewing the state-of-the-art algorithms for the
hypervolume-related problems described in Section~\ref{ch:hvstoa:probs}.
The techniques and paradigms used in such algorithms are described first in this section.
The following two sections provide a detailed overview of the existing algorithms for
computing the hypervolume indicator (Section~\ref{ch:hvstoa:algs:hv}), 
and hypervolume contributions (Section~\ref{ch:hvstoa:algs:hvc}). The last of these four sections
(Section~\ref{ch:hvstoa:algs:hssp}) briefly overviews the existing exact and approximation algorithms
for the HSSP.
For a quick overview of the available and the fastest algorithms both with respect
to runtime and to asymptotic complexity, 
see the ``Remarks" sections: \ref{s:rmk:hv}, \ref{s:rmk:hvc}, \ref{s:rmk:hssp}, and
\ref{s:rmk:greedy}.
Links to available implementations are provided in Appendix~\ref{s:hvstoa:links}.

To make
the referencing of algorithms easier, the state-of-art algorithms are identified
by a name and also alphanumerically. A letter and a number (\eg [A2]) are assigned to each algorithm.
The letter identifies to which problem the algorithm relates to (A -- hypervolume indicator, B -- hypervolume
contributions, C -- HSSP, D -- (greedy) approximation to the HSSP)
and the number is assigned according to the
order in which they are introduced in the following sections. For example, algorithm [B4] is the
fourth algorithm related to hypervolume contributions (Section~\ref{ch:hvstoa:algs:hvc}).

\subsection{Paradigms}\label{ch:hvstoa:algs:parad}

\begin{figure}[t]
  \center
\begin{tabular}{ccc}
  \subfigure[3D example]{\label{fig:hvstoa:DS:3D}
    \begin{overpic}[width=3.8cm,height=3.8cm]{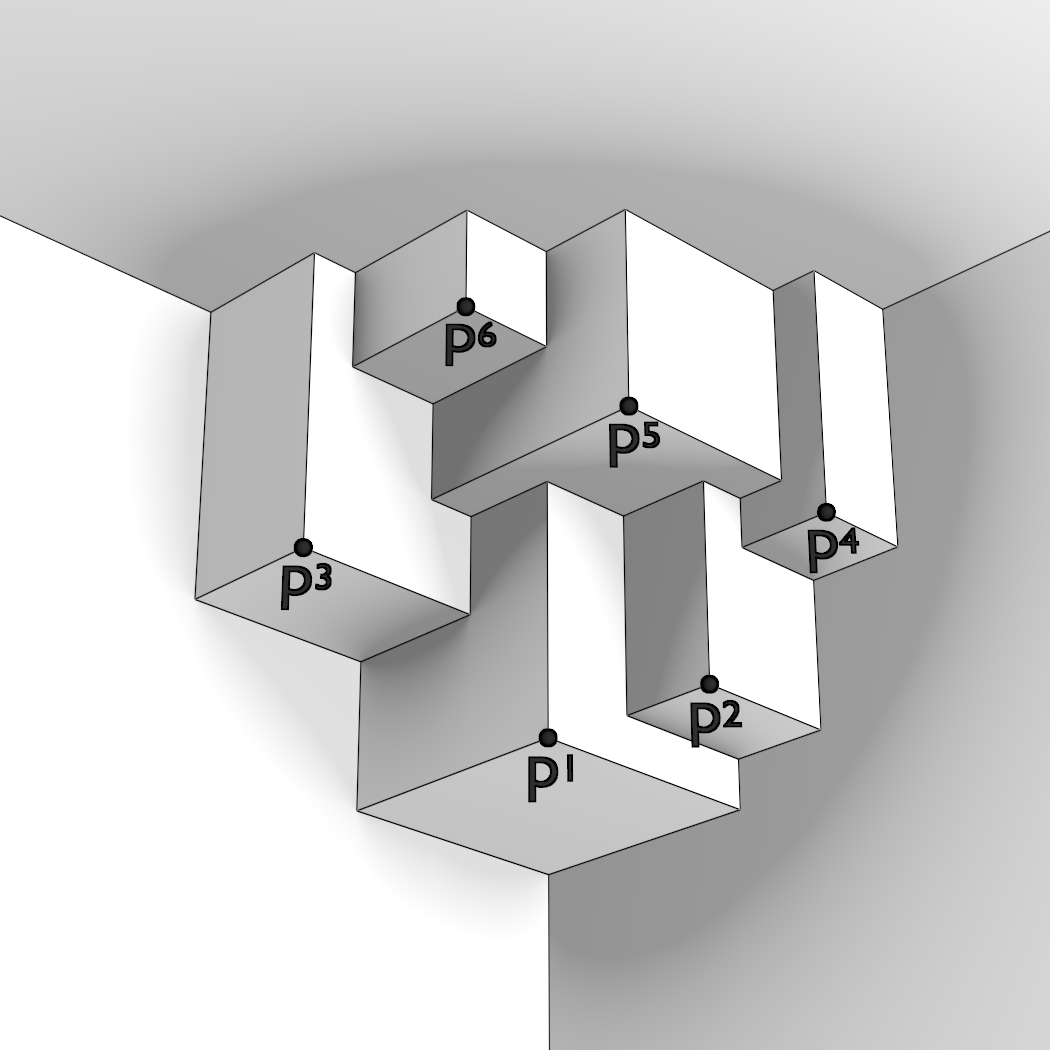}
        \put(3,3){\includegraphics[width=0.85cm,height=0.8cm]{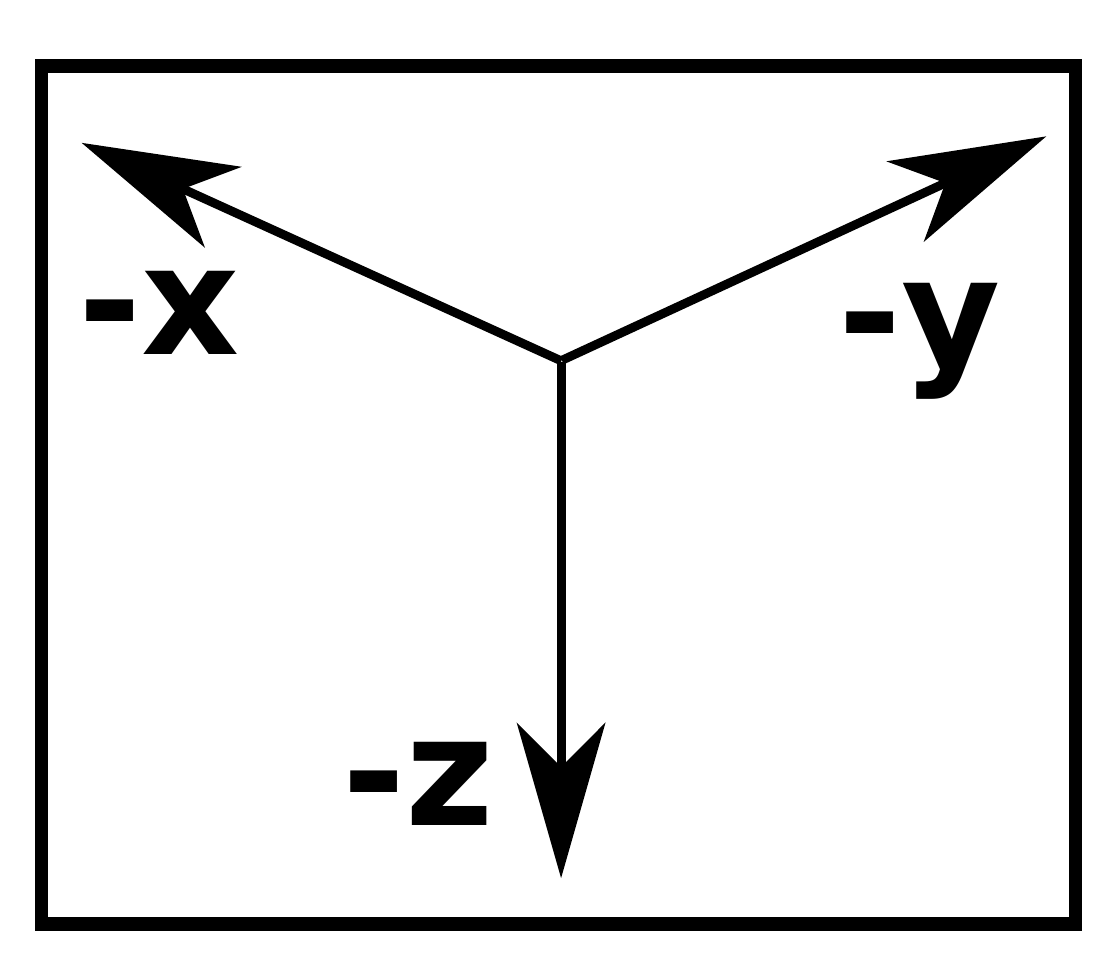}}  
    \end{overpic}
  }
  \hspace{0.5cm}
  &\subfigure[2D projection at $z=\cz{r}$]{\label{fig:hvstoa:DS:2D}\includegraphics[width=3.8cm,height=3.8cm]{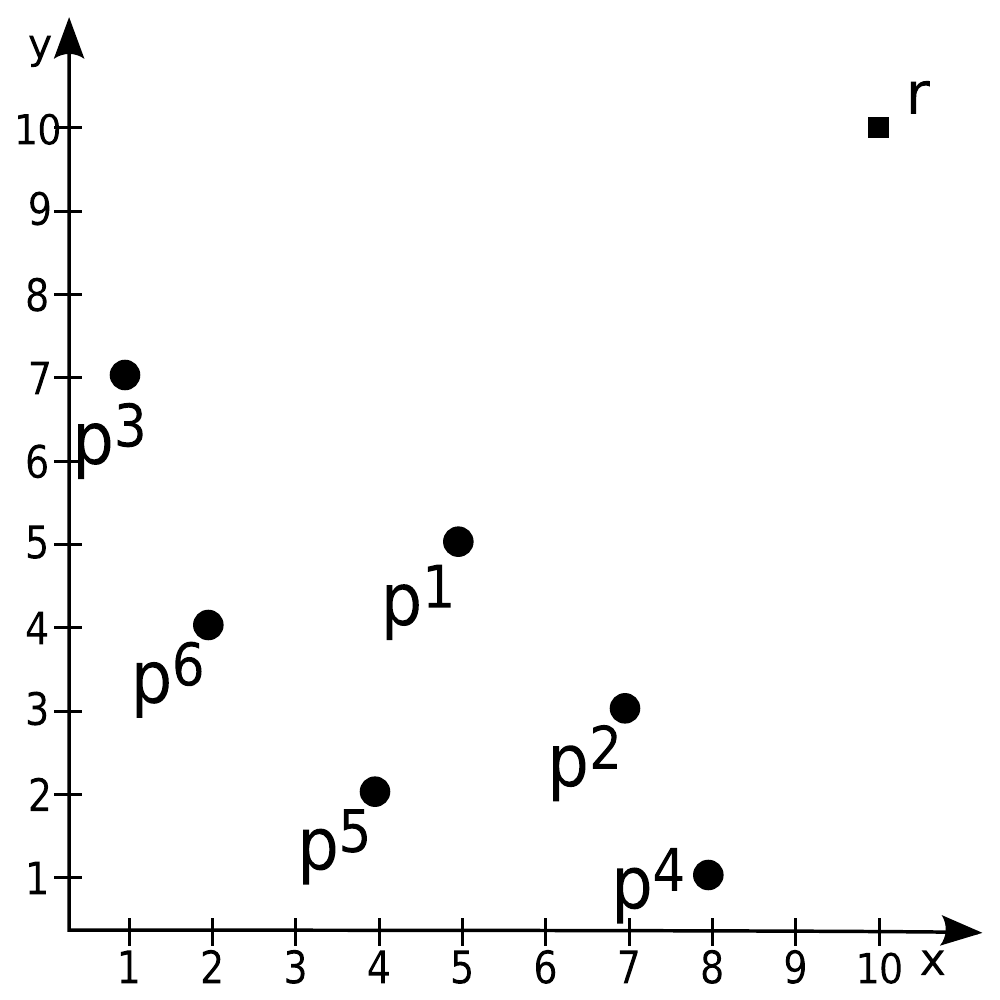}}
  \hspace{0.5cm}
   &
\raisebox{1.7cm}{
   \begin{tabular}{c|c}
        \small
        $p$ & point \\
        \hline
        $p^1$ & $(5,5,1)$ \\
        $p^2$ & $(7,3,2)$ \\
        $p^3$ & $(1,7,4)$ \\
        $p^4$ & $(8,1,5)$ \\
        $p^5$ & $(4,2,6)$ \\
        $p^6$ & $(2,4,8)$ \\
        \hline
    \end{tabular}
    }
\end{tabular}
  \caption[3D hypervolume: base example]
  {Three-dimensional base example and the corresponding projection on the $(x,y)$-plane. The reference point is $r=(10, 10, 10)$.}
  \label{fig:hvstoa:DS}
    \vspace{-12pt}
\end{figure}

Figure~\ref{fig:hvstoa:DS} shows a $3$-dimensional example that is used for illustration purposes
throughout this section. Moreover, $n$ is used to represent the input size.

\subsubsection{Inclusion-Exclusion Principle}
\label{algs:par:iep}
The inclusion-exclusion principle is a technique consisting of sequentially iterating
over an inclusion step followed by an exclusion step, where the $i^{th}$ step
involves some computation for every combination of $i$ points.
This technique is discussed in~\cite{IEHV} for the $\hvp$ problem:
The hypervolume indicator of a set $\Sr\subset\Rd$ of $n$ points is the sum
of the hypervolume (indicator) of every subset with a single point, minus the
hypervolume of the component-wise maximum of each pair of points,
plus the hypervolume of the component-wise maximum of each subset with three points,
minus the hypervolume of the component-wise maximum of each subset of
four points, and so on. This technique can be very inefficient,
if applied as explained above it is exponential in the number of points, $\Theta(2^n)$.

\subsubsection{Dimension Sweep}\label{ss:stoa:DS}
\label{algs:par:ds}

Dimension sweep~\citep{Preparata} is a paradigm which has been widely used in the development of algorithms for hypervolume-related problems (\eg~\ref{HV:HSO},
\ref{HV:WFG}, \ref{HV:FPL}, \ref{HV:HV3D}, \ref{HV:HV4D}).
A problem involving $n$ points in $\R^d$ is solved with this paradigm
by visiting all points in
ascending (or descending) order of one of the coordinates, solving a
$(d-1)$-dimensional subproblem for each point visited, and combining the
solutions of those subproblems. The subproblems themselves can often be solved
using dimension sweep as well, until a sufficiently low-dimensional
base case
is reached, which can be solved easily by a dedicated algorithm.
However, the time complexity of the resulting algorithms typically increases by an $O(n)$
factor per dimension.

A typical dimension-sweep algorithm for $\hvp$ problem works as follows. 
Input points are sorted and visited in ascending order of the last
coordinate. The $d$-dimensional dominated region is partitioned into $n$ slices
by axis-parallel cut hyperplanes defined by the last coordinate value of each
input point and the reference point. The desired hypervolume indicator value is
the sum of the hypervolumes of all slices, and the hypervolume of a slice is the
hypervolume of its $(d-1)$-dimensional base multiplied by its height. The base
of a slice is the $(d-1)$-dimensional region dominated by the projection of the
points below it according to dimension $d$ onto the corresponding cut
hyperplane. The height of a slice is the difference between the values of
the last coordinate of two consecutive points.

\begin{figure}[h]
  \center
  \subfigure[]{\label{fig:allslices}
                                \includegraphics[width=3.5cm,height=3.5cm]{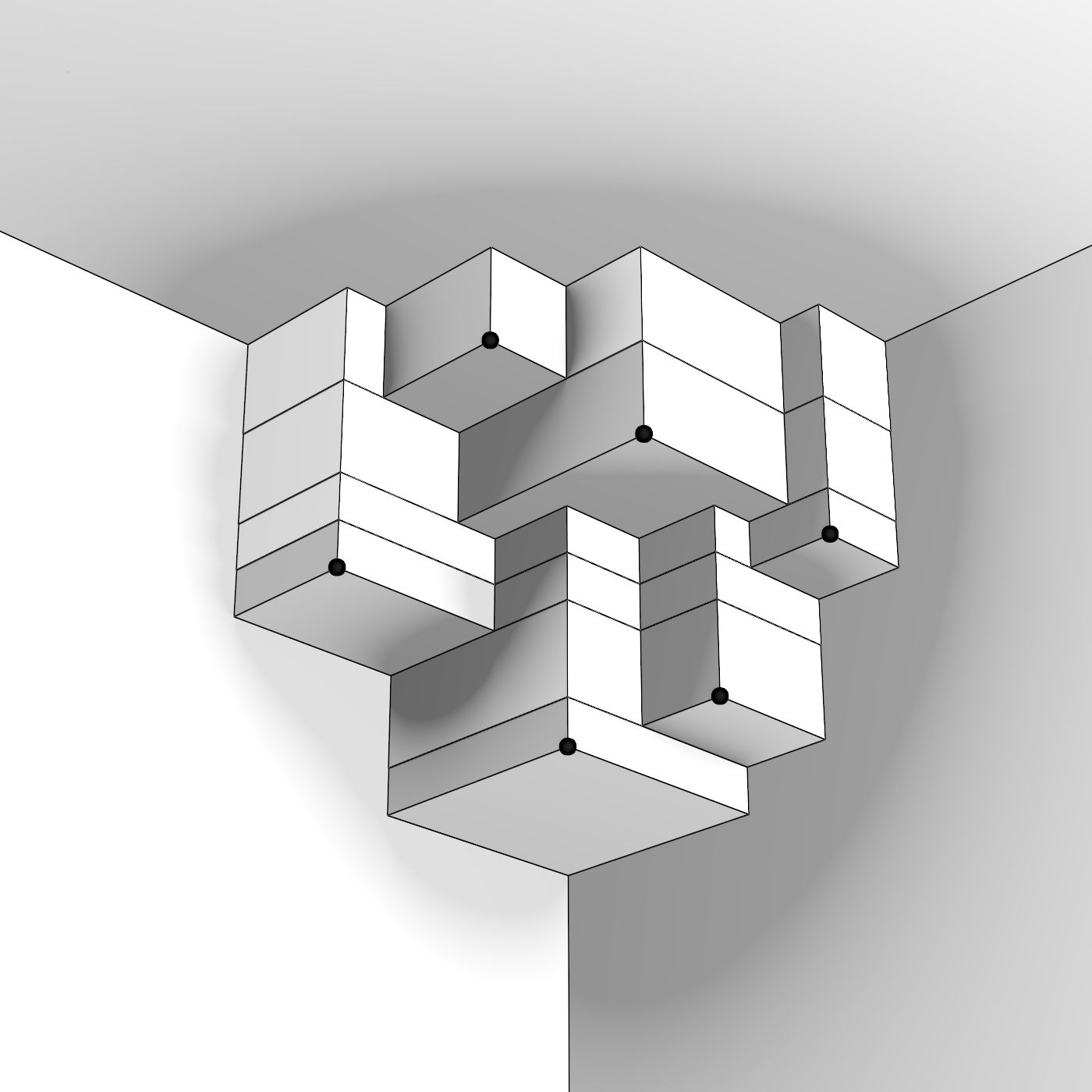}}
  \subfigure[]{\label{fig:allslices:s6}
                                \includegraphics[width=3.5cm,height=3.5cm]{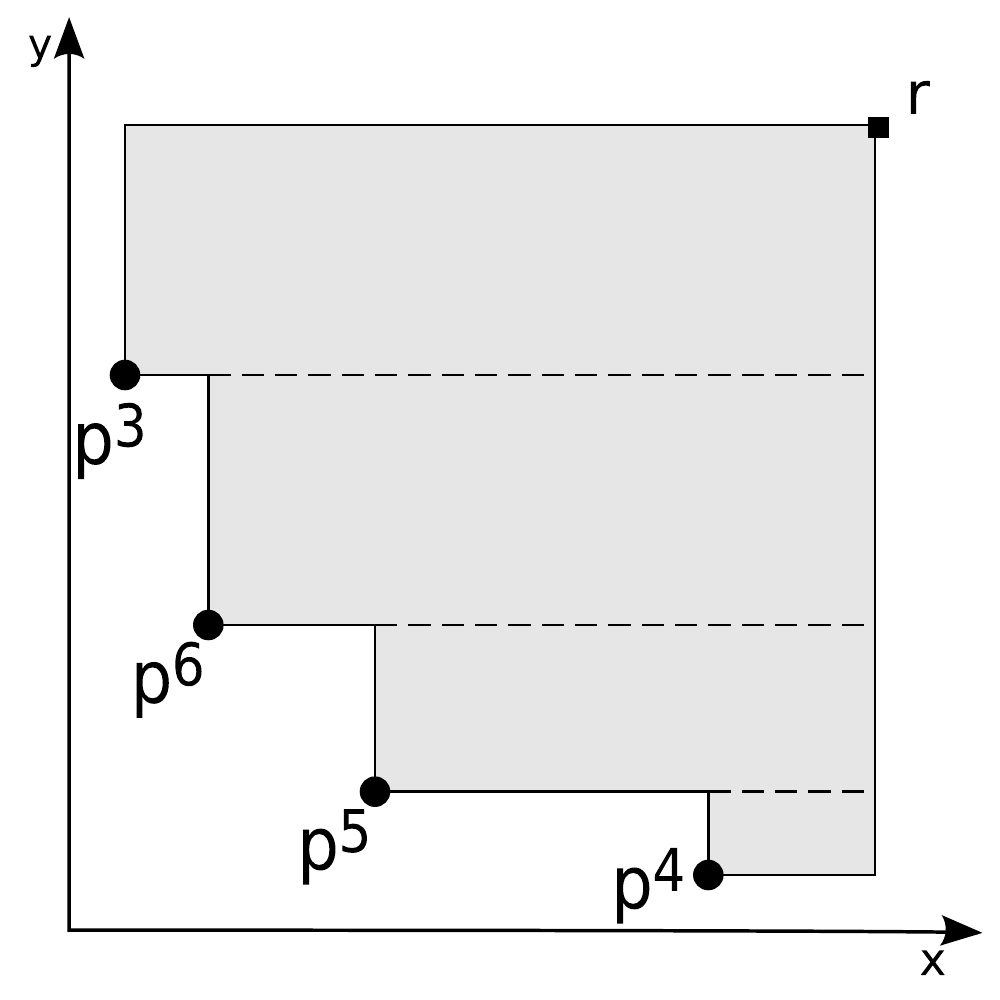}}
  \caption[Slice division of a volume]{Example of: \subref{fig:allslices} the slice division of
  the volume in Figure~\ref{fig:hvstoa:DS:3D} and; ~\subref{fig:allslices:s6} of the
  area of the corresponding topmost slice.
    }
    \vspace{-6pt}
\end{figure}

Figure~\ref{fig:allslices} exemplifies how the volume in the example of Figure~\ref{fig:hvstoa:DS}
could be split into $n=6$ (horizontal) slices. 
This splitting and the computation of the volume for
this example is further detailed in the explanation of HV3D (see~\ref{HV:HV3D}) in
Section~\ref{ch:hvstoa:algs:hv}.
Figure~\ref{fig:allslices:s6} shows a splitting of the base area of the topmost slice in Figure~\ref{fig:allslices}.

The algorithms for hypervolume-based problems using dimension sweep differ mostly 
in the $(d-1)$-dimensional subproblem considered. For example, for the $\hvp$ problem,
one option would be to compute
the $(d-1)$-dimensional hypervolume indicator of the base of the slice from scratch,
while
another would be to
avoid the full computation by updating the hypervolume of the base of the previous slice.

\subsubsection{Spatial Divide-and-Conquer}
\label{algs:par:sdc}

\begin{figure}[h]
  \center
  \subfigure[First subproblem]{\label{fig:hvstoa:sdc:lower}\includegraphics[width=3.4cm,height=3.4cm]{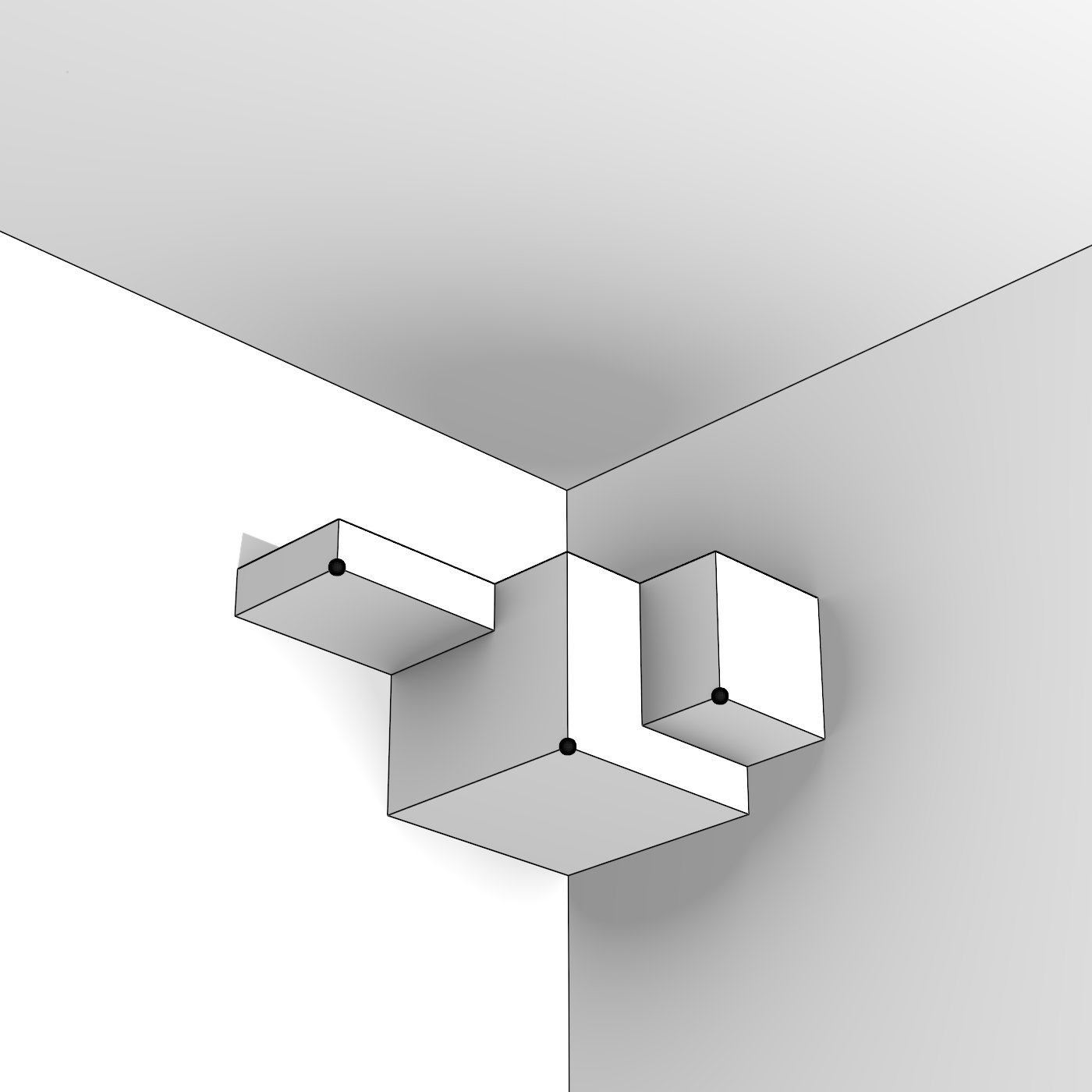}}
  \hspace{0.5cm}
  \subfigure[Second subproblem]{\label{fig:hvstoa:upper}\includegraphics[width=3.4cm,height=3.4cm]{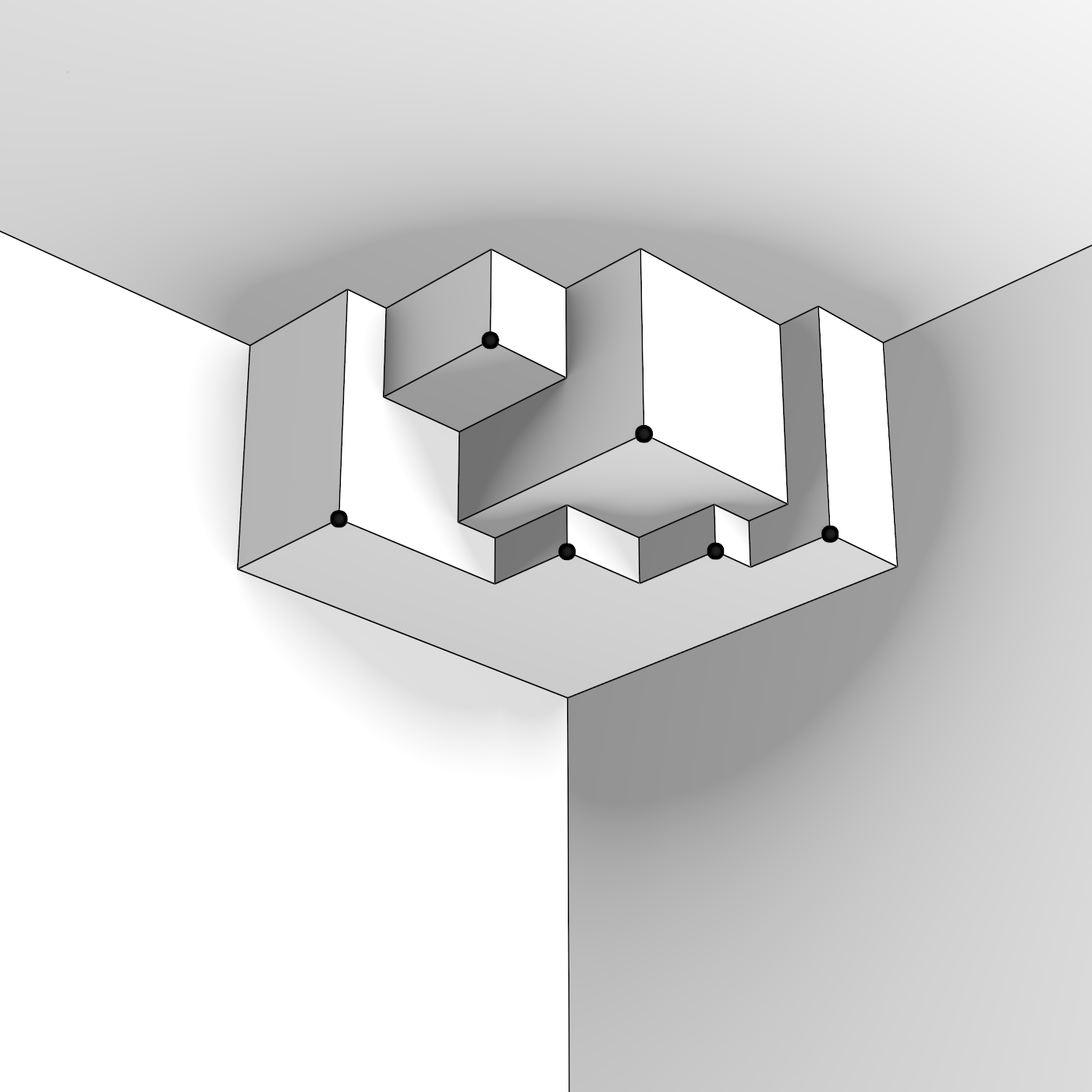}}
  \hspace{0.5cm}
  \caption[Volume division with SDC]{Example of the volume division in Figure~\ref{fig:hvstoa:DS:3D} using the spatial divide-and-conquer approach.}
  \label{fig:hvstoa:sdc}   
\end{figure}

This technique consists of splitting the $d$-dimensional
hypervolume into two (or more)
parts and recursively solving each part until a problem easy to solve is reached.
For example, the problem may be
split into two subproblems according to the median point, $p$, of a given
dimension $i$, \ie, the $\lceil (n+1)/2\rceil^{th}$ point with lowest coordinate $i$.
The axis parallel hyperplane at the value $\cd{p}{i}$ of coordinate $i$
divides the hypervolume in two parts.
The first part refers to a subproblem containing the $\lfloor n/2\rfloor$ points below $p$ in
the $i^{th}$ coordinate and the reference point is the one of the current problem but projected
on the splitting hyperplane.
The second subproblem contains all $n$ points, but the $\lfloor n/2\rfloor$ points
below $p$ in the  $i^{th}$ coordinate are projected onto the splitting hyperplane.
Figure~\ref{fig:hvstoa:sdc} shows an example of these two subproblems where coordinate $i=3$
is used for splitting and the median point $p^4$ is the splitting point $p$.

This approach is used, for example, by HOY algorithm (see~\ref{HV:HOY} in Section~\ref{ch:hvstoa:algs:hv}) to compute $\hvp$ and by Bringmann and Friendrich's algorithm
(see~\ref{HVC:BF1} in Section~\ref{ch:hvstoa:algs:hvc}) to compute \allcp.
An orthogonal partition tree is typically used as the underlying data structure,
as in the mentioned algorithms.
In that case, the hyperrectangle bounded below by the component-wise minimum of the initial point set
and above by the reference point is recursively partitioned in axis parallel regions,
and each one of them is associated to a node.
Moreover, the typical base case of the recursion (a leaf of the partition tree)
occurs when a partition consisting of a trellis is reached.
A trellis is a region (hyperrectangle) where every point in the subproblem dominates that region in all coordinates except one~\citep{HOY}.

The Multidimensional Divide-and-Conquer~\cite{MDC} is a different type of divide-and-conquer that divides
the problem into two $d$-dimensional subproblems of size $n/2$ and one $(d-1)$-dimensional subproblem
of size $n$ (the merge step).
This paradigm is mentioned only for completeness, as
the $O(n\log n)$ HVDC3D~\citep{Guerreiro:MSc} algorithm for the $d=3$ case of the $\hvp$ problem
appears to be the only algorithm for hypervolume-based problems based on this paradigm.

\subsection{Techniques}
\label{ch:hvstoa:algs:tech}

\subsubsection{Bounding Technique}
\label{algs:tec:bt}

This technique consists of
projecting points onto the surface of an axis
parallel $d$-dimensional box~\citep{WFG09:contrib,Bringmann2009pre2010}
(bounding step)
and discarding the
points that become dominated (filtering step), in $d$-dimensional space.
In practice, it consists of determining the auxiliary set $\Jr$ in the definition of delimiter
(see Definition~\ref{def:dlmtr}), where $\Jr$ is the smallest set of points weakly dominated by $p$ that
delimits its contribution.
It is used in several algorithms~\cite[\eg][]{WFG2011, IWFG, EMO15:GPUHV},
in particular, to compute hypervolume contributions.
Note that, to explicitly compute $\Jr$, the bounding step requires $O(n)$ time,
and the filtering step can be performed in $O(n \log^{\max(1,d-2)} n)$ time~\citep{MDC}.

\begin{figure}[h]
  \center
  \subfigure[Hypervolume Contribution]{\label{fig:hvstoa:bd:pre}\includegraphics[width=3cm,height=3cm]{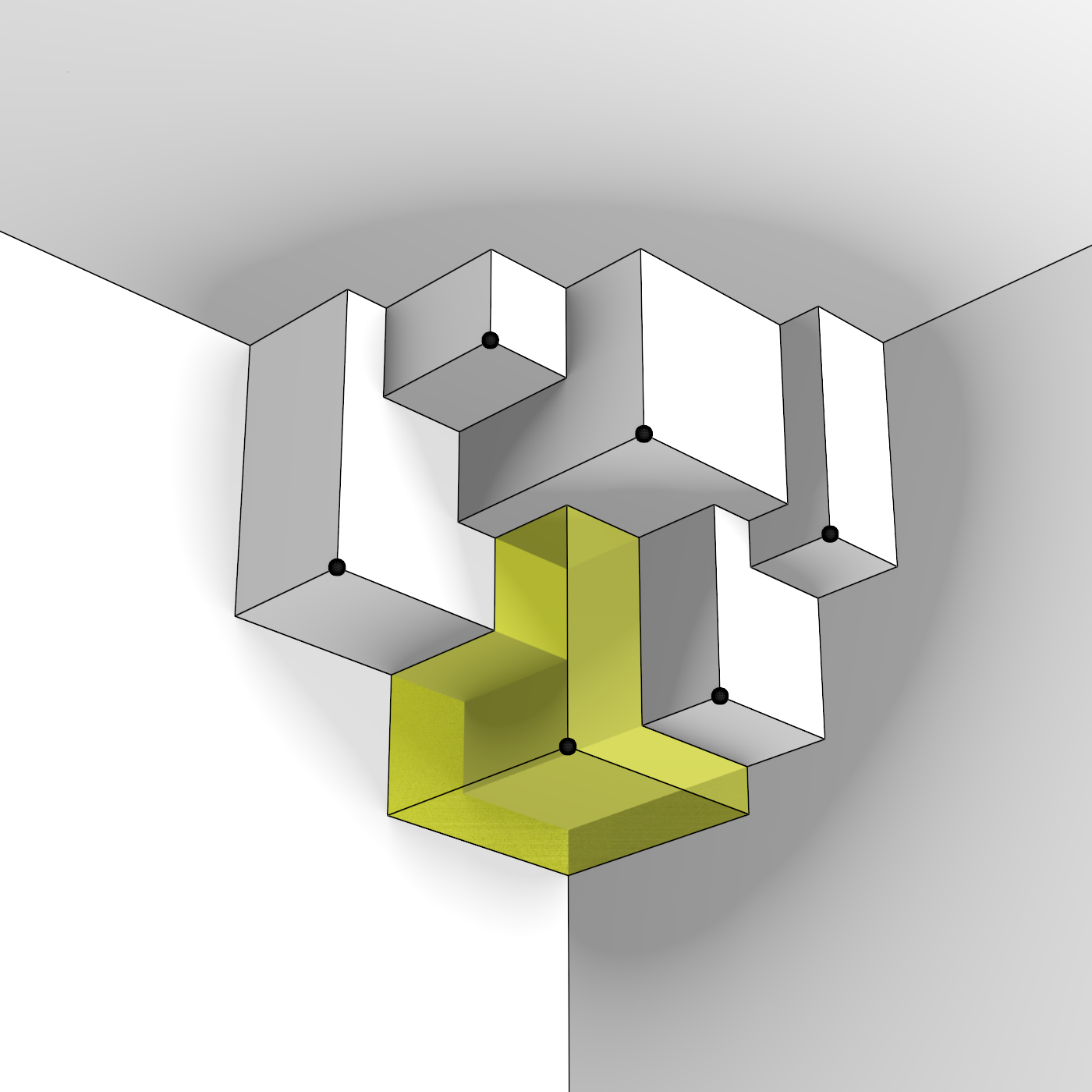}} 
  \hspace{0.5cm}
  \subfigure[Bounding]{\label{fig:hvstoa:bd:proj}\includegraphics[width=3cm,height=3cm]{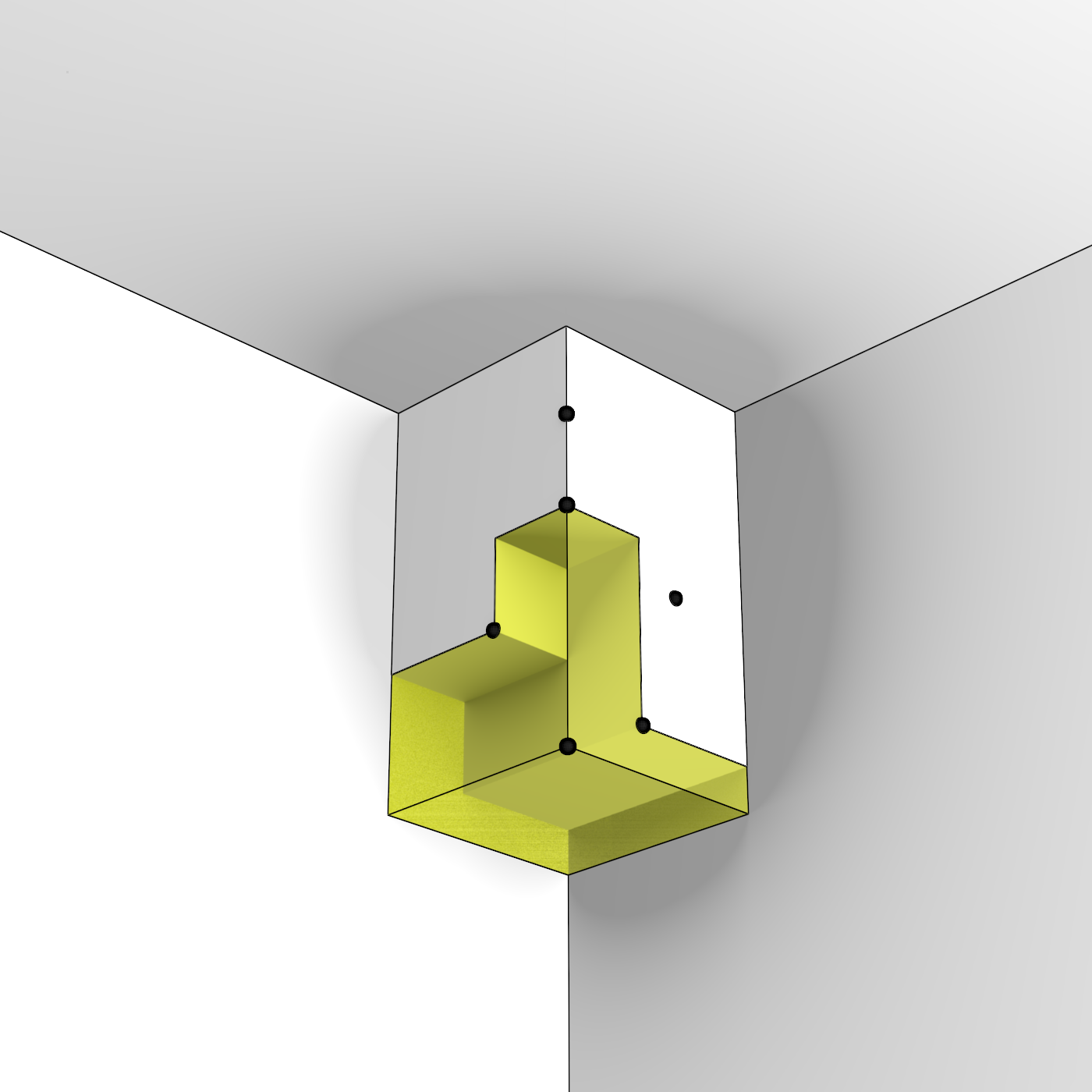}}
  \hspace{0.5cm}
  \subfigure[Filtering]{\label{fig:hvstoa:nd}\includegraphics[width=3cm,height=3cm]{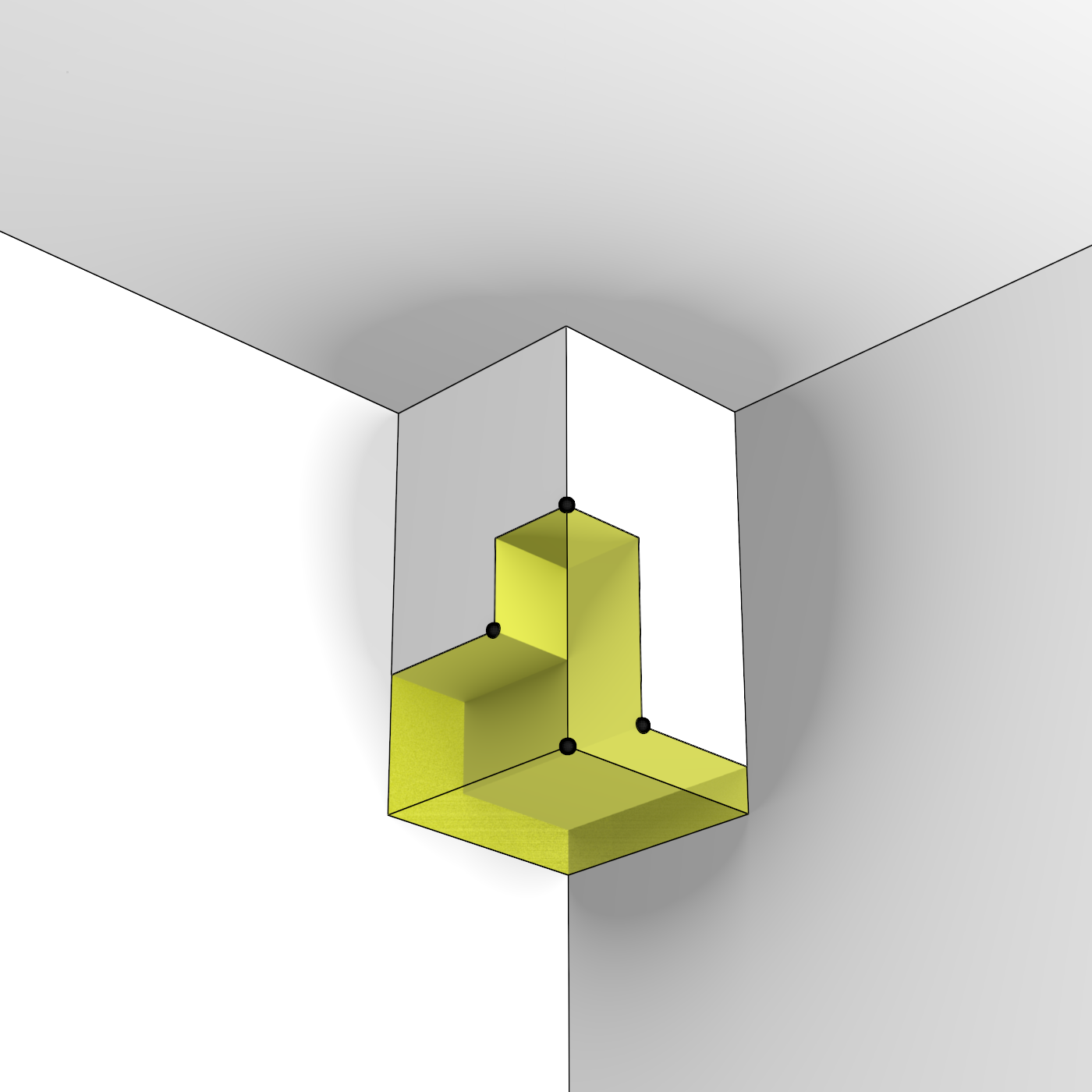}}
  \vspace{-0.2cm}
  \caption[Example of the bounding technique]
  {Example of the bounding technique for the contribution of $p^1$, in Figure~\ref{fig:hvstoa:DS:3D}.}
  \label{fig:hvstoa:bound}
\end{figure}

See Figure~\ref{fig:hvstoa:bound} for an example. Figure~\ref{fig:hvstoa:bd:pre} shows the hypervolume of
the point set
$\Sr=\{p^2,\ldots,p^6\}$ from Figure~\ref{fig:hvstoa:DS} and the
hypervolume contribution of $p^1$ (in transparent yellow).
Figure~\ref{fig:hvstoa:bd:proj} shows the bounding step, 
where all points (not dominated by $p^1$) are projected on the surface of the region dominated by $p^1$,
and where the gray axis-parallel box shown is the volume dominated by such projections.
Only the nondominated points among those projected points are kept (see Figure~\ref{fig:hvstoa:nd}). 

This technique is used in the computation of hypervolume contributions because these projections are
enough to delimit the contribution of $p^1$ and because the absolute position of the delimiters
resulting in those projections is irrelevant, \ie, have no influence on the contribution of $p^1$.
Moreover, this bounding technique allows to further discard the points that are not delimiters
of the contribution of $p^1$ as they are unnecessary to compute it.

\subsubsection{Objective Reordering}
\label{algs:tec:or}

\begin{figure}[h]
    \vspace{-8pt}
  \center
  \subfigure[$(x,y,z)$]{\label{fig:hvstoa:ord:xyz}
      \begin{overpic}[width=0.24\linewidth]{{HV-labeled-xyz}}
     \put(3,3){\includegraphics[width=0.95cm,height=0.9cm]{coordSystem}}  
    \end{overpic}
    }
  \hspace{0.5cm}
\subfigure[$(y,z,x)$]{\label{fig:hvstoa:ord:yzx}
      \begin{overpic}[width=0.24\linewidth]{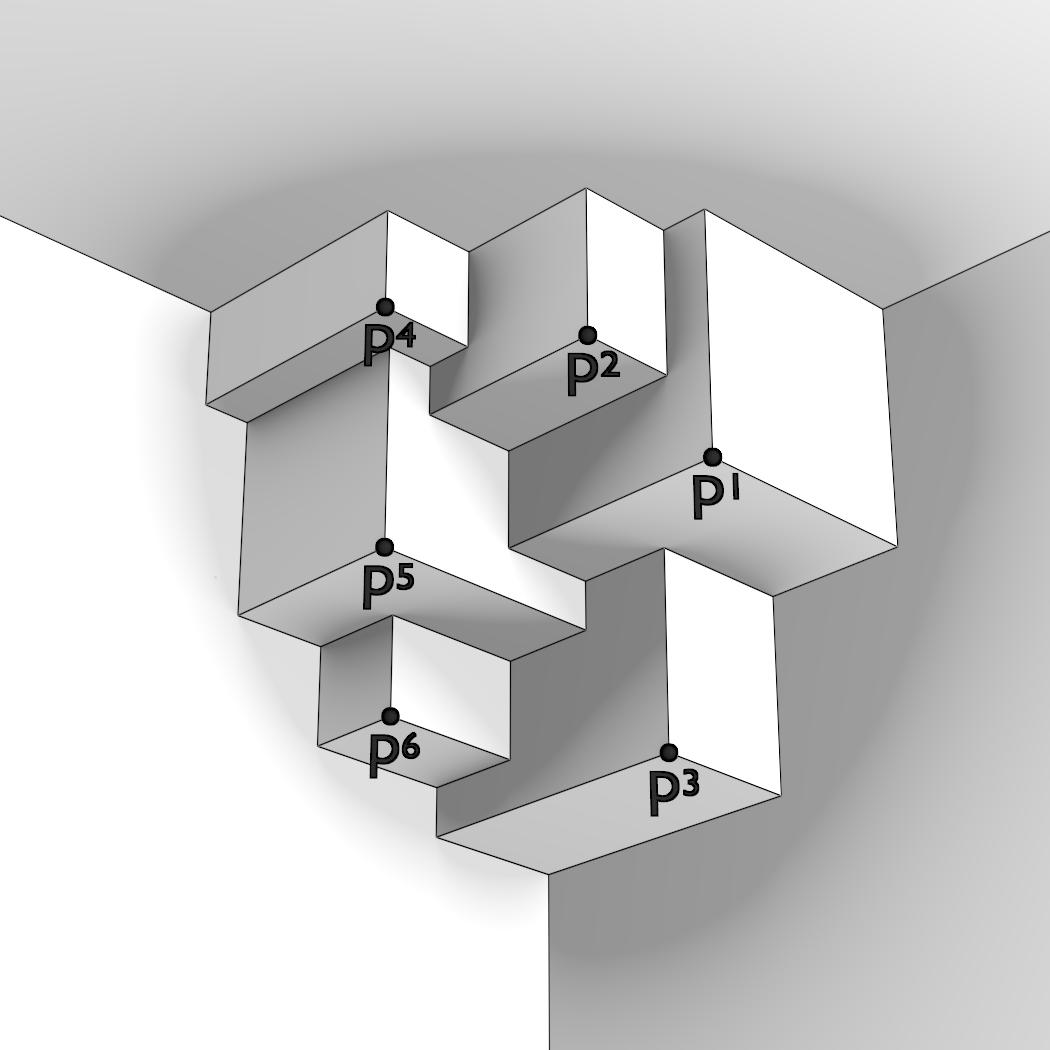}
     \put(3,3){\includegraphics[width=0.85cm,height=0.8cm]{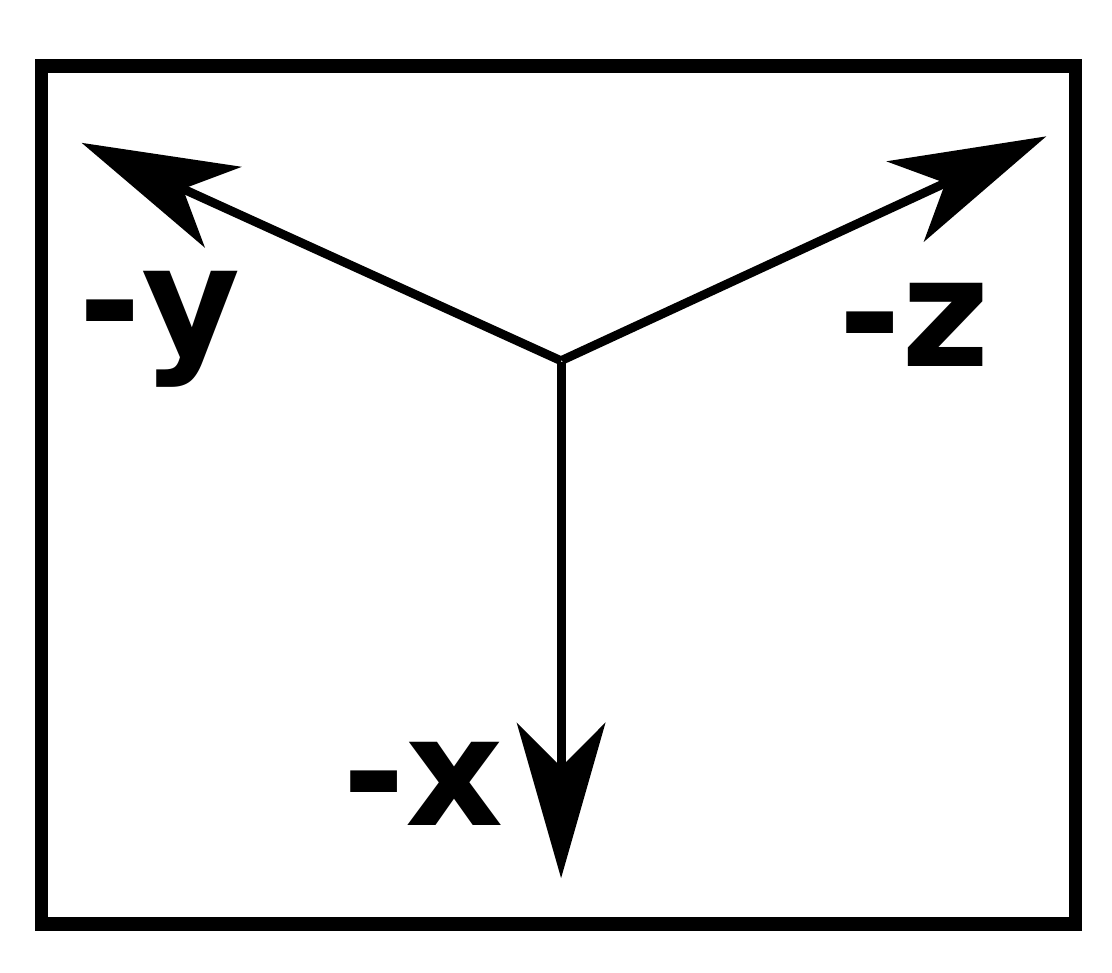}}  
    \end{overpic}
    }
  \hspace{0.5cm}
  \subfigure[$(z,x,y)$]{\label{fig:hvstoa:ord:zxy}
      \begin{overpic}[width=0.24\linewidth]{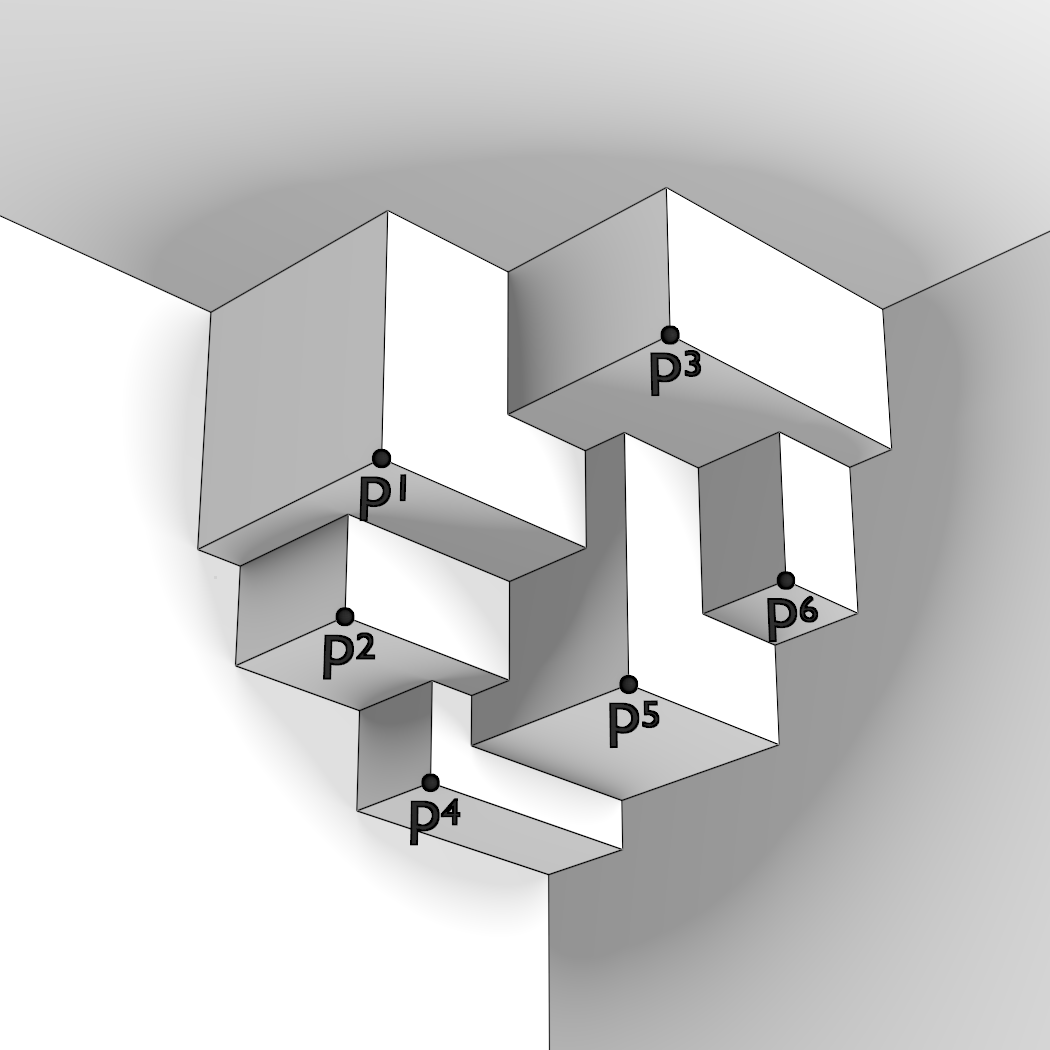}
     \put(3,3){\includegraphics[width=0.85cm,height=0.8cm]{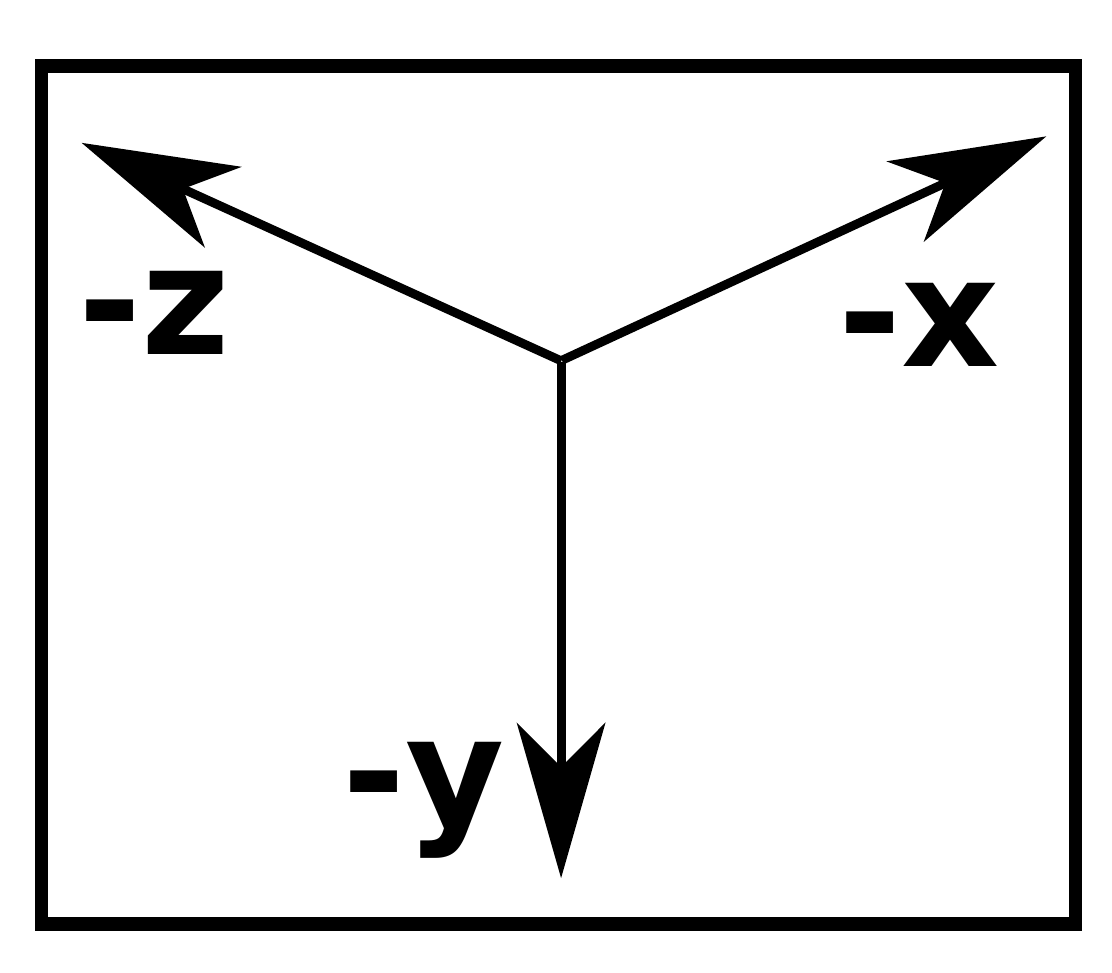}}  
    \end{overpic}
    }
  \vspace{-0.3cm}
  \caption[Objective reordering]
  {Example of Figure~\ref{fig:hvstoa:DS:3D} considering three of the six objective order possible.}
  \label{fig:hvstoa:ord}
    \vspace{-10pt}
\end{figure}

Problems in computational geometry such as those related to the hypervolume indicator are invariant
with respect to objective reordering and, in particular, to the case where it is equivalent
to rotation.
See the example in Figure~\ref{fig:hvstoa:ord} of a volume 
in $(x,y,z)$ coordinate system (see Figure~\ref{fig:hvstoa:ord:xyz})
and its counter-clockwise (see Figure~\ref{fig:hvstoa:ord:yzx}) and
clockwise rotations (see Figure~\ref{fig:hvstoa:ord:zxy}), corresponding to the $(y,z,x)$ and $(z,x,y)$ coordinate systems, respectively.
Reordering objectives is important since, although the result of the computation does not change,
it can have an impact on the implementation runtime~\citep{WFGreorder}.
For example, dimension-sweep approaches are usually sensitive to objective ordering as a particular
order may result in more dominated points further in the recursion than others.
While \etal~\cite{WFGreorder} proposed several heuristics to determine the best objective order to consider.
Algorithms such as WFG (see Section~\ref{ch:hvstoa:algs:hv}) benefit with the integration of such heuristics,
by repeating it in several steps of the computation of the hypervolume indicator, 
which results in fast algorithms in practice in many data set instances.

\subsubsection{Local Upper Bounds}
\label{algs:tec:lub}
\begin{figure}[h]
  \vspace{-0.2cm}
  \center
  \includegraphics[width=0.20\linewidth]{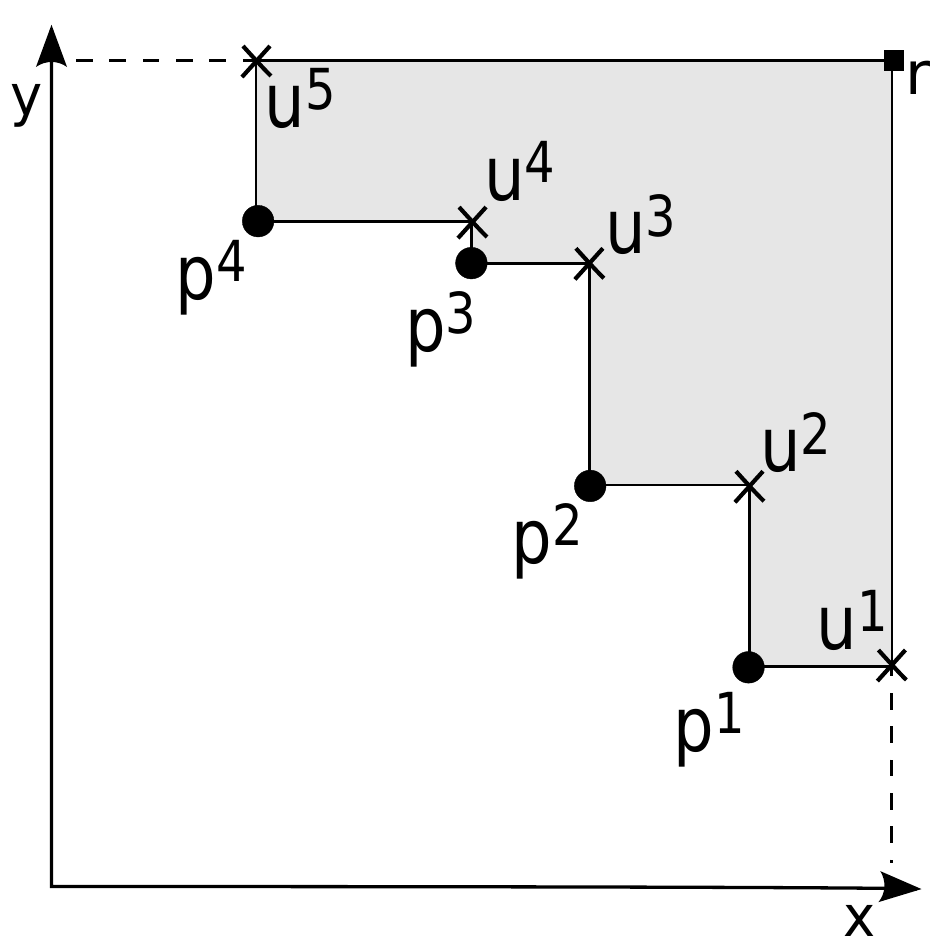}
    \vspace{-0.2cm}
  \caption[Example of local upper bounds]
  {Example of the points which are the local upper bounds (points $u^1,\ldots,u^5$ represented by crosses, $\times$)
  given the nondominated point set $\{p^1,\ldots,p^4\}$.}
  \label{fig:hvstoa:ubs}
    \vspace{-10pt}
\end{figure}

In the context of multiobjective optimization, the \emph{search region}~\citep{Klamroth2015} is
understood as a promissing regions in the objective space where more optimal solutions can be found.
It is therefore disjoint with respect to the region dominated by a set of already found nondominated
points and bounded from above by a reference point, $r\in\Rd$.
Recall that the hypervolume indicator can be defined as the union of
boxes bounded below by a point in the nondominated point set and above by the
reference point. Analogously, the search space is also defined by the union of boxes
which are unbounded below and bounded
above by \emph{local upper bounds}~\citep{Kaplan2008}.
The set of local upper bounds can be roughly defined as the nondominated point set of the search region
considering maximization (see Figure~\ref{fig:hvstoa:ubs}).
These local upper bounds can be computed from the given set of $n$ nondominated
points and the reference point.
In $d=2$ there are $n+1$ local upper bounds and in $d=3$ there are $2n+1$~\citep{Kerstin2015}. 
In the general $d\geq 2$ case there are
$\Theta(n^{\lfloor d/2 \rfloor })$
upper bounds~\citep{Kaplan2008, Klamroth2015}.
Such local upper bounds can be determined and used for the computation of
the hypervolume indicator (see~\ref{HV:HBDA} in Section~\ref{ch:hvstoa:algs:hv}).

\subsubsection{Dealing With Dominated Points}
\label{algs:tec:dom}
Given a point set $\Sr\subset\Rd$, if there is a dominated point $q\in\Sr$ then it has 
zero hypervolume contribution to $\Sr$. However, if it is dominated by a single point, $p\in\Sr$,
then $q$ is an inner delimiter of the contribution of $p$ to $\Sr$ which implies that
if $q$ is removed from $\Sr$ then the contribution of $p$ increases.
Therefore, point $q$ is needed when computing the contribution of $p$ to $\Sr$.
There are mainly two ways to deal with dominated points in this scenario: 
one that preserves the structure of the contribution of $p$ to $\Sr$ and one that
destroys it. 
The latter is typically a work-around that allows to use algorithms
that assume a nondominated point set as input.
\begin{figure*}[t]
  \center
\begin{tabular}{llll}
\subfigure[\scriptsize{$H(p^{5},\{p^{1},\ldots,p^{4}\})$}]
            {\label{fig:dpts:v1}\includegraphics[width=3.5cm,height=3.5cm]{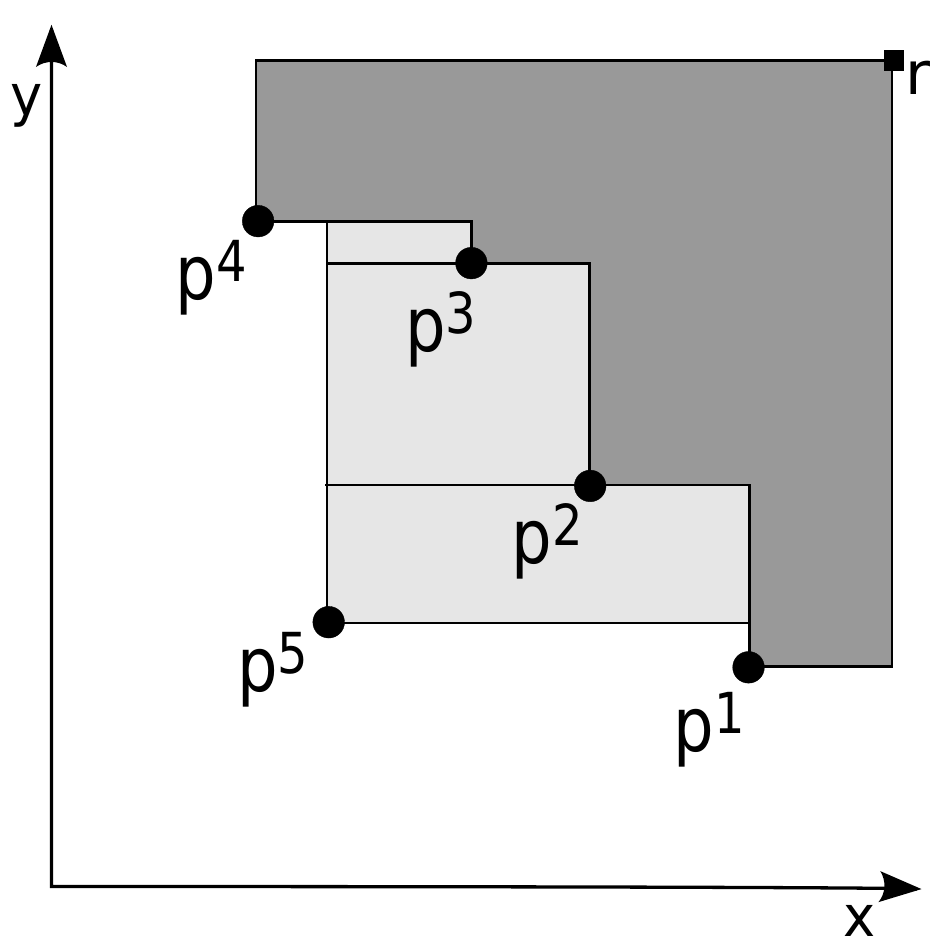}}
  \hspace{-0.3cm}
  &\subfigure[\scriptsize{$H(p^{2},\{p^{1},p^{3},p^{4}\})$}]
            {\label{fig:dpts:v2:1}\includegraphics[width=3.5cm,height=3.5cm]{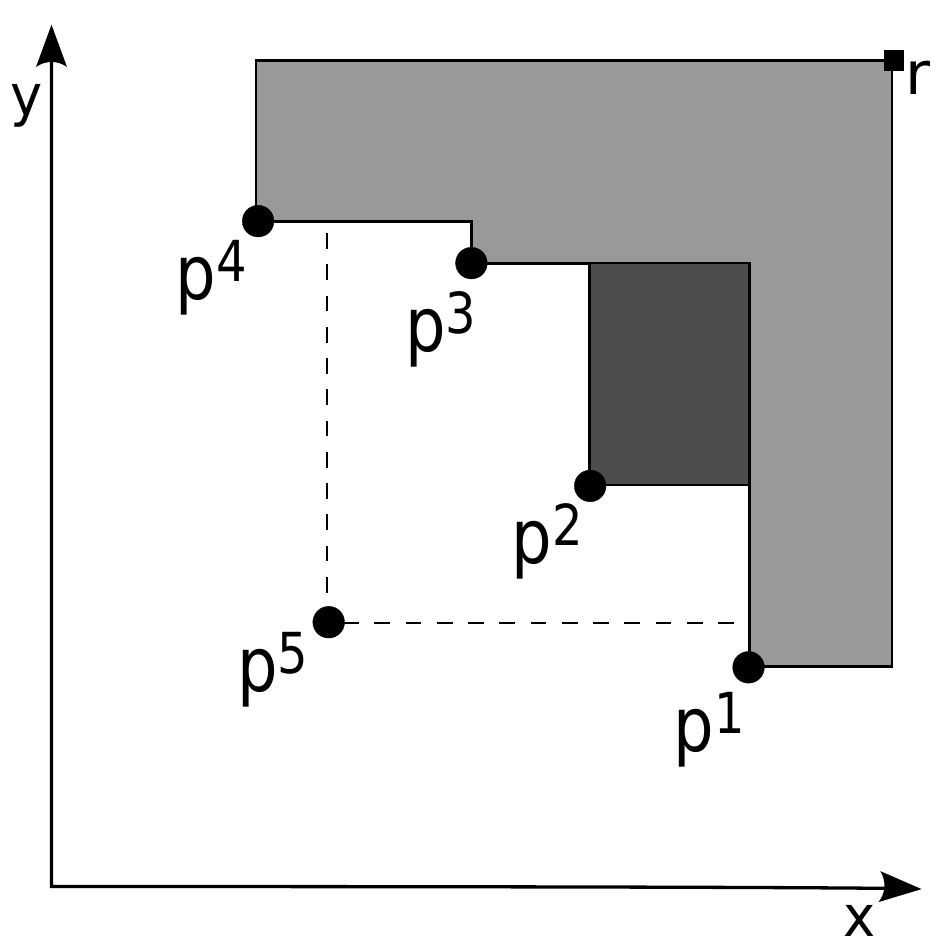}}
  \hspace{-0.3cm}
  &\subfigure[\scriptsize{$H(p^{3},\{p^{1},p^{4}\})$}]
            {\label{fig:dpts:v2:2}\includegraphics[width=3.5cm,height=3.5cm]{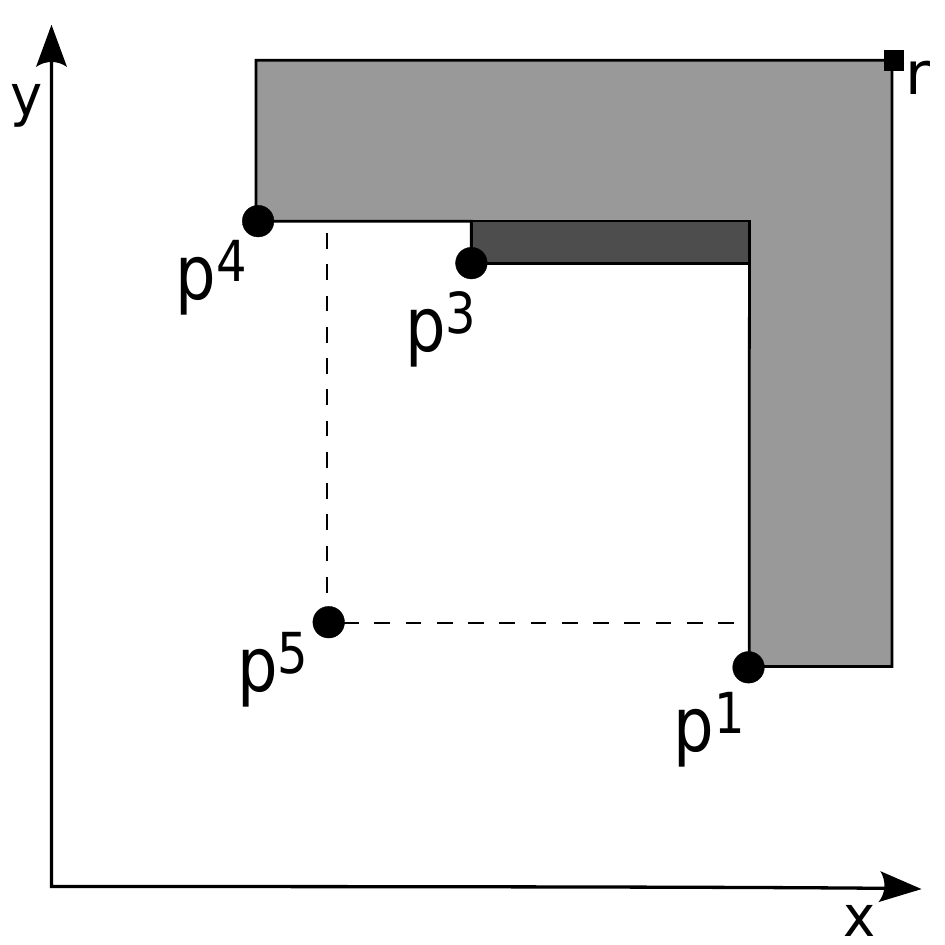}} 
  \hspace{-0.3cm}
  &\subfigure[\scriptsize{$H(p^{5},\{p^{1},p^{4}\})$}]
            {\label{fig:dpts:v2:3}\includegraphics[width=3.5cm,height=3.5cm]{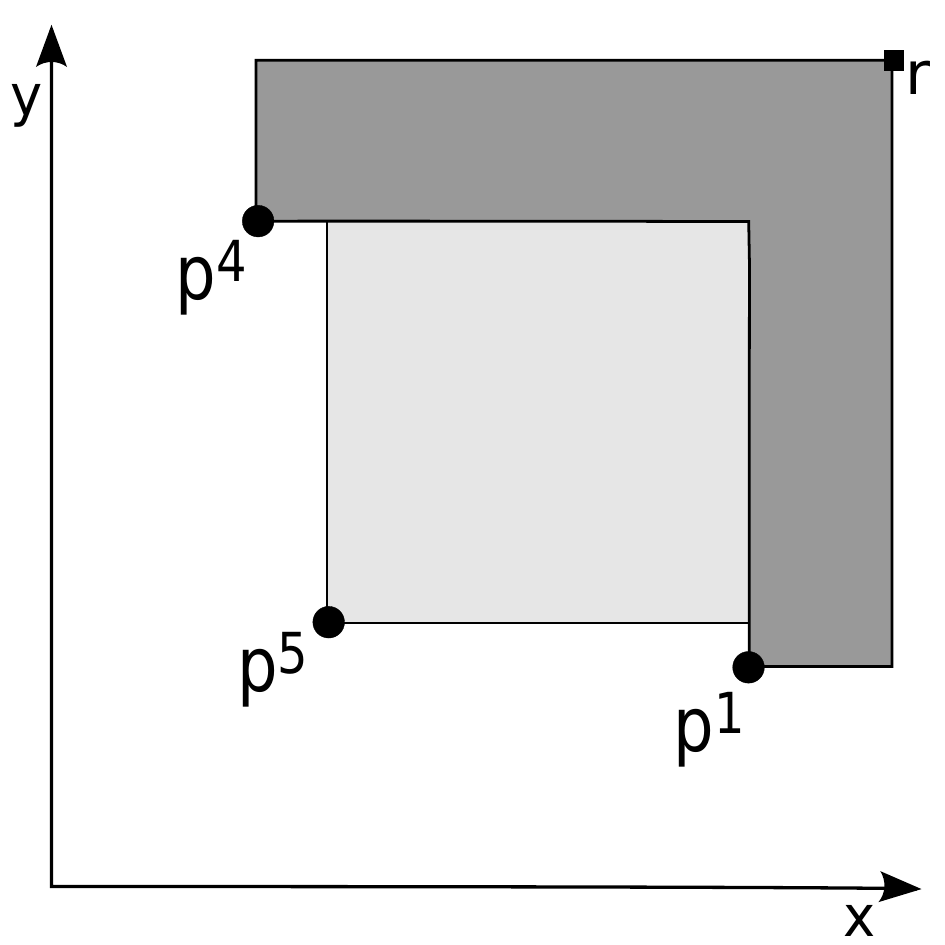}}
\end{tabular}
    \vspace{-0.2cm}
  \caption[Contribution]{Computation of the contribution of $p^5$ to $\{p^1,\ldots,p^4\}$
        either by: (a) splitting it in slices; (b-d) or by previously removing the contribution of
        the points dominated by $p^5$.
    }
  \label{fig:dpts}
    \vspace{-8pt}
\end{figure*}

Figure~\ref{fig:dpts} shows an example where $\Sr=\{p^1,\ldots,p^5\}$.
Assume that the point set $\{p^1, \ldots, p^4\}\subset\Rtwo$
is kept sorted in a linked list.
The contribution of $p=p^5$ to $\Sr$ can be computed in a structure-preserving manner by
splitting the contribution of $p$ in three horizontal (or vertical) slices,
as in Figure~\ref{fig:dpts:v1},
and summing up their areas.
A structure-destructive manner  
would consist of computing 
the contribution of $p^2$ to $\{p^1, p^3, p^4\}$, removing $p^2$ from the list,
computing 
the contribution of $p^3$ to $\{p^1,p^4\}$, removing $p^3$ from
the list (see Figures~\ref{fig:dpts:v2:1} and~\ref{fig:dpts:v2:2}), 
then computing the contribution of $p$ to $\{p^1, p^4\}$ (see Figure~\ref{fig:dpts:v2:3}).
Subtracting the first two contributions from the latter one yields the contribution of
$p$ to $\Sr$:
\begin{equation*}
H(p^5, \Sr) = H(p^5,\{p^{1},p^{4}\})-H(p^{3},\{p^{1},p^{4}\})-H(p^{2},\{p^{1},p^{3},p^{4}\})
\end{equation*}

Note that the described structure-destructive method requires
solving a sequence of $t+1$ $\onecp$ problems, where $t$ is the number of points dominated by $p$.
In addition, the information concerning the inner delimiters,
and consequently the shape, of the contribution of $p$ to $\Sr$ is lost (see
Figure~\ref{fig:dpts:v2:3}).
Thus, structure-preserving methods may be more efficient, and more adequate,
than structure-destructive ones particularly when it comes to update contribution(s).
While the latter would likely have to recompute them from scratch,
the former may save time by taking advantage of its data structures.

\setcounter{paraa}{1}
\section{Hypervolume Indicator}\label{ch:hvstoa:algs:hv}

There are many algorithms to compute the $\hvp$ problem.
This section overviews only the most conceptually distinct ones
(that are based on a different paradigm/technique)
and the fastest ones according either to their asymptotic complexity
or to their runtime efficiency.
The list of algorithms excluded from this overview include some
algorithms based on problems for which the $\hvp$ is a
particular case~\cite[\eg][]{FatBoxes,chan2008slightly,YildizSuri2012}
and others specific to the
$\hvp$ computation~\cite[\eg][]{IEHV, WFG:IIHSO},
approximation algorithms
(\eg, HypE~\cite{Bader2010}) and algorithms based on parallelization~\cite[\eg][]{EMO15:GPUHV, exQHV}).

\subsection{Algorithms}
\label{ch:hvstoa:algs:hv:desc}

\myparagraph{Lebesgue Measure Algorithm (LebMeasure)}\label{HV:LebM}%
The LebMeasure algorithm~\cite{Fleischer03} was one of the first algorithms proposed
to compute $\hvp$ for any number of dimensions ($d>1$).
The algorithm maintains a list of points, $\Sr\subset\Rd$, initially set to the input point set. 
The first step is to remove a point $p$ from $\Sr$ and to compute  and accumulate the hypervolume of an hypercube inside the
region exclusively dominated by $p$.
This is the hypercube
bounded below by $p$ and bounded above in every coordinate $i\in\{1,\ldots,d\}$
by $\cd{b}{i}$, the lowest $i$-th coordinate value in the point set $\{q\in\Sr \:|\: p_i\leq q_i\}$.
To account for the part of the contribution of $p$ to $\Sr$ not yet computed,
$d$ new points are added to $\Sr$, 
each one is a projection of $p$ onto an axis-parellel hyperplane at
$\cd{b}{i}$ in coordinate $i$. 
Then, dominated points are removed from $\Sr$.
These steps are repeated until $\Sr$ is empty.
Although LebMeasure was presented as a polynomial-time algorithm, 
its time complexity was later shown to be exponential in $d$:
$O(n^d)$~\citep{WhileLebM05}.

\myparagraph{Hypervolume by Slicing Objectives (HSO)}\label{HV:HSO}%
The HSO algorithm~\citep{Knowles2002PhD,WFG:HSO}%
\footnote{The algorithm was proposed in~\cite{Knowles2002PhD} and was later
named and studied in more detail in~\cite{WFG:HSO}. Note that~\cite{WFG:HSO}
also assign independent authorship of such an algorithm to Zitzler and provide a reference to
source code (\url{ftp://ftp.tik.ee.ethz.ch/pub/people/zitzler/hypervol.c}).} 
is a direct application of the
dimension sweep approach to the computation of $\hvp$ problem. HSO works exactly as explained
in Section~\ref{ss:stoa:DS}
and has $O(n^{d-1})$-time complexity.
Several algorithms were proposed based on HSO, considering different $(d-1)$-dimensional
subproblems, data structures and/or base cases or by combining it with other techniques
(\eg, FPL~\ref{HV:FPL} and WFG~\ref{HV:WFG}).

\myparagraph{Fonseca, Paquete and L\'opez-Iba\~nez's algorithm (FPL)}%
\label{HV:FPL}%
The FPL algorithm~\citep{FPL} is based on HSO and has $O(n^{d-2}\log n)$ time complexity
and $O(n)$ space complexity. 
FPL improves upon HSO through the use of more efficient data structures, 
the caching of previous computations
and the use of a better base case for the recursion.
In particular, FPL maintains points sorted according to every dimension in circular doubly linked lists.
In an FPL recursion call, points are removed from (some) lists in decreasing order of a given dimension
and are reinserted in reverse order. This behavior enables constant time insertions. 
Moreover, when a new point is visited in an $i$-th dimensional subproblem, it adds contribution only
to the $(i-1)$-th dimensional slices above it in dimension $i-1$.
Thus, if $(i-1)$-th dimensional slices below that point have been previously computed then
FPL has that information stored and does not need to recompute them as HSO does.
Finally, FPL stops the recursion at $d=3$ where it uses HV3D (see~\ref{HV:HV3D}).
Note that FPL may be improved with more efficient base cases.
For example, if HV4D$^+$ (see~\ref{HV:HV4D}) is used as a base case for $d=4$, the
time complexity of FPL would improve to $O(n^{d-2})$ for $d\geq 4$.

\myparagraph{Hypervolume Overmars and Yap (HOY)}\label{HV:HOY}%
HOY algorithm~\citep{HOY} has $O(n^{d/2} \log n)$ time complexity, although it was initially
thought to have $O(n^{d/2})$ time complexity~\citep{nopHOY} due to a gap in the analysis~\cite{HOY}.
It is based on an algorithm for Klee's measure problem by Overmars and Yap
and uses a streaming variant of an orthogonal partition tree as the underlying data structure.
Only $O(n)$ space complexity is required by constructing the referred tree on-the-fly
instead of storing it completely as in the classical variant
which would require $O(n^{d/2})$ space complexity.

HOY is based on the Spatial Divide-and-Conquer paradigm (see Section~\ref{algs:par:sdc}).
It measures the hypervolume dominated by a given point set $\Ar\subset\Rd$
inside the region bounded by a lower and an upper reference point, $\ell$ and $u$, respectively.
The set $\Ar$ is initialized as the set of input points, $u$ to the given reference point and
$\ell$ to the component-wise minimum of all input points.
HOY recursively partitions in two parts the considered region,
firstly using dimension one.
Then, HOY checks if dimension $i$ satisfies certain criteria for partitioning, 
otherwise, it checks dimension $i+1$ and so on.
The recursion base case is achieved when the ($d$-dimensional) 
dominated region of the current partition forms a trellis.
Moreover, before HOY is used,
points are previously sorted according to dimension $d$.
This allows to shrink the space partition considered in each recursive call 
by visiting points in $\Ar$ in ascending order of dimension $d$ until
a point $q$ that dominates the $(d-1)$ projection of the 
space partition is found. In such case, the hypervolume of the space region between $q_d$ and $u_d$
is computed right away and $u_d$ can be set to $q_d$.

\myparagraph{Walking Fish Group algorithm (WFG)}\label{HV:WFG}%
WFG algorithm~\citep{WFG2011} is currently one of the fastest
for computing the hypervolume indicator in
many dimensions, particularly for $d>7$ (see the experimental results in~\cite{Renaud2017}),
even though it is not asymptotically the fastest. WFG was initially reported to have 
$O(2^{n+1})$ time complexity~\citep{WFG2011} but Lacour~\etal~\cite{Renaud2017} recently tightened
this upper bound to $O(n^{d-1})$ and presented a lower bound of $\Omega(n^{d/2} \log n)$.

WFG is an algorithm mainly based on the bounding technique and on the inclusion-exclusion principle,
and is further optimized by integrating ideas of dimension sweep and objective reordering.
WFG works as follows, given the point set $\Xr\subset\Rd$ and a reference point $r\in\Rd$.
Points in $\Xr$
are sorted and visited in ascending order of dimension $d$.
For each point $p\in\Xr$ visited, the $(d-1)$-dimensional
contribution of $p$ to the set of already visited points, $\Sr$ (\ie, $H(p^*,\Sr^*)$),
is then computed and multiplied by the difference ($r_d-p_d$).
The hypervolume of $\Sr$ is the sum of all such multiplications.
The contribution $H(p^*,\Sr^*)$ is computed by subtracting to the hypervolume of $\{p^*\}$
the hypervolume of the set $\Sr^*$ but bounded by $p^*$,
\ie, $H(p^*,\Sr^*)=H(p^*,\Jr)=H(\{p^*\}) - H(\Jr)$, where $\Jr$ is the set obtained
with the bounding technique (see Section~\ref{algs:tec:bt}).
In its turn, $H(\Jr)$ is computed recursively with WFG.
In practice, WFG alternates between $\onecp$ and $\hvp$ problems.

The bounding technique in combination with dimension sweep
allows many points to become dominated and, thus, it
plays an important role in reducing the required computational effort.
This places WFG among the fastest algorithms for many dimensions ($d>4$).

\myparagraph{Chan's algorithm (Chan)}\label{HV:Chan}%
Chan's algorithm~\citep{Chan2013} has $O(n^{d/3} \text{ polylog } n)$ time complexity,
which is currently the best time complexity to compute $\hvp$ for $d \geq 4$.
Chan's paper proposes an algorithm for the general case of Klee's Measure problem and then derives
an algorithm for the special case of the hypervolume indicator (referred as the union of arbitrary
orthants).
Chan's algorithm is a spatial
divide-and-conquer algorithm
combined with a procedure to simplify partitions
that is repeated in every few levels of cutting, reducing the number of boxes
contained in a partiton.
One of the main differences to HOY (see~\ref{HV:HOY}) is the method used for selecting the dimension
to be used for partitioning, which changes at every level of recursion: first, it uses dimension one,
then dimension two, and so on.
Despite its good time complexity, no implementation is found available online.

\myparagraph{Quick Hypervolume (QHV and QHV-II)}\label{HV:QHV}%
\label{HV:QHVII}%
QHV~\citep{RusFra2014:QHV} is an algorithm for the general case of $\hvp$ problem
that is based on quicksort. This algorithm
can be viewed as a type of spatial divide-and-conquer but instead of dividing the
hypervolume in two, as discussed in the example given in Section~\ref{algs:par:sdc},
it splits the region into $2^d$ regions.
To compute $\hvp$ for a point set $\Sr\subset\Rd$, first a point $p\in\Sr$
is selected as a pivot. The point in $\Sr$ with the greatest contribution
to the empty set is selected for this purpose. The pivot splits the $d$-dimensional space into
$O(2^d)$ hyperoctants sharing a corner at $p$ and $p$ is discarded.
The two hyperoctants referring to the regions that contain the points dominating $p$
or the points that $p$ dominates are discarded.
QHV is called recursively for each of the remaining hyperoctants (still partially dominated by $\Sr$)
with the points inside that region and the projection on the hyperplanes delimiting the hyperoctant of
the points that partially dominate it.
QHV includes a subrotine to discard dominated points from each hyperoctant.
The hypervolume of $\Sr$ is the sum of the hypervolume
returned from the recursive calls plus the contribution of $p$ to the empty set.
In base cases with up to $10$ points, a simple algorithm as HSO or based in the Inclusion-Exclusion
principle is used.

In the worst-case, QHV has $O(n^dn^2)$-space complexity, and an initially reported time complexity of $O(n(d+\log^{n-2})2^{nd})$~\citep{RusFra2014:QHV}, which was later tightened to $O(2^{d(n-1)})$~\cite{QHVII}.
However, its performance depends on the characteristics of the data set considered. For example,
the authors showed that the time complexity on a data sets where points are uniformly distributed over
a hypersphere is $O(dn^{1.1}\log^{n-2}n)$.
In practice, QHV was observed to be competitive with WFG.
Although a parallel version of QHV exists~\citep{exQHV},
only the sequential version is taken
into account in this paper.

A modified version of QHV, called QHV-II, was recently proposed~\cite{QHVII}.
QHV-II uses a different partitioning scheme, where the pivot is used to split the
$d$-dimensional space in $d$ hypercuboids. Moreover, points whose projection became dominated,
may not be immediately removed, because the dominance check of each new projection is
performed only against the pivot.
QHV-II has a $O(d^{n-1})$ time complexity. However, similarly to QHV, its performance
depends on the data set, for example, there is a given problem for which QHV-II has
$\Theta(n\log^{d-1}n)$ time complexity~\citep{QHVII}. Overall, Jaszkiewicz~\cite{QHVII}
showed that QHV-II has better worst-case time complexity than QHV and empirically showed
that QHV-II performs less operations. The author points out that, unlike the original
implementation of QHV that uses low-level code optimizations and different algorithms
for small subproblems, the implementation of QHV-II is simple, and conjectures that
faster performance than QHV can be achieved by using the same low-level code optimizations.

\myparagraph{Hypervolume Box Decomposition Algorithm (HBDA)}\label{HV:HBDA}%
Lacour~\etal~\cite{Renaud2017} proposed an algorithm to
compute $\hvp$ for any number of dimensions. This algorithm, HBDA,
computes the
hypervolume indicator by partitioning the dominated region into
$O(n^{\lfloor\frac{d}{2}\rfloor})$ axis-parallel boxes and adding up the
corresponding hypervolumes.
The partitioning results from computing all local upper bounds,
where each box is associated to one local upper bound.
The incremental version of the algorithm (HBDA-I)
runs in $O(n^{\lfloor\frac{d}{2}\rfloor+1})$ time
and is characterized by computing a sequence of $n$ $\hvupdp$ problems,
allowing
input points to be processed in any order.
This update consists of updating the set of local upper bounds and then
recomputing the hypervolume indicator from the resulting box decomposition.
Since the current box decomposition must be stored across
iterations, $O(n^{\lfloor\frac{d}{2}\rfloor})$ space is required. By processing
input points in ascending order of any given coordinate, the memory
requirements are reduced to $O(n^{\lfloor\frac{d-1}{2}\rfloor})$, and the
time complexity is improved to $O(n^{\lfloor\frac{d-1}{2}\rfloor+1})$.
HBDA-NI (the non-incremental version)
has been shown to be competitive in $d\geq 4$ dimensions, but its
memory requirements are a limiting factor for large $d$.

Note that HBDA-I can be easily adapted to recompute the hypervolume
in $O(n^{\lfloor\frac{d}{2}\rfloor})$ time
when the reference point is changed to a new location (and is still strongly
dominated by every point in the input set), provided that the data structures
are set up as in HBDA-I.
It is enough to identify the local upper bounds with a coordinate of
the old reference point, replace it with the coordinate of the new one,
and then recompute the hypervolume from the updated set of local upper bounds.

\myparagraph{Dimension-sweep algorithms for $d=3$ (HV3D and HV3D$^+$)}%
\label{HV:HV3D}%
\label{HV:HV3D+}%

\begin{figure}[t]
  \center
  \begin{tabular}{c|c}
  \subfigure[Slice 1 (between $\cz{p}^1$ and $\cz{p}^2$)]{\label{fig:HV3D:s1}
            \includegraphics[width=3cm,height=3cm]{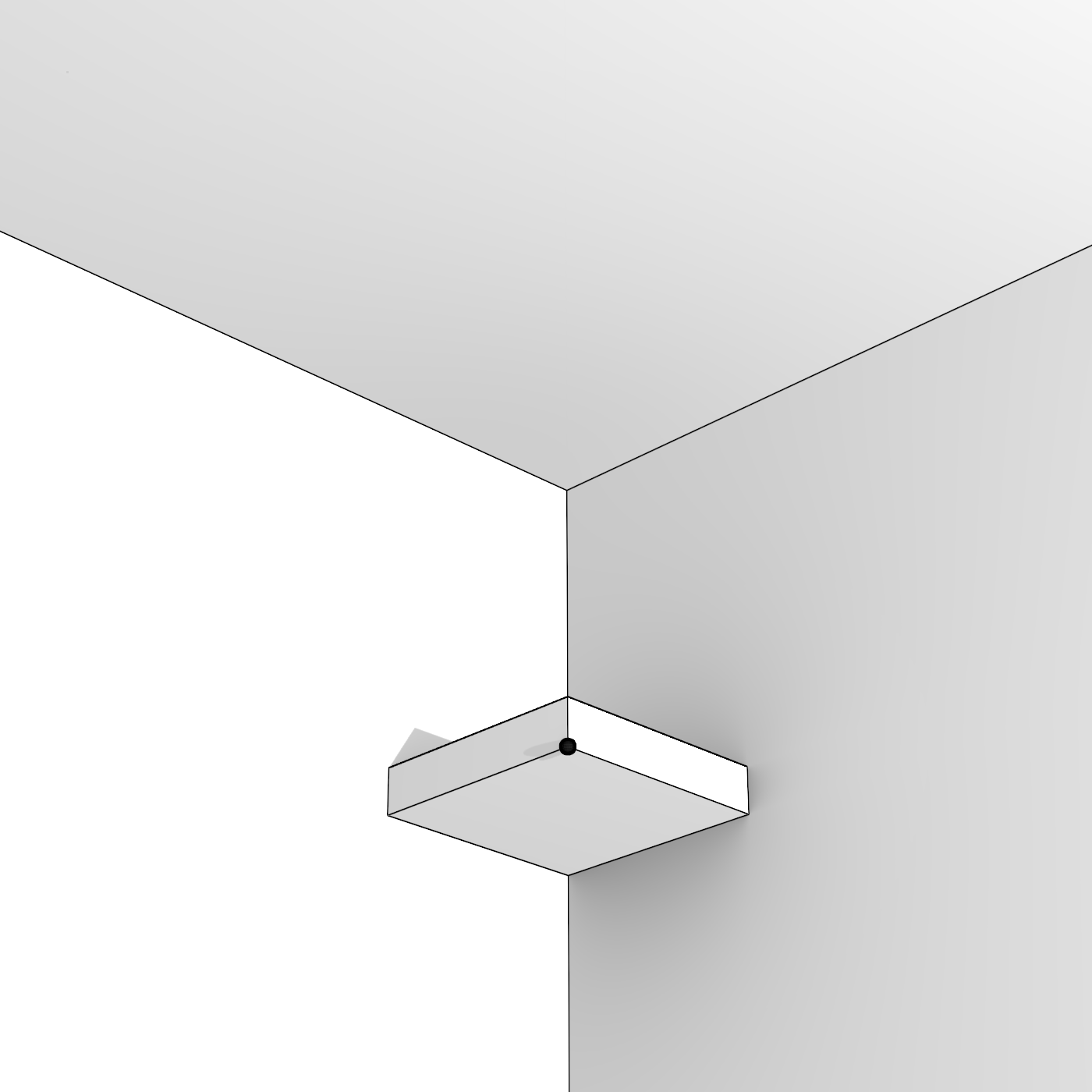}
            \hspace{0.4cm}
            \includegraphics[width=3cm,height=3cm]{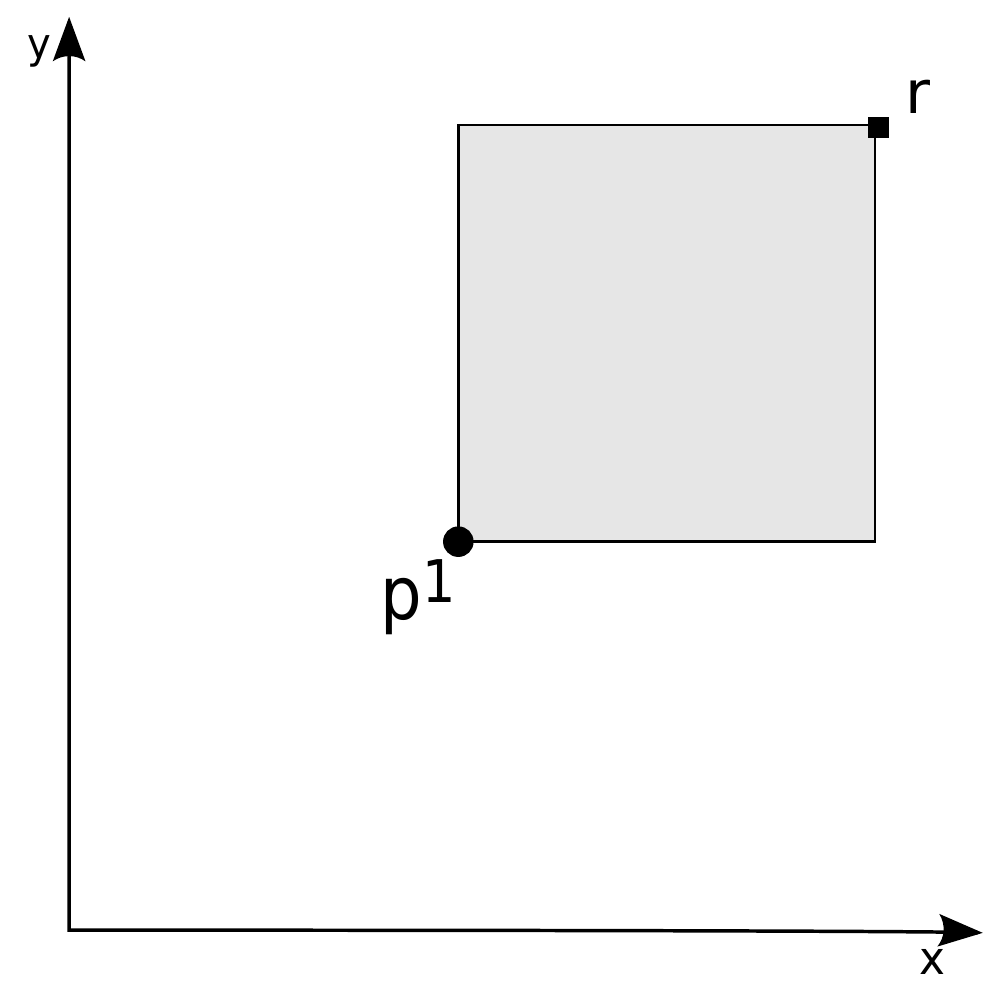}} \hspace{0.2cm} 
        & \hspace{0.2cm}
  \subfigure[Slice 2 (between $\cz{p}^2$ and $\cz{p}^3$)]{\label{fig:HV3D:s2}
            \includegraphics[width=3cm,height=3cm]{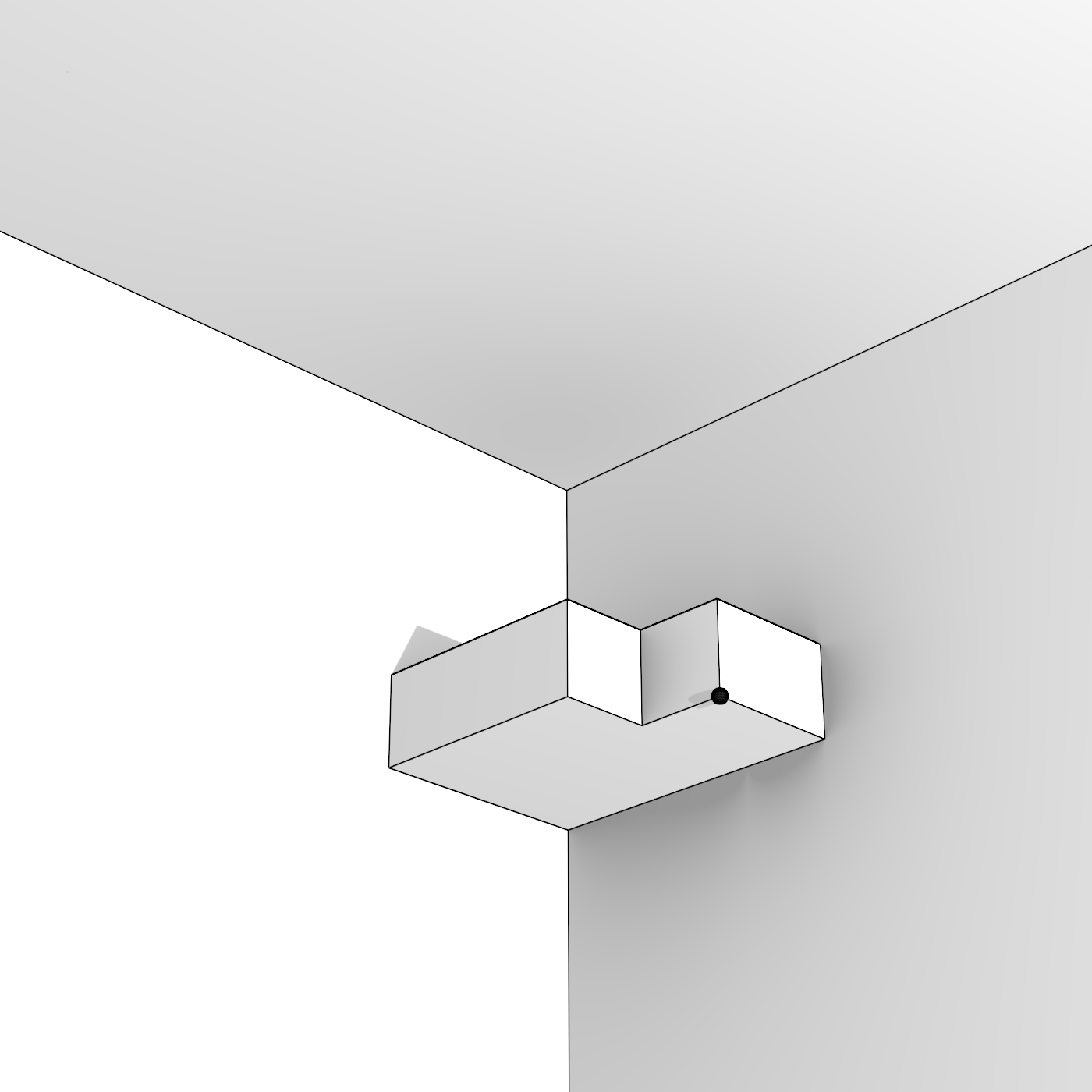}
            \hspace{0.4cm}
            \includegraphics[width=3cm,height=3cm]{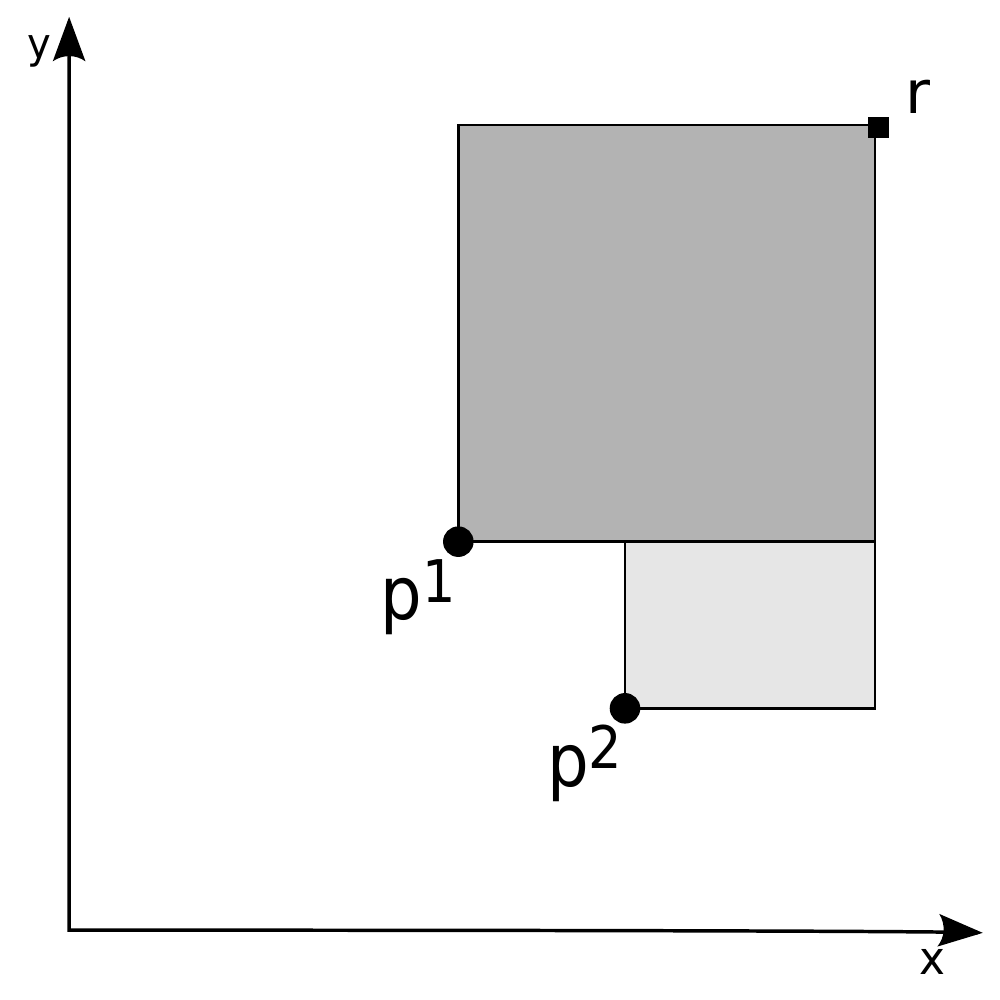}}\\
  \subfigure[Slice 3 (between $\cz{p}^3$ and $\cz{p}^4$)]{\label{fig:HV3D:s3}
            \includegraphics[width=3cm,height=3cm]{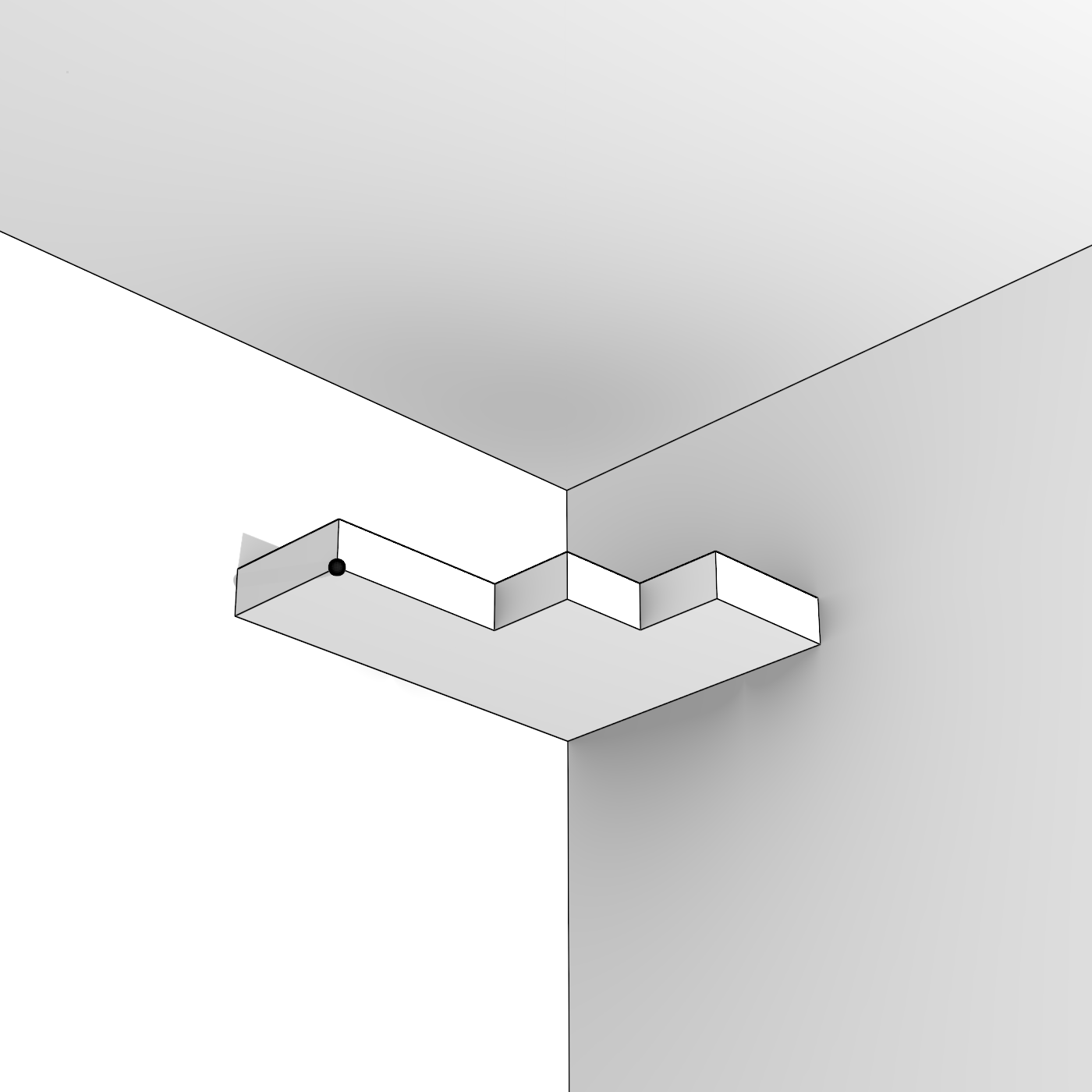}
            \hspace
            {0.5cm}\includegraphics[width=3cm,height=3cm]{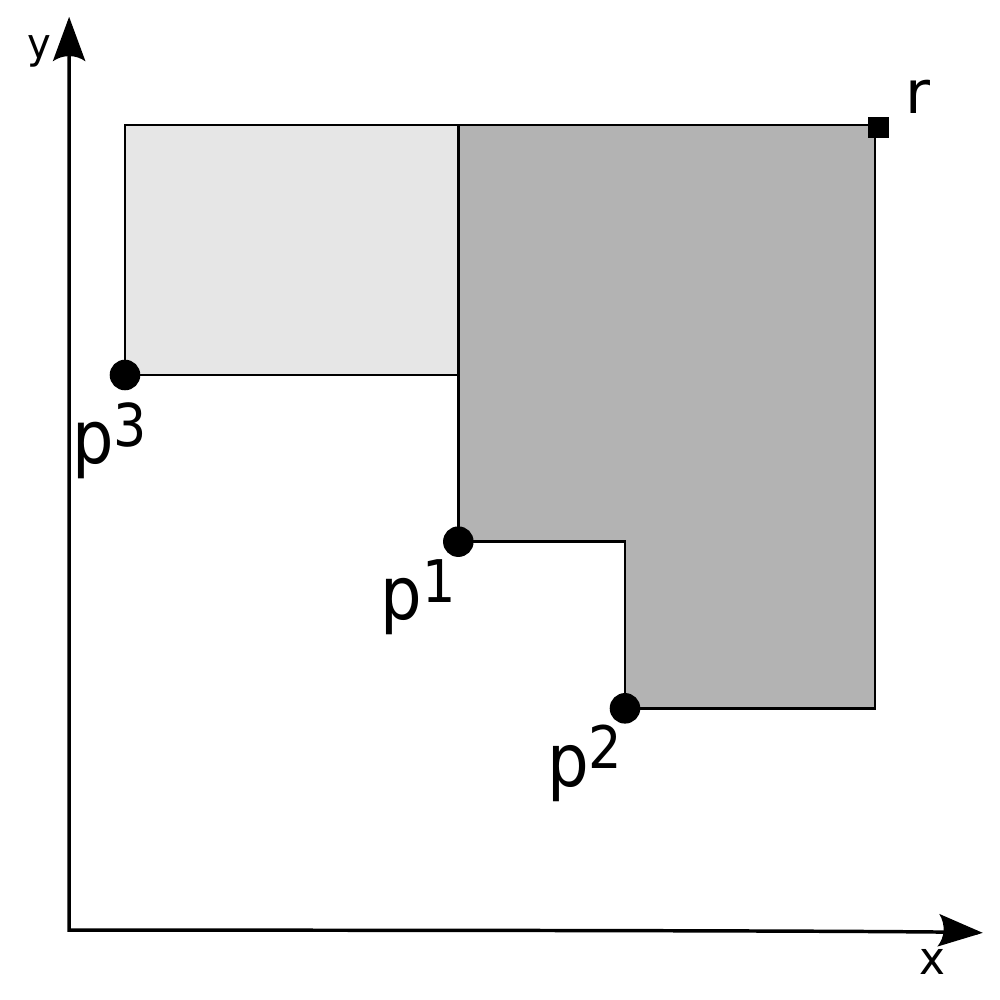}} \hspace{0.2cm} 
        & \hspace{0.2cm}
  \subfigure[Slice 4 (between $\cz{p}^4$ and $\cz{p}^5$)]{\label{fig:HV3D:s4}
            \includegraphics[width=3cm,height=3cm]{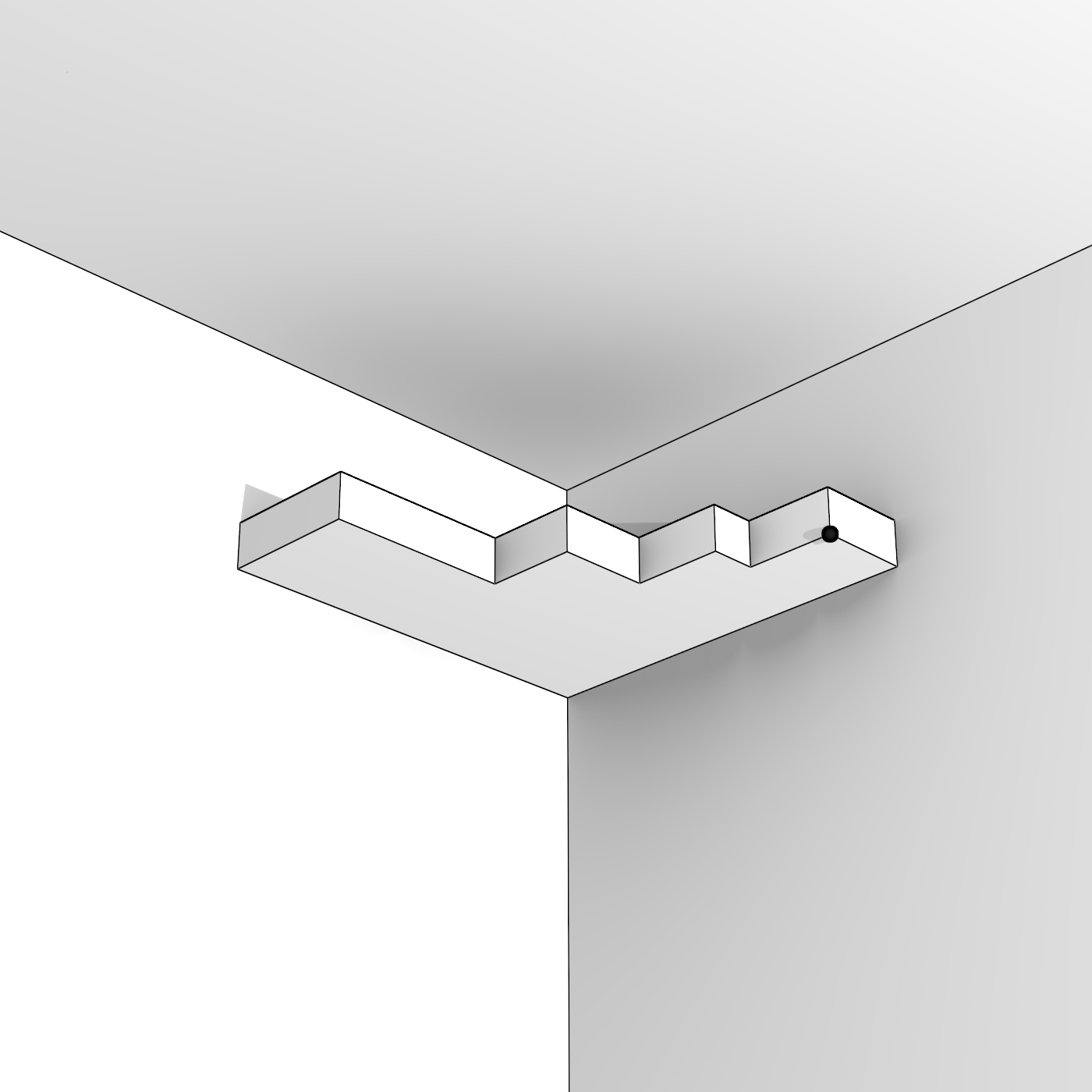}
            \hspace{0.4cm}
            \includegraphics[width=3cm,height=3cm]{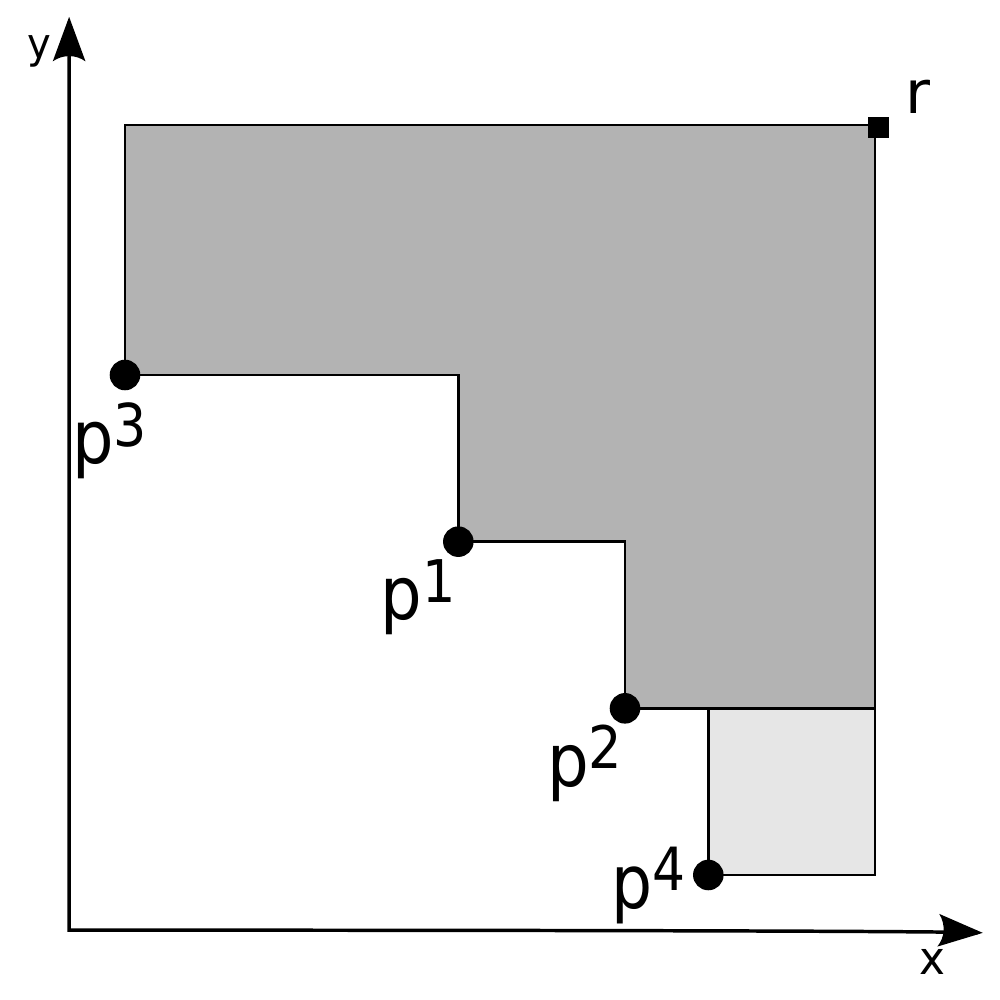}}\\
  \subfigure[Slice 5 (between $\cz{p}^5$ and $\cz{p}^6$)]{\label{fig:HV3D:s5}
            \includegraphics[width=3cm,height=3cm]{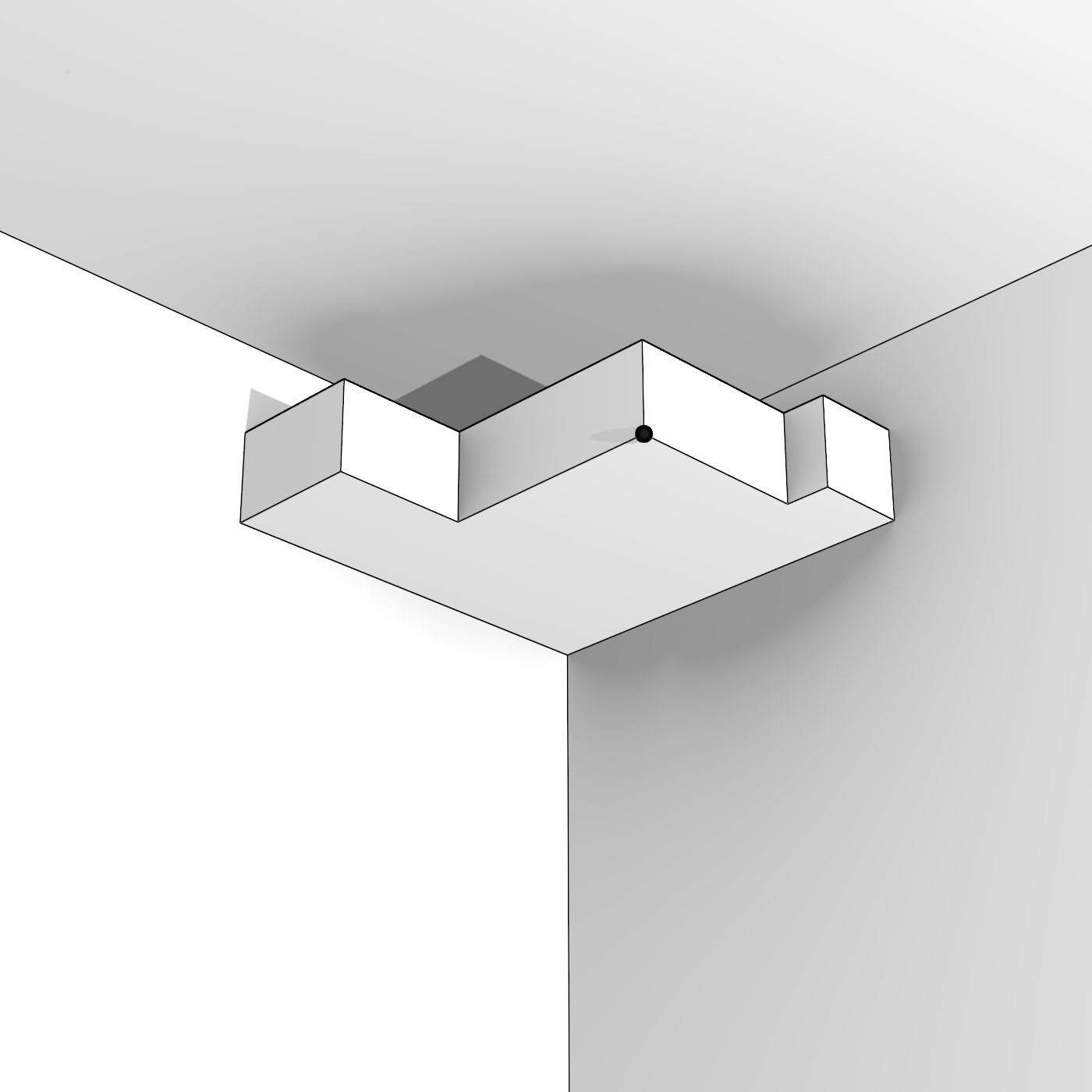}
            \hspace{0.4cm}
            \includegraphics[width=3cm,height=3cm]{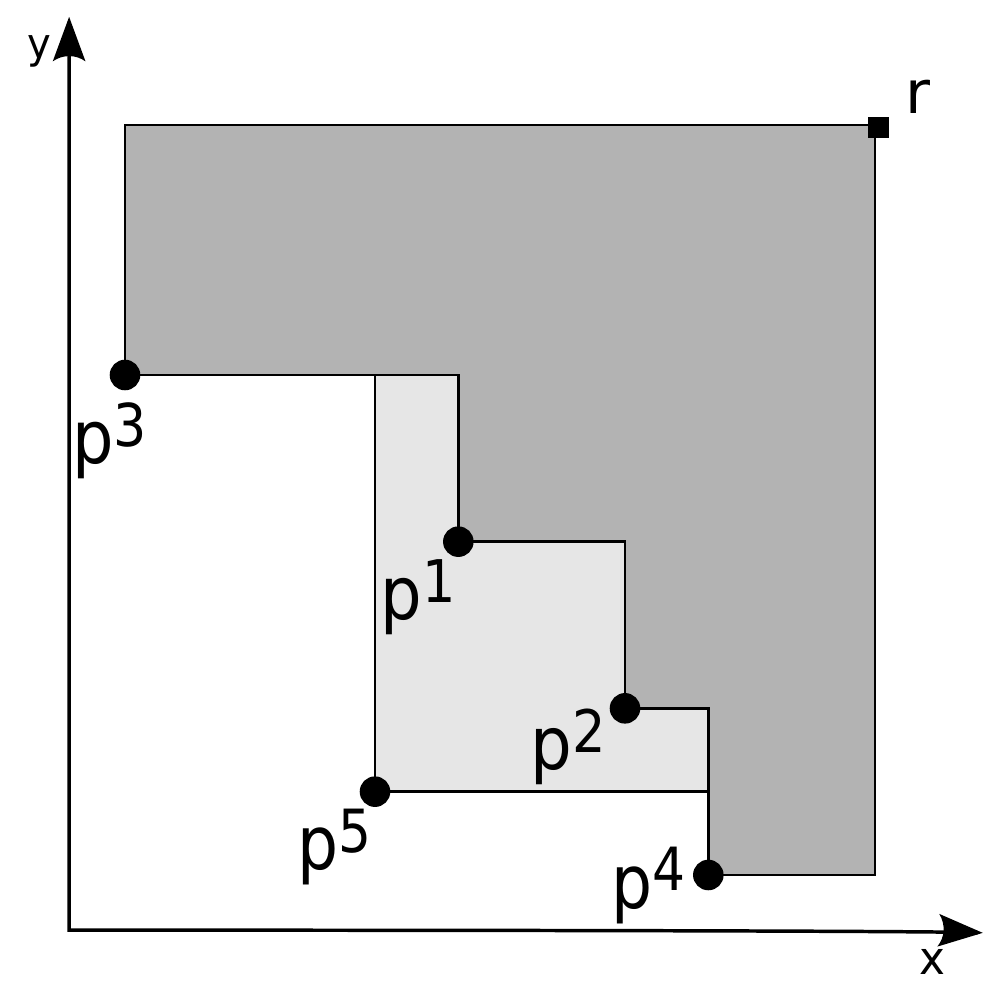}} \hspace{0.2cm} 
        & \hspace{0.2cm}
  \subfigure[Slice 6 (between $\cz{p}^6$ and $\cz{r}$)]{\label{fig:HV3D:s6}
            \includegraphics[width=3cm,height=3cm]{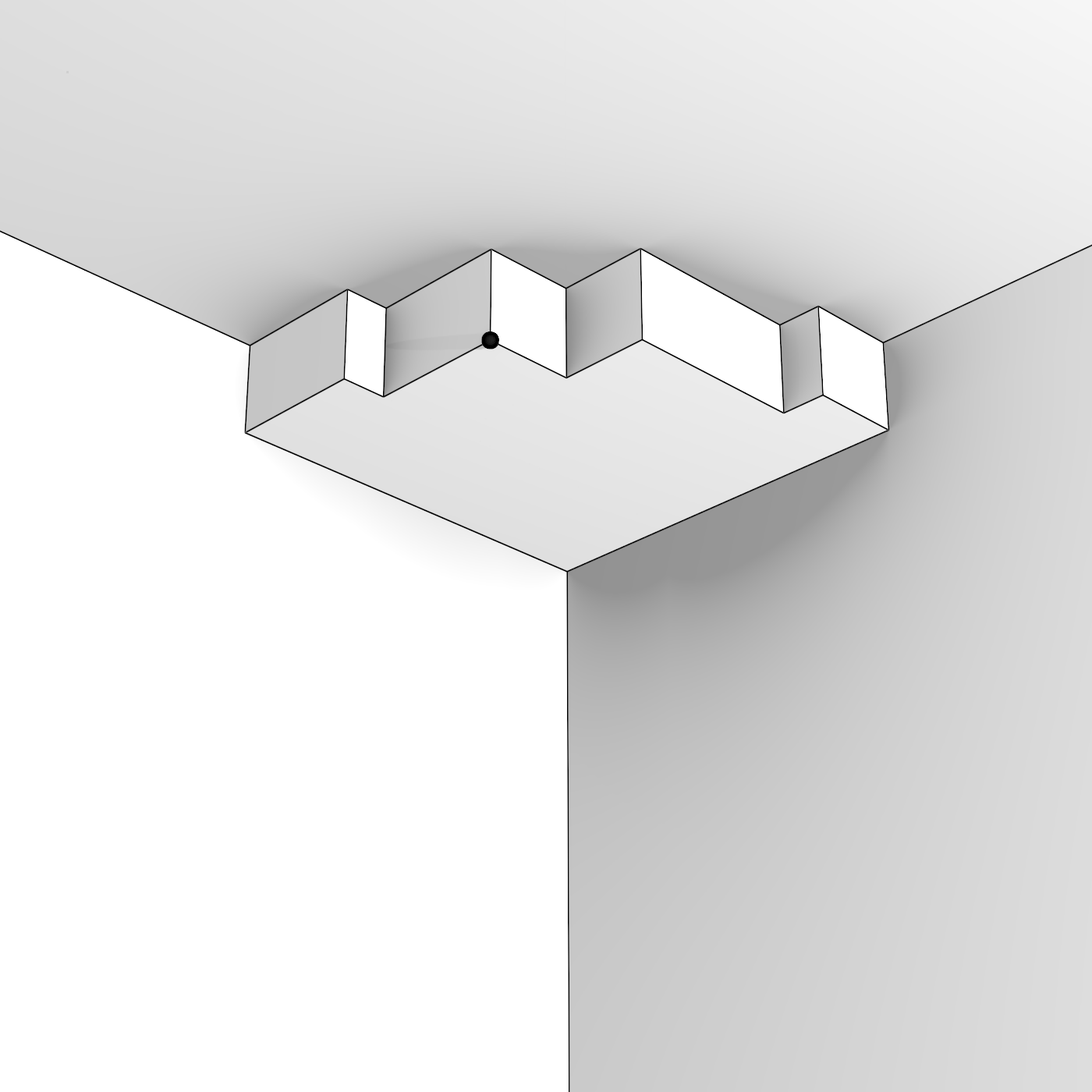}
            \hspace{0.4cm}
            \includegraphics[width=3cm,height=3cm]{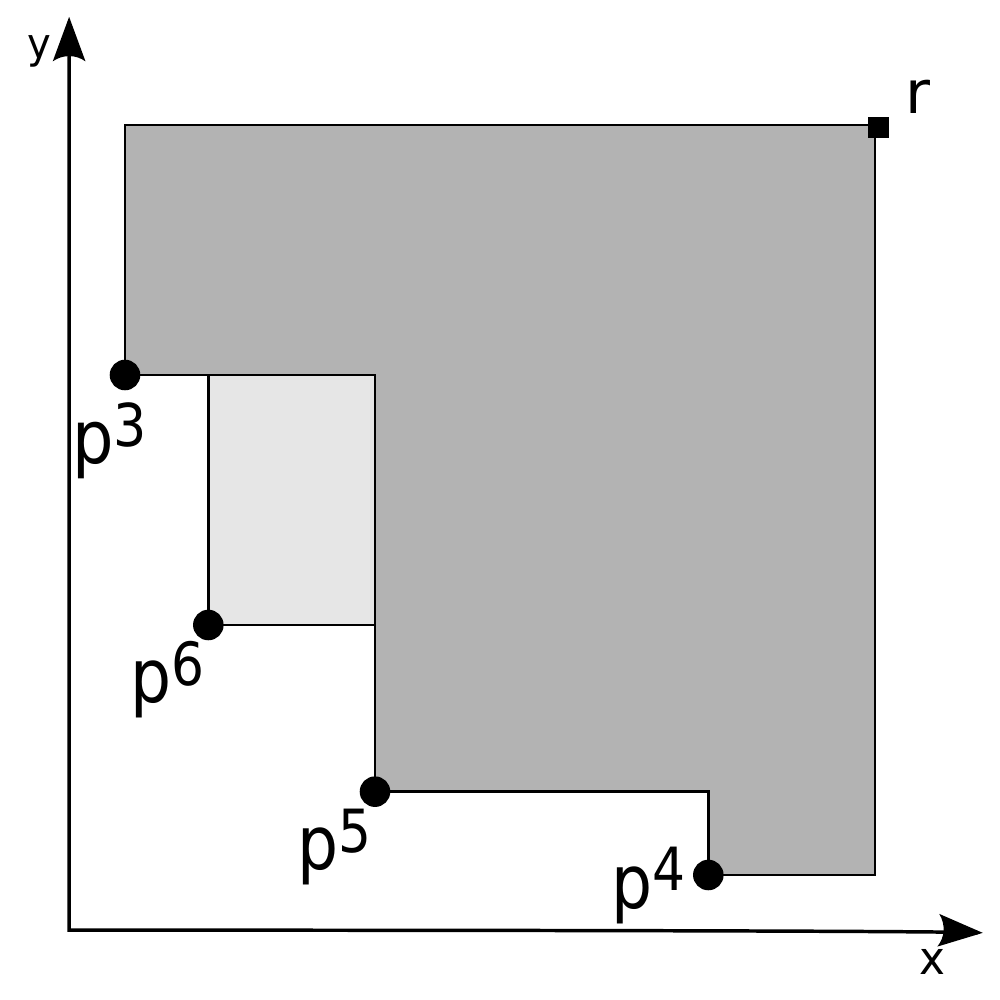}}\\
  \end{tabular}

    \vspace{-0.2cm}
  \caption[Example of slices of a volume]{The splitting of the volume in Figure~\ref{fig:hvstoa:DS:3D} into 6 slices and the corresponding $2$-dimensional bases.
  The base area of the previous slice is depicted in dark gray in the $2$-dimensional figures.}
  \label{fig:HV3D}
\end{figure}

HV3D~\citep{HV3D} is a dimension-sweep algorithm for the $d=3$ case of $\hvp$ problem
with $\Theta(n \log n)$
time complexity and $O(n)$ space complexity.
Given a point set $\Sr=\{p^1,\ldots,p^n\}\subset\Rt$, the volume dominated by $\Sr$ is
divided into slices 
(see the example in Figure~\ref{fig:allslices}) from bottom up.
Each pair of figures in Figure~\ref{fig:HV3D} 
shows the three-dimensional representation of a slice and its two-dimensional base.

HV3D works by solving a sequence of $2$-dimensional $\hvupdp$ problems, as follows.
Points in $\Sr$ are sorted
and visited in ascending $z$-coordinate order. Each point $p\in\Sr$ marks the
beginning of a new slice, the base area of which is computed by updating the
area of the base of the previous slice (if it exists). This is illustrated in
Figure~\ref{fig:HV3D}
where the darker gray region represents
the base of the previous slice to be updated.
To that end, the points visited so far whose projections on the $(x,y)$-plane
are mutually nondominated are kept sorted in ascending order
of the $y$ coordinate using a height-balanced binary tree, $\Tr$.
In the example of Figure~\ref{fig:HV3D:s4}, $\Tr$ contains $\{p^4,p^2,p^1,p^3\}$
after computing the fourth slice.
For each $p\in\Sr$, the outer delimiter to the right
($q=p^4$ in the example of slice 5 where $p=p^5$)
is determined in $O(\log n)$ steps by searching $\Tr$.
Then, the contribution of $p^*$ to $\Tr^*$
is computed by visiting the successors of $q$ in $\Tr$ in ascending order of $y$
until the outer delimiter to the left is found (in the example, $p^3$). 
Such contribution is computed in a structure-destructive manner as explained
in Section~\ref{algs:tec:dom}. In this case, each point in $\Tr$ weakly dominated by $p$ on the
$(x,y)$-plane is removed from $\Tr$ in $O(\log n)$, and $p$ is added to $\Tr$ next to its outer
delimiters in $O(\log n)$ as well.
The base area of the new slice is computed by summing the contribution of $p^*$ and
the base area of the previous slice, and the volume of the slice is computed
by multiplying its base area by its height. In the example of slice 5, after computing
its volume, $\Tr$ contains the points delimiting its base, \ie, $\Tr$ contains $\{p^4, p^5, p^3\}$.

In the above, each point in $\Sr$ is visited twice: once when it is added to
$\Tr$ and again when it is removed from $\Tr$. Since all of the corresponding
operations are performed in $O(\log n)$ time, the algorithm has amortized
$O(n \log n)$ time complexity.

A modified version of HV3D
that supports linear-time updates was recently proposed~\cite{HVCTEC2017}. It
is called HV3D$^+$, it has the same time and space complexities
but it is faster than HV3D~\citep{HVCTEC2017}.
The key difference between the two is that in HV3D$^+$ the data structure setup is performed
as a preprocessing step, in $O(n \log n)$ time, previous to the computation step,
that is performed in
$O(n)$ time,
and not altogether as in HV3D.
This data structure keeps track, for each point $p\in\Sr$,
the successor point according to coordinate $z$, and the outer delimiters of $p$ at
$z=p_z$.
By performing the same sweep and the same operations in the binary tree $\Tr$ as in HV3D,
and recalling the two neighbour points in the tree next to which each point is inserted to
(the two outer delimiters of $p$), allow HV3D$^+$ to then subsequently perform the same sweep more
efficiently, in $O(n)$ time. 
The key idea is to avoid having to query (again) the tree $\Tr$ to find the two points next to which
each point is going to be inserted to in $\Tr$, when the same sweep is repeated.
Consequently,
the binary tree can be replaced by a linked list, and all $O(\log n)$-time operations become
$O(1)$ time instead.

An advantage of HV3D$^+$ is that when a point is added/removed to the set
($\hvupdp$ problem), the data structure can be updated in linear time and the hypervolume
is then recomputed in linear time as well. This update procedure is called HV3D$^+$-R.
There is an alternative update procedure, called HV3D$^+$-U, that only computes the contribution
of the point being added/removed (see~\ref{HVC:HV3D+U}).
As HV3D$^+$-U performs less computations than HV3D$^+$-R (it skips a few steps),
the former computes $\hvupdp$ problem faster. In practice,
it can be up to three times faster than HV3D$^+$-R~\cite{HVCTEC2017}.
An advantage of the HV3D$^+$-R version is that it can be used to recompute the hypervolume
in linear time after changing the reference point.

In the decremental scenario both update versions, HV3D$^+$-R and HV3D$^+$-U, assume that $\Sr$
is a nondominated point set but admit that points in $\Sr$ may be dominated
by $p$ in the incremental case. Admitting dominated points in the latter case
is only possible because, unlike HV3D, the contribution of a point to the base of a 
slice is computed in a structure-preserving way.
This is the key factor that makes possible the extension of HV3D$^+$ to the
all contributions problem (see~\ref{HVC:HVC3D}).

\myparagraph{Dimension-sweep algorithms for $d=4$ (HV4D and HV4D$^+$)
}\label{HV:HV4D}%
\label{HV:HV4D+}%
HV4D~\citep{HV4D} and HV4D$^+$~\cite{HVCTEC2017} are amortized $O(n^2)$-time and $O(n)$-space algorithms for the particular case of $d=4$ of the $\hvp$ problem.
Although they have worse time complexity than Chan's algorithm (see~\ref{HV:Chan}),
they are currently the fastest ones among the algorithms with available implementations.
HV4D algorithm is an extension of HV3D to four dimensions where a sequence of
three-dimensional $\hvupdp$ problems is solved via the corresponding $\onecp$
problems using similar techniques to those in the EF algorithm (see~\ref{HVC:EF}).
Points in the input set
$\Sr\subset\Rf$ are visited in ascending order of the last coordinate, partitioning
the dominated region into four-dimensional slices. For each $p\in\Sr$, the base
volume of the new slice is computed by updating the volume of the base of
the previous slice with the contribution of $p^*$ to the projection on
$(x,y,z)$-space of the points visited so far.

\begin{figure*}[t]
  \center
\begin{tabular}{llll}
  \subfigure[Contribution]{\label{fig:UHV3D:3D:ex}\includegraphics[width=3.5cm,height=3.5cm]{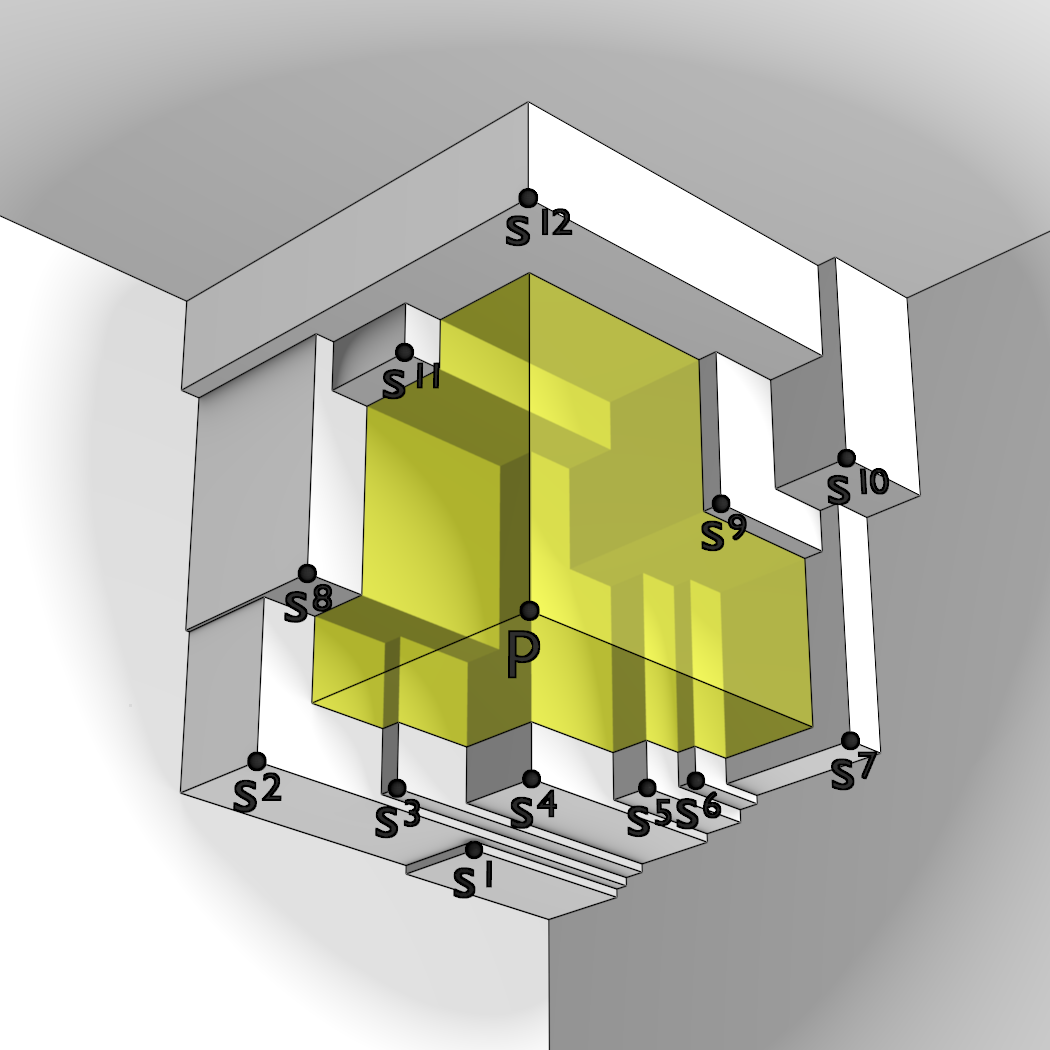}}
  \hspace{1cm}
  &\subfigure[Box partitioning]{\label{fig:UHV3D:3D:boxes}\includegraphics[width=3.5cm,height=3.5cm]{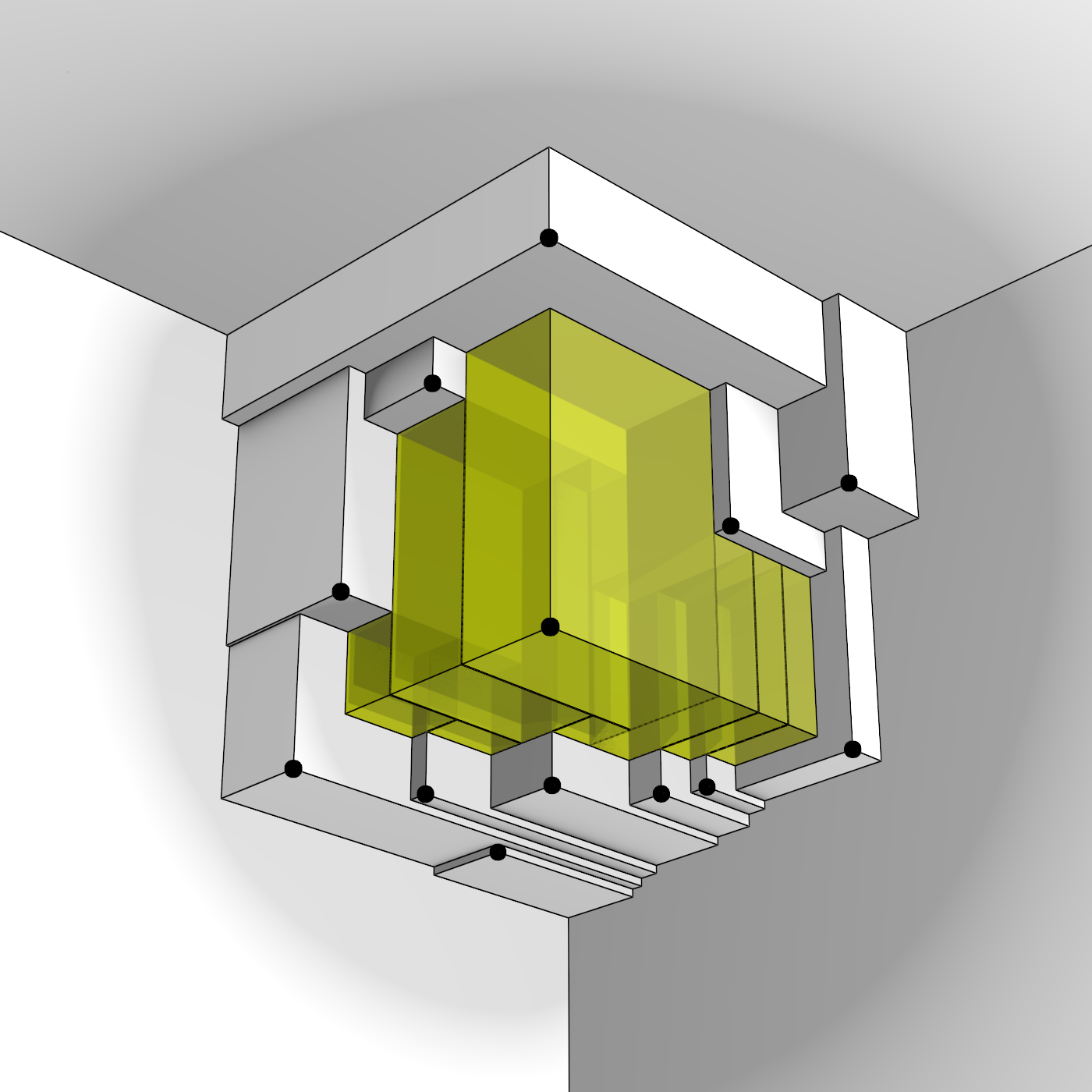}} 
\end{tabular}
  \vspace{-0.3cm}
  \caption[Box partitioning of a contribution in $d=3$]{Example of a contribution,
  $\Ctr(p^*, \Lr^*)$, and of its division in boxes, where $\Lr^*=\{s^1,\ldots,s^{12}\}$.}
  \label{fig:UHV3D:3D}
    \vspace{-10pt}
\end{figure*}

For that purpose, the points visited so far whose projections are nondominated
are stored in a data structure, $\Lr$.
The contribution of each $p^*\in\Sr^*$ to $\Lr^*$ is computed
using a procedure that partitions the contribution in boxes (see example in
Figure~\ref{fig:UHV3D:3D})
and takes linear time
provided that points in $\Lr$ are sorted in two lists 
in ascending order of the $y$ and $z$ coordinates, respectively and,
provided that $\Lr^*\cup\{p^*\}$ is a nondominated point set.
However, even though $\Sr$ is a nondominated point set, $\Sr^*$ may not be
and consequently, $p^*$ may dominate some points in $\Lr^*$.
To fulfill the last requirement, 
the contribution of $p^*$ to
$\Lr^*$ is computed in a structure-destructive manner by removing the points
dominated by $p^*$ one by one as explained in Section~\ref{algs:tec:dom}.
Since a three-dimensional contribution is computed at most twice for each input point,
once when it is added to $\Lr$ and once in case it is removed from $\Lr$,
then $O(n)$ calls to the procedure to compute
a three-dimensional contribution are performed
and consequently,
the time complexity of HV4D amortizes to $O(n^2)$.

HV4D$^+$ is a modified and faster~\cite{HVCTEC2017} version of HV4D
which uses the $O(n)$-time HV3D$^+$-U algorithm (see~\ref{HVC:HV3D+U})
to compute the $3$-dimensional contribution.
Because HV3D$^+$-U computes the contribution in a structure-preserving way,
\ie, dominated points do not have to be explicitly removed previously,
only $n$ calls to HV3D$^+$-U are performed.

\subsection{Remarks}
\label{s:rmk:hv}

Table~\ref{res:HV} summarizes the algorithms described in Section~\ref{ch:hvstoa:algs:hv:desc}.
It indicates for how many dimensions
they can be used for ($d$) and for which they are recommended ($d'$)
based on their runtime performance and/or time complexity.
Moreover, it indicates whether an
implementation of such algorithms is available online and which are the paradigms/techniques used
(see Sections~\ref{ch:hvstoa:algs:parad} and~\ref{ch:hvstoa:algs:tech}).
The bottom part of Table~\ref{res:HV} summarizes the best algorithms
regarding time complexity and/or runtime performance (based on the experimental results presented
in~\cite{RusFra2014:QHV, Renaud2017, HVCTEC2017}).

HV3D$^+$ is the recommended algorithm for $d=3$ since it is optimal and the fastest in practice.
For $d\geq 4$, Chan's algorithms is recommended since it has the best time complexity,
but, because no implementation is available and there is no information on how it performs in practice,
other algorithms are recommended as alternatives.
For $d=4$, HV4D$^+$ is the fastest one.
For $d=5,6$, HBDA-NI~\citep{Renaud2017} held the best runtimes while for $d\geq7$ it is not always
clear which is the fastest one among WFG, QHV/QHV-II and HBDA-NI,
particularly because their ranking seems to be strongly dependent of the input data,
see~\cite{RusFra2014:QHV,Renaud2017,QHVII}.
Since HBDA-NI only outperformed the other two in the experiments in~\cite{Renaud2017} in a specific data
set that is particularly difficult for WFG 
and because its memory requirements grow exponentially with $d$, HBDA-NI is not recommended for $d\geq 7$.
In such cases, WFG and QHV/QHV-II are preferable.

Note that HBDA-I and (one of) the update variants of HV3D$^+$, HV3D$^+$-R (see~\ref{HV:HV3D+}),
can be used to efficiently update the hypervolume
under reference-point changes, in $O(n^{\lfloor d/2\rfloor})$ and $O(n)$ time,
respectively.
Both algorithms can also be used for the $\hvupdp$ problem, although the former only
for the incremental scenario.
However, the hypervolume indicator can be updated faster with HV3D$^+$-U (see~\ref{HVC:HV3D+U})
to solve the $\onecp$ problem for $d=3$~\cite{HVCTEC2017},
and an adapted version of WFG as in~\cite{Renaud2017} for $d>3$.

\begin{table}
\small
\center
  \caption[Algorithms for $\hvp$]
  {Algorithms for the $\hvp$ problem. (DS - Dimension sweep, IE - Inclusion-Exclusion,
  B - Bounding Technique, SDC - Spatial Divide-and-Conquer, 
  LUBs - Local Upper Bounds).}
\begin{tabular}{|r|c|c|c|c|c|c|}
    \hline
    Algorithm & $d$ & $d'$ & Time complexity & Available & Characteristics \\
    \hline
    LebMeasure~\ref{HV:LebM} & $\geq 2$ & - & $O(n^d)$ & ? & - \\
    HSO~\ref{HV:HSO}    & $\geq 2$ & - & $O(n^{d-1})$ & Yes & DS \\ 
    HOY~\ref{HV:HOY} & $\geq 2$ & - & $O(n^{d/2} \log n)$ & Yes & SDC, DS \\
    \hline
    \hline
    FPL~\ref{HV:FPL} & $\geq 2$ & 2 & $O(n^{d-2}\log n)$ & Yes & DS \\
    HV3D$^+$~\ref{HV:HV3D+} & 3 & 3 & $\Theta(n\log n)$ & Yes & DS \\
    HV4D$^+$~\ref{HV:HV4D+} & 4 & 4 & $O(n^2)$ & Yes & DS \\
    Chan~\ref{HV:Chan} & $\geq 4$ & $\geq 4$ & $O(n^{d/3}\text{polylog } n)$ & No & SDC \\
    HBDA-NI~\ref{HV:HBDA} & $\geq 2$ & $5,6$ & $O(n^{\lfloor\frac{d}{2}\rfloor})$ & Yes & LUBs\\ 
    WFG~\ref{HV:WFG} & $\geq 2$ & $\geq 7$ & $\Omega(n^{d/2}\log n)$ & Yes & IE, B, DS \\ 
    QHV~\ref{HV:QHV} & $\geq 2$ & $\geq 7$ & $O(2^{d(n-1)})$ & Yes & SDC \\ 
    QHV-II~\ref{HV:QHVII} & $\geq 2$ & $\geq 7$ & $O(d^{n-1})$ & Yes & SDC \\ 
    \hline
  \end{tabular}
  \label{res:HV}
\vspace{-0.12cm}
\end{table}

Note that, most of the currently fastest algorithms available (HV3D$^+$, HV4D$^+$ and WFG)
are all dimension-sweep based.
Each one solves a sequence of subproblems smaller in the number of dimensions and in a way that avoids
recomputing everything from scratch and/or that tries to reduce the problem size.
For example, HV3D$^+$ and HV4D$^+$ solve $\hvp$ by iterating over $\hvupdp$ 
(incremental scenario) while WFG alternates between $\onecp$ and $\hvp$ problems and takes advantage
of the bounding technique.
By doing so, they all reduced the computational costs required when compared to HSO.

\stepcounter{paraa}
\section{Hypervolume Contributions}\label{ch:hvstoa:algs:hvc}

Hypervolume-based selection is typically related to the $\allcp$ and/or the
$\hssp$ problems,
while the $\hvp$ problem is more related to the performance evaluation of EMOAs.
The computation of hypervolume contributions is frequently required either directly by the EMOA
selection method (\eg, in the special case of $k=n-1$ of $\hssp$, or for ranking),
or indirectly, in the inner steps of hypervolume-related algorithms
(\eg, in algorithms to approximate the $\hssp$).
Although algorithms for the $\hvp$ problem can also be used to compute contributions
by solving a sequence of $\hvp$ problems, 
it is typically more advantageous to use algorithms particular to such problems.
Thus, the algorithms in Section~\ref{ch:hvstoa:algs:hv}
should be used only as last resort, when there is no problem-specific alternative.

This section focuses on algorithms for problems related to the exact computation of
hypervolume contributions, in particular: 
$\allcp$, $\onecp$, $\mincrp$, $\allcupdp$, and the $\allcupdps$ problem.
Only the state-of-the-art algorithms are described. Those (non-competitive) algorithms 
purely based on HSO~\cite[\eg][]{ZhouHVC}
and approximation algorithms~\cite[\eg][]{Bader2010,Bringmann2009approx}) were left out.

\subsection{Algorithms for {\mdseries$\allcp$} problem}

\myparagraph{Bringmann and Friedrich algorithm (BF1)}\label{HVC:BF1}%
Bringmann and Friedrich~\cite{Bringmann2010} proposed an algorithm for the $\hsspc$ problem,
here referred as BF (see~\ref{HSSP:BF}). 
Given a point set $\Sr\subset\Rd$ of $n$ points and a subset size $k$,
the algorithm computes the contribution to $\Sr$ of every subset of $n-k$ points. 
In the particular case of $k=n-1$, the algorithm computes the contribution of
every subset of size $1$, \ie, the $\allcp$ problem.
For this particular case, the algorithm has $O(n^{d/2} \log n)$ time complexity and will be here
referred to as BF1.

\myparagraph{Exclusive (Contribution) Quick Hypervolume (exQHV)}\label{HVC:exQHV}%
The authors of QHV (see~\ref{HV:QHV}) extended the algorithm for the
$\allcp$ problem in any number of dimensions~\citep{exQHV}.
The main differences to QHV are: 1) the need to store all contributions instead of a single
value; 2) the point used as pivot and; 3) the exclusion of dominated points.
In the case of QHV, the point $p$ with greatest contribution to the empty set is the pivot.
Because the contribution of $p$ also has to be computed,
the authors decided not to use $p$ as pivot, 
but to use instead the point $p'$ with greatest contribution to the empty set among the
points obtained from the coordinate-wise maximum
between $p$ and each point in the hyperoctant.
In QHV, all dominated points in an hyperoctant are discarded.
However, in exQHV, points dominated by a single point
are needed for computing the contribution of the point dominating it and, therefore,
only the points dominated by two or more points are excluded from each hyperoctant.
As in QHV, an algorithm such as HSO or based on the Inclusion-Exclusion principle is
used for small cases with up to 10 points. Such algorithms were also adapted to
compute contributions~\citep{exQHV}.

\myparagraph{Dimension-sweep algorithms for $d=2,3$ (EF2D, EF and HVC3D)}
\label{HVC:EF2D}
\label{HVC:EF}%
\label{HVC:HVC3D}%
Emmerich and Fonseca~\cite{EF2011} proposed a $\Theta(n \log n)$ algorithm for the $d=2$ case
of the $\allcp$ problem,
here referred to as EF2D,
that computes all (box-shaped) contributions in linear time after sorting the input set.
EF~\cite{EF2011} is
a dimension-sweep algorithm for the $d=3$ case that has
$\Theta(n \log n)$ time complexity and $O(n)$ space complexity.
This
algorithm extends HV3D to the computation of the
contributions of all points in a nondominated point set $\Sr\subset\Rt$.
As in HV3D, EF visits all points in $\Sr$ in ascending order of $z$-coordinate
while maintaining a balanced binary tree $\Tr$.
This tree plays the same role as in HV3D, that is, to maintain the visited points
(nondominated on the $(x,y)$-plane) that delimit the base of the current slice.
These are now also the points whose contributions are being computed, and when the projection of
a point becomes dominated and the point is removed from $\Tr$, it also means that its contribution is
fully computed.
Therefore, in addition to the data structure used in HV3D,
for each point $p\in\Tr$, a partition of the contribution of $p^*$ to $\Tr^*$
is stored in a doubly-linked list of boxes and the contribution of $p$ is computed
with the update of such list,
which is updated along with the update of $\Tr$.
A total of $O(n)$ boxes are created, each operation on a box
has a $O(1)$ cost, and all operations on boxes amortize to $O(n)$. The time complexity of
EF is dominated by the operations in $\Tr$, with an overall amortized $O(n \log n)$ time complexity,
as HV3D.

HVC3D~\cite{HVCTEC2017} is a $O(n\log n)$ time algorithm that is an extension
of HV3D$^+$ (see~\ref{HV:HV3D+}) to the $d=3$ case of the $\allcp$ problem.
HVC3D works in a similar way to EF but allows linear-time updates (see~\ref{HVC:HVC3D-U}).
By using the data structure of HV3D$^+$ that takes $O(n \log n)$ time in the pre-propressing step,
the tree $\Tr$ is replaced by a doubly-linked list, which allows to compute all contributions
in amortized $O(n)$ time. Moreover, instead of a list of boxes, each point $p$ in
$\Tr$ (now stored in a linked list) has an associated sorted doubly-linked list
with the delimiters of its contribution on the $(x,y)$-plane. Such list of delimiters
is updated in a strucuture-preserving way as new points from $\Sr$ are visited,
even if a point dominated by $p$ is visited.
Moreover, unlike EF, the fact that HVC3D can be extended to efficiently update contributions,
even in the presence of dominated points,
allows it to be further used in an algorithm for the $d=4$ case (HVC4D).

\myparagraph{Dimension-sweep algorithm for $d=4$ (HVC4D)}
\label{HVC:HVC4D}%
HVC4D~\citep{HVCTEC2017} is a dimension-sweep algorithm for the $d=4$ case of
the $\allcp$ problem, and it has $O(n^2)$-time and $O(n)$-space complexity.
It is currently the asymptotically and the empirically fastest algorithm for this
case~\citep{HVCTEC2017}.
It can be viewed as an extension of HV4D$^+$ (see~\ref{HV:HV4D+}) for the all contributions case.
It solves a sequence of $3$-dimensional cases of the incremental scenario
of $\allcupdp$ problem in linear time using HVC3D-U (see~\ref{HVC:HVC3D-U}).
Points in the input set $\Sr\subset\Rf$ are visited in ascending order of the last
coordinate, partitioning the dominated region into 4-dimensional slices.
For each new visited point $p\in\Sr$, the $3$-dimensional contributions to the
projection on the $(x,y,z)$-space of $p$ and of the points delimiting the base of the previous slice
are updated, given that $p$ is added to the new slice.
The $4$-dimensional contributions to the new slice of such points are computed by multiplying their $3$-dimensional contribution by the height of the slice.

\subsection{Algorithms for {\mdseries$\onecp$} problem}

\myparagraph{Incremental HSO (IHSO)}\label{HVC:IHSO}%
IHSO~\citep{WFG:IHSO,WFG:IHSO:Fixed} is an algorithm for computing the contribution of
a single point, $p\in\Rd$, to a set of points $\Xr\subset\Rd$, \ie, for the $\onecp$ problem.
It is based on HSO, but only the contribution of $p$ to $\Xr$ is split into slices whose hypervolumes
are summed up at the end.
Points that are not delimiters of the contribution of $p$ to $\Xr$ are ignored.
Additionally, an objective reordering heuristic is used to choose a good order to process the
objectives.
IHSO works even if some points in $\Xr$ are dominated by $p$.

There is an improved version of IHSO
that uses FPL's data structure (see~\ref{HV:FPL}).
It was proposed as part of IIHSO~\citep{WFG:IIHSO} which is an algorithm for
the $\hvp$ problem that uses IHSO to iteratively solve $\onecp$ problems.
\myparagraph{HV3D$^+$ update version (HV3D$^+$-U)}
\label{HVC:HV3D+U}%
HV3D$^+$-U~\cite{HVCTEC2017} is a version of HV3D$^+$ (see~\ref{HV:HV3D+}) to compute the
contribution of a point $p\in\Rt$ to a set $\Sr\subset\Rt$ in linear time,
even if $p$ dominates points in $\Sr$,
provided that the data structure is already set up as in HV3D$^+$.
HV3D$^+$-U performs the same computations as HV3D$^+$
but restricted to the region dominated by $p$,
\ie, only the contribution of $p$ to $\Sr$ is partitioned in slices.
Thus, it performs the same sweep and management of the data structure,
but as only the base area of the slices where $p^*$ has contribution have to be computed/updated,
it skips the computation of the volume of slices below
$z=p_z$, and stops as soon as a point in $\Sr$ that dominates $p$ on the $(x,y)$-plane
is visited.

HV3D$^+$-U has resemblances to the procedure
used by HV4D to computes the $d=3$ case of $\onecp$ problem in linear time (see~\ref{HV:HV4D}).
Apart from how they represent the contribution of $p$,
the former through its delimiters and the latter through boxes,
both require the data structures to be previously set up.
Moreover, they work in a similar way when $\{p\}\cup\Sr$ is a nondominated point set,
however, this is actually a requirement of HV4D
but not of HV3D$^+$-U.

\subsection{Algorithms for {\mdseries$\mincrp$}}

\myparagraph{Incremental HSO* least contributor algorithm (IHSO*)}\label{HVC:IHSOs}%
IHSO*~\citep{WFG:IHSO} is an algorithm based on IHSO (see~\ref{HVC:IHSO}) for the $\mincrp$ problem.
In the search for the least contributor, IHSO* tries not to fully compute all contributions.
Of all schemes tested in~\cite{WFG:IHSO},
the best approach to avoid full computations was
to use an objective reordering heuristic combined with Best-First-Queueing (BFQ).
At each step of BFQ, the point with the currently lowest (partially) computed contribution is picked and
a ``bit" more of its contribution is computed. A parameter of BFQ defines how much a ``bit" more is.
This process continues until a point with its contribution fully computed is picked and this is
the least contributor.
However, the notion of a ``bit of contribution" has to be reasonably defined, \ie,
if the considered granularity is too large, then all contributions are computed, if too small,
then an excessive number of iterations is required.
The time complexity of IHSO* is not reported.
IHSO* authors also proposed an algorithm to update the least contributor under single-point
changes~\citep{WFG:IHSOupd}, \ie, to identify the (new) least contributor after adding or removing a point.

\myparagraph{Incremental WFG least contributor algorithm (IWFG)}\label{HVC:IWFG}%
IWFG~\citep{IWFG,IWFG2016} is an algorithm for the computation of the $\mincrp$ and is the result of
a combination between WFG (see~\ref{HV:WFG}) and IHSO* (see~\ref{HVC:IHSOs}).
It uses the best-first queuing mechanism of IHSO* to gradually update contributions
until the least contributor is found.
Therefore, in IWFG, the contribution of each point is divided into slices and the first slice
associated to each point is computed using WFG.
Then, repeatedly, the point with the lowest (partially) computed contribution is picked and its next
slice is computed also using WFG.
The algorithm stops when the selected point is one whose contribution is
fully computed (is the least contributor).
Cox and While~\cite{IWFG2016} proposed an improved version of IWFG that uses a new slicing scheme
and reordering of objectives for each point, which led to better runtimes
and to outperform IHSO*.
The time complexity of IWFG is not reported.

\subsection{Algorithms for {\mdseries$\allcupdp$}}

\myparagraph{Hupkens and Emmerich update algorithm for $d=2$ (UHVC2D)}\label{HVC:UHVC2D}%
Hupkens and Emmerich~\cite{HVC2D2013}
proposed an algorithm, here called UHVC2D, to update the contributions of every
point in a set $\Sr\subset\Rtwo$ to the set $\Sr$ itself under single-point changes to $\Sr$
(incremental and decremental cases of $\allcupdp$),
in the $d=2$ case.
This algorithm has $O(n)$ space complexity and
$O(\log n)$ time complexity provided that $\Sr$ is a nondominated set and is previously
stored sorted along a coordinate in a balanced binary tree.
Igel~\etal~\cite{CMA-MO} had previously proposed an update method very similar to UHVC2D.
However, it was proposed as a part of a procedure for ranking solutions and
only the decremental scenario was considered.

Note that there are at most two delimiters of the contribution of a point $p\in\Sr$
to a nondominated set $\Sr\subset\Rtwo$ and each point in $\Sr$ only delimits the contribution of its
delimiters. 
Therefore, UHVC2D does a $O(\log n)$-time search in the tree that stores $\Sr$
to find the delimiters of the contribution of the point $p$ being added/removed.
The contribution  of $p$ and each delimiter is then updated in $O(1)$ time.
The incremental case has an extra step to remove points in $\Sr$
dominated by the new point before the contribution update.
As the tree insertion/deletions are performed in $O(\log n)$, 
UHVC2D takes amortized $O(\log n)$ time per point.
Moreover, note that the algorithm could be easily applied to the $\hvupdp$ problem for $d=2$.

\myparagraph{Guerreiro and Fonseca algorithms for $d=3$ (HVC3D-R and HVC3D-U)}
\label{HVC:HVC3D-R}%
\label{HVC:HVC3D-U}%
Guerreiro and Fonseca~\cite{HVCTEC2017} proposed two algorithms based on the HVC3D algorithm (see~\ref{HVC:HVC3D}) to solve the $d=3$ case of $\allcupdp$ problem, namely HVC3D-R and HVC3D-U.
Both algorithms have $O(n)$-time and $O(n)$-space complexity provided that for a point set $\Sr\subset\Rt$,
the data structures are already set up (as in HVC3D), and
the corresponding contributions are known (only needed by HVC3D-U).
HVC3D-R and HVC3D-U also correspond to extension of HV3D$^+$-R and HV3D$^+$-U algorithms (see~\ref{HV:HV3D+} and ~\ref{HVC:HV3D+U}),
respectively, to the case of all contributions.
Hence, HVC3D-R consists of adding/removing a point to $\Sr$, updating the data structure
in linear time similiarly to HV3D$^+$-R, and then recomputing all contributions as HVC3D, in linear time. HVC3D-R can also be used to recompute all contributions in linear time
after the reference point is changed.
HVC3D-U works like HVC3D-R, the main difference being that it only computes the
(part of the) contributions that are inside the region dominated by the point $p\in\Rt$
being added/removed, \ie, the joint contributions. Similarly to HV3D$^+$-U, this allows
to skip some steps that HVC3D-R has to perform.
Namely, the contributions in the slices below $z=p_z$ are not computed,
and the algorithm stops as soon as a second point in $\Sr$ that dominates the projection
of $p$ on the $(x,y)$-plane is found.
The point set $\Sr\cup\{p\}$ is assumed to be nondominated only for the decremental
scenario, while this is not required for the incremental case.
HVC3D-U can be up to two times faster than HVC3D-R~\cite{HVCTEC2017}.

\subsection{Algorithms for {\mdseries$\allcupdps$}}
\myparagraph{Guerreiro~\etal~algorithms for $d=2,3$ (gHSS2D-u and gHSS3D-u)}
\label{HVC:gHSS2D-u}%
\label{HVC:gHSS3D-u}%
Integrated in the (incremental) greedy algorithms gHSS2D and gHSS3D
to approximate the $\hssp$ (see Section~\ref{ch:hvstoa:algs:hssp:gr}),
Guerreiro~\etal~\cite{gHSSECJ2016} proposed update procedures for the $d=2$ and
the $d=3$ cases of $\allcupdps$ problem, 
respectively.
Recall that in such case, given the sets of points $\Sr,\Rr\subset\Rd$,
the goal is to update the
contributions to $\Rr$ of the points in $\Sr$ when a point $p\in\Rd$ is added/removed
from $\Rr$.
The update procedures in gHSS2D and gHSS3D 
work for both incremental and decremental scenarios provided that
$\Sr\cup\Rr$ is a nondominated point set.
Let us call these procedures gHSS2D-u and gHSS3D-u, respectively.
The gHSS2D-u algorithm updates such contributions
in $O(|\Sr|+|\Rr|)$ time provided that the data structure is already set up,
\ie, points are previously sorted and a flag identifies to each set, $\Rr$ or $\Sr$,
each point belongs to. By visiting points in order once, each box-shaped contribution is
updated in constant time.

The gHSS3D-u is a dimension sweep algorithm that uses objective reordering.
Assuming that $\Sr\cup\Rr\cup\{p\}$ is a nondominated point set,
and that $\Sr\cup\Rr\backslash\{p\}$
is previously sorted in linked lists according to all dimensions,
gHSS3D-u works as follows.
It first considers the objective order $(x,y,z)$, updates the contributions of points
in $\Sr$ that are in the octact containing the points dominated by $p$
on the $(x,y)$-plane (and have lower $z$-coordinate than $p$),
and then those in the octact containing the points that dominate $p$
on the $(x,y)$-plane (and have higher $z$-coordinate than $p$).
Then it repeats this procedure for the other points in $\Sr$ considering
the objective orders $(y,z,x)$, and $(z,x,y)$.
A single call to gHSS3D-u has $O(|\Sr|·\cdot|\Rr|)$ time complexity.
When used sequentially for the incremental scenario (as
in gHSS3D) the number of operations amortize to linear time per point added to $\Rr$.

\subsection{Remarks}
\label{HVC:rmks}
\label{s:rmk:hvc}

Table~\ref{res:HVC} summarizes the information regarding the algorithms described in this section.
It reports the best algorithms for each problem and the number of dimensions for which
they can be used, the corresponding time complexities (if known) and indicates
whether an implementation is available online. In the latter case, a ``Yes*" means that
an implementation is not exactly available but is easy to implement/obtain.
In the case of UHVC2D and EF2D, the algorithms are easily implemented,
and in the case of gHSS2D-u and gHSS3D-u, they can be extracted from the code of
gHSS2D and gHSS3D, respectively.
Although WFG was originally proposed for the $\hvp$ problem,
with some (simple) modifications/adaptations it can be used for efficiently computing contributions, and such version is here referred as WFG-c.
Recall that WFG
explicitly solves several $\onecp$ problems in its inner steps (see~\ref{HV:WFG}).
The code for WFG available online can be easily modified to make use of that inner step to compute just
the $\onecp$ and even to compute the $\allcp$ problem
(as it was the done in~\cite{HVCTEC2017} with version 1.11 of WFG for this latter case).
This is more advantageous than iterating over the original WFG and use it to
compute every contribution as the difference between two hypervolumes (see Definition~\ref{def:hvc}).

\begin{table}
\small
\center
  \caption{Algorithms for computing hypervolume contributions related problems.}
\begin{tabular}{|c|r|c|c|c|c|}
    \hline
    Problem & Algorithm & $d$ & Time complexity & Available \\
    \hline
    \multirow{6}{*}{$\allcp$} & BF1~\ref{HVC:BF1} & $\geq 2$ & $O(n^{d/2} \log n)$ & No \\
     & EF2D~\ref{HVC:EF2D} & $2$ & $O(n \log n)$ & Yes* \\
     & HVC3D~\ref{HVC:HVC3D} & $3$ & $O(n \log n)$ & Yes \\
     & HVC4D~\ref{HVC:HVC4D} & $4$ & $O(n^2)$ & Yes \\
     & exQHV~\ref{HVC:exQHV} & $\geq 2$ & - & Yes \\
     & WFG-c~\ref{HV:WFG} & $\geq 2$ & $\Omega(n^{d/2}\log n)$ & Yes* \\
    \hline
    \multirow{3}{*}{$\onecp$} & IHSO~\ref{HVC:IHSO} & $\geq 2$ & - & No \\ 
     & HV3D$^+$-U~\ref{HVC:HV3D+U} & $3$ & $O(n)$ & Yes \\
     & WFG-c~\ref{HV:WFG} & $\geq 2$ & $\Omega(n^{d/2}\log n)$ & Yes* \\
    \hline
    \multirow{2}{*}{$\mincrp$} & IHSO*~\ref{HVC:IHSOs} & $\geq 2$ & - & No \\
     & IWFG~\ref{HVC:IWFG} & $\geq 2$ & - & Yes \\
    \hline
    \multirow{2}{*}{$\allcupdp$} & UHVC2D~\ref{HVC:UHVC2D} & $2$ & $O(\log n)$ & Yes* \\
     & HVC3D-U~\ref{HVC:HVC3D-U} & $3$ & $O(n)$ & Yes \\
    \hline
    \multirow{2}{*}{$\allcupdps$} & gHSS2D-u~\ref{HVC:gHSS2D-u} & $2$ & $O(n+m)$ & Yes* \\
    & gHSS3D-u~\ref{HVC:gHSS3D-u} & $3$ & $O(nm)$ & Yes* \\
    \hline
  \end{tabular}
  \label{res:HVC}
\vspace{-0.1cm}
\end{table}

The reported time complexities of UHVC2D, HV3D$^+$-U, gHSS2D-u, and \mbox{gHSS3D-u}
assume that the data structures were previously set up. 
In the case of UHVC2D, the reported time complexity represents the worst-case
if $\{p\}\cup\Sr$ is a nondominated point set. 
Otherwise, it represents the amortized time complexity per point for a sequence of $n$ calls,
\ie, with $n$ incremental updates to an initially empty set $\Sr$.
As proposed, gHSS2D-u and \mbox{gHSS3D-u} require $\Sr\cup\Rr\cup\{p\}$ to be a nondominated point set.
The reported time complexity of gHSS3D-u amortizes to $O(|\Sr|+|\Rr|)$ per point in $\Rr$,
if used for a sequence of incremental cases, where points in $\Rr$ are added one by one.

For the $\allcp$ problem, EF2D, HVC3D, and HVC4D are the recommended algorithms
for $d=2$, $d=3$, and $d=4$,
respectively. For $d>4$, as there is no available implementation of BF1, exQHV and WFG-c are both
recommended as alternatives. 

The $\onecp$ problem is trivially solved for the $d=2$ case in $O(n \log n)$.
For the $d=3$ case, the recommendation is to use HV3D$^+$-U to compute it in linear time
provided that the data structure is already set up, which can be done in $O(n \log n)$ time
(see~\ref{HV:HV3D+}) if required.
For $d=4$, it is asymptotically faster to solve the $\hvp$ problem with HV4D$^+$ than to use WFG-c.
As there are no experimental comparisons, both are recommended.
For $d>4$, WFG-c is the recommended algorithm.
Note, however, that any algorithm for the $\hvp$ problem can also be used to solve 
the $\onecp$ problem by taking advantage of the bounding technique (see Section~\ref{algs:tec:bt})
as WFG does (see~\ref{HV:WFG}), \ie, by computing $H(\Jr)$ after a preprocessing step
to determine the auxiliar point set $\Jr$ (see Definition~\ref{def:dlmtr}). 

IWFG is the only algorithm available online specific for the $\mincrp$
problem. However, in the experiments in~\cite{HVCTEC2017},
using HVC3D and HVC4D to solve the $\allcp$ problem for the $d=3,4$ cases, and then identifying the least contributor, was much faster (up to 456 times faster) than to use IWFG.
The same is not observed for $d>4$ in comparison to exQHV~\citep{exQHV}.
Therefore, for the $d=3$ and the $d=4$
cases of $\mincrp$, HVC3D and HVC4D are recommended, respectively, and for $d>5$,
IWFG is recommended.

For the $\allcupdp$ problem, the best is to use UHVC2D for $d=2$,
and HVC3D-U for $d=3$, while for $d>3$
the best alternative is to use the algorithms recommended for the $\allcp$ problem.
In the case of the update of contributions under reference-point changes,
the $d=2$ case is trivial, for the $d=3$ case the recommendation is to use
use HVC3D-R (see~\ref{HVC:HVC3D}), and for the $d>3$ case, is to use the recommended
algorithms for $\allcp$.

Due to the absence of dedicated algorithms for the $\allcps$ problem and for
the $\allcupdps$ problem for $d>3$,
the recommendation is to iterate over algorithms for the $\onecp$ problem.
For the $d=2$ and $d=3$ cases of $\allcupdps$ problem,
the recommendation is to use gHSS2D-u and gHSS3D-u, respectively.

\stepcounter{paraa}
\section{Hypervolume Subset Selection}\label{ch:hvstoa:algs:hssp}

Although the computation of the hypervolume indicator raised a lot of interest in the last decade
and better algorithms have successively been proposed (see Section~\ref{ch:hvstoa:algs:hv}), 
there have been fewer contributions for the HSSP, which is a \textbf{NP}-hard problem~\cite{Bringmann2017}.
There is currently no general algorithm for $d>3$ that does not check every 
combination of $k$ points. 
Only for $d=2,3$, or for small $k$ or $k$ close to $n$,
are there a few algorithms that avoid that much computation.
The existing algorithms to compute the HSSP exactly are briefly described next.

\subsection{Exact Algorithms for the $\hssp$}\label{ch:hvstoa:algs:hssp:ex}

Bringmann and Friedrich~\citep{Bringmann2010} proposed the first algorithm for the
general $d$-dimensional case for the exact computation of the HSSP in 2010. This algorithm,
which will be here referred as BF~{\myparref\label{HSSP:BF}},
explicitly checks every combination of $(n-k)$ points to discard.
As HOY (see~\ref{HV:HOY}), BF algorithm
is an adaptation of Overmars and Yap's algorithm.
BF has a $O(n^{d/2} \log n + n^{n-k})$ time complexity and
$O(\min(n^{d/2},n^{n-k}))$-space complexity.
Most of the remaining algorithms 
are focused on the particular cases of 2 and 3 dimensions, and are based on dynamic programming.
In 2009, a $O(kn^2)$-time dynamic programming algorithm was proposed for the $d=2$ case~\citep{HSSP2DWHypE2009}
which will be referred here as DPHSS~{\myparref\label{HSSP:DPHSS}}.
A faster algorithm as proposed in 2014 by
Bringmann~\etal~\cite{Bringmann:2014},
and two other in 2016 by Kuhn~\etal~\cite{KFPetal2016}.
The algorithm by Bringmann~\etal~\cite{Bringmann:2014}, HypSSP~{\myparref\label{HSSP:HypSSP}},
is also a dynamic programming algorithm that relies on modelling the
contribution of each point to a set as a linear function of a reference point.
The algorithm was initially shown to have $O(nk + n\log n)$ time complexity~\citep{Bringmann:2014},
but with slight modifications it can be tightened to $O(k(n-k) + n\log n)$~\citep{HSSPsmallk},
leading to a new version called extreme2d~{\myparref\label{HSSP:extreme2d}}.
The algorithms by Kuhn~\etal~\cite{KFPetal2016} rely on two alternative formulations of the HSSP/$\hsspc$:
a formulation of the $\hsspc$ as a $k$-link shortest path ($k$-LSP) problem~{\myparref\label{HSSP:k-LSP}},
for which the authors propose a dynamic programming algorithm that has $O(k(n-k) + n\log n)$ time complexity;
and a formulation of the HSSP as an Integer Linear Programming (ILP) problem that relies in a partitioning of
the dominated region in $O(n^d)$ boxes, and
which can be solved with any general-purpose ILP solver~{\myparref\label{HSSP:ILP}}.
These are interesting formulations even though the former works only for the 2-dimensional case and
the generalization of the latter to any number of dimensions~\citep{Kuhn2015PhD,BBeILP2018}
has a number of constraints exponential in $d$.
The space complexity of the above dynamic programming algorithms is $O(n)$ if only the hypervolume
value of the optimal subset(s) is required, and have either $O(nk)$ (DPHSS, HypSSP, extreme2d)
or $O(n^2)$ space complexity ($k$-LSP) if such subset is also needed.

More recently, Bringmann~\etal~\cite{Bringmann2017} proposed a  $O(n^{\sqrt{k}})$ time
dynamic programming algorithm for $d=3$, which exploits the fact that the boundary of the volume of
a point set $\Sr$ can be described by a planar graph with $O(|\Sr|)$ vertices, and the use of separators to split
the problem. This algorithm will be referred here as BCE3D~{\myparref\label{HSSP:sqrtk}}.
For the general
$d$-dimensional case, Gomes~\etal~\cite{BBeILP2018} 
proposed a branch-and-bound algorithm (B\&B)~{\myparref\label{HSSP:BnB}}, and a few methods to compute
upper bounds. These upper bounds rely on hypervolume contributions ($\allcp$ and $\allcps$ problems),
and consequently, the runtime efficiency of the branch-and-bound algorithm relies on the efficiency of the update
algorithms used for the $\allcupdp$ and $\allcupdps$ problems.
The most recent paper on HSSP algorithms focused on a different perspective, 
Groz and Maniu~\cite{HSSPsmallk} proposed algorithms for the cases where either $k$ or $n-k$ is small, which avoid enumerating all subsets by relying on data structures
for extreme-point queries, range trees, hash maps, and other techniques.
In particular, the authors propose a $O(n \log^d n)$ algorithm for $k=2$. For $d=3$, they propose algorithms with the following time complexities: $O(n \log^2 n)$ for $k=2$,
$O(n \log n)$ for $k=n-2$,
$O(n^{3/2})$ for $k=n-3$, 
and $O(n^{n-k-0.62\lfloor\frac{n-k}{3}\rfloor})$ for the case where $n-k$ is small.
These algorithms will be referred to as GM-sc~{\myparref\label{HSSP:GM-sc}}.

\subsubsection{Remarks}
\label{s:rmk:hssp}

\begin{table}
\small
\center
  \caption{Exact Algorithms for the HSSP}
\begin{tabular}{|r|c|c|c|c|c|c|}
    \hline
    Algorithm & $d$ & $d'$ & Time complexity & Available & Paradigm \\
    \hline
    BF \ref{HSSP:BF} & $\geq 2$ & $\geq 4$ & $O(n^{d/2} \log n + n^{n-k})$ & No & SDC \\
    DPHSS \ref{HSSP:DPHSS} & 2 & - & $O(kn^2)$ & ? & DP \\ 
    HypSSP \ref{HSSP:HypSSP} & 2 & - & $O(kn+n\log n)$ & Yes & DP \\ 
    extreme2d \ref{HSSP:extreme2d} & 2 & 2 & $O(k(n-k)+n\log n)$ & Yes & DP \\ 
    $k$-LSP \ref{HSSP:k-LSP} & 2 & 2 & $O(k(n-k)+n\log n)$ & Yes & digraph-based, DP \\
    ILP \ref{HSSP:ILP} & 2,3 & - & $\Omega(n^d)$ & Yes* & ILP \\ 
    BCE3D \ref{HSSP:sqrtk} & $3$ & $3$ & $n^{O({\sqrt{k}})}$ & No & DP \\
    B\&B \ref{HSSP:BnB} & $\geq2$ & $\geq3$ & - & Yes & Branch-and-Bound \\
    GM-sc \ref{HSSP:GM-sc} & $3$ & $3$ & $O(n^{n-k-0.62\lfloor\frac{n-k}{3}\rfloor})$ & Yes & SDC  \\
    \hline
  \end{tabular}
  \label{res:HSSP:exact}
\vspace{-0.1cm}
\end{table}

Table~\ref{res:HSSP:exact} summarizes the existent algorithms for solving HSSP exactly,
indicating the number of dimensions for which they can be used ($d$),
and whether or not there is an implementation available online
(``Available" column).
Column $d'$ indicates for which number of dimensions each algorithm is the most indicated/faster.

For the 2-dimensional case, the currently fastest algorithms are extreme2d
(the improved version of HypSSP~\ref{HSSP:HypSSP}) and $k$-LSP. 
It is worth emphasizing that, for a fixed $n$-point set and increasing $k$, 
the runtime of both algorithms (and of B\&B)
is progressively slower the closer $k$ is from $n/2$ and 
then becomes faster as $k$ approximates $n$~\cite{KFPetal2016,BBeILP2018,HSSPsmallk}. 

For the 3-dimensional case, there are currently five algorithms: BF, ILP, BCE3D, B\&B,
and GM-sc. Regarding time complexity, the best choice depends on $n$ and $k$.
For example, for small $k$, BCE3D and B\&B should be the best choices, while for
$k$ close to $n$, BF, B\&B and GM-sc should be the best choices.
Although B\&B was shown to be faster than solving the ILP with 
scip~\citep{BBeILP2018}, there are no other comparisons
between exact algorithms for the $d=3$ case,
and implementations of only B\&B and GM-sc can be found online. 

For special cases of small $k$ or small $n-k$, more efficient algorithms
can be used. For the $k=1$ case, the solution can be trivially computed in $O(n)$ time.
For $k=n-1$ the problem can be efficiently solved by solving the $\mincrp$ (or $\allcp$)
problem (see the recommended algorithms in Section~\ref{HVC:rmks}).
For $k\geq n-3$, there are the specific algorithms GM-sc, but B\&B is recommended as well.
For more general cases where $k$ becomes closer to $\frac{n}{2}$ or for large $n$,
the algorithms for $d\geq 3$ may not be practical.
For example, in~\cite{BBeILP2018}, for $d=3$, $k=\frac{n}{2}$ and
$n$ close to 100, B\&B took 100 seconds for some data sets,
and, in ~\cite{HSSPsmallk}, for $d=3$, $k=n-3$ and $n\geq50000$, GM-sc took more than $100$ seconds.
Therefore, the existing exact algorithms for $d\geq3$ are recommended only for specific cases
such as small $n$, or $k$ close to either $1$ or $n$. 
For any other cases in $d\geq 3$, if an approximation to the HSSP is enough,
a greedy algorithm should be preferable.

For the $d>3$ case, the B\&B algorithm is likely the best alternative, otherwise
it would be necessary to consider all possible subsets of size $n-k$ (with BF)
which can be very time consuming.

\paragraph{On the extension of the $d=2$ algorithms to $d\geq 3$}
In the study of optimal distributions for the 3-dimensional case, there are up to $n$
points that influence the optimal placement of a point and its contribution, \ie,
there are up to $n$ delimiters of the contribution of a point,
while for the $d=2$ case, there are only two~\citep{Auger10HV}.
Therefore, the choice of a point may depend on all other points. Consequently,
that makes the extension of dynamic programming algorithms for the $2$-dimensional case
to the $3$-dimensional case more difficult.

\stepcounter{paraa}
\setcounter{para}{0}
\subsection{Approximation algorithms for the HSSP}\label{ch:hvstoa:algs:hssp:gr}

The greedy approximation of HSSP is an alternative to the computationally expensive exact algorithms,
in particular for $d>2$.
Several greedy approaches were proposed in the literature. Most of these approaches are generic approaches
that iterate over hypervolume-based problems and thus, implementing them with different
algorithms for these problems leads to different
instances of such approaches. This means that the time complexity and runtime of an instance
of a generic greedy approach depends on the underlying hypervolume-based algorithm used to implement it, but
the approximation result does not.
In the following, let $\Xr\subset\Rd$ be
a nondominated point set such that $|\Xr|=n$ and that an optimal subset, $\Sr^{opt}\subseteq\Xr$,
of size $k\in\{1,\ldots,n\}$ needs to be selected.
Let $r\in\Rd$ be a reference point such that every point in $\Xr$ strongly dominates it.

There are four generic greedy approaches: the decremental greedy approach 
(called gHSSD)~{\myparref\label{HSSPg:GD}},
the incremental greedy approach (called gHSS)~{\myparref\label{HSSPg:GI}}, the local search-based approach (here referred to as LS~{\myparref\label{HSSPg:LS}}),
and the Global Simple Evolutionary Multiobjective Optimizer (GSEMO)~{\myparref\label{HSSPg:GSEMO}}.
The decremental greedy approach (also known as ``Greedy Front Reduction" approach)\cite{WFG2007, WFG2006}
consists of discarding $n-k$ points from $\Xr$ one at a time so as to maximize the hypervolume retained at each step.
It may be computed as a sequence of $\allcupdp$ problems. 
The greedy incremental approach (also known as ``Greedy Front Addition" approach)~\cite{WFG2007} 
consists of selecting $k$ points from $\Xr$ one at a time so as to maximize
the hypervolume gained at each step.
It may be computed as a sequence of $\allcupdps$ problems, although it is not imperative to know the
contributions to the set of selected points, $\Sr'$, of all point in $\Xr\backslash\Sr'$ in order
to select the one that contributes the most. That is the case of the $O(n \log n)$ algorithm proposed
in~\cite{HSSPsmallk} for the $d=2$ case, here referred to as gHSSEx~{\myparref\label{HSSPg:GI:gHSSEx}},
that relies on a data structure for extreme point queries
to query for the maximal contributor between two adjacent points in $\Sr'$
in logarithmic time. 
Both gHSS and gHSSD approaches provide an approximation ratio to the $\hssp$, 
even though the approximation ratio to the $\hsspc$ with either approaches can be arbitrarily
large (see Section~\ref{ch:hvstoa:props}).

The LS approach~\cite{WFG2006, WFG2007} is based on the premise
that for large $n$ and small values of $k$ it may be advantageous to try out several sets of size $k$, \ie,
perform several cheap computations instead of the more demanding ones of updating all contributions.
In LS, a $k$-size subset $\Sr$ of $\Xr$ is initially selected randomly. Then,
a small number of points are (randomly) replaced by others in $\Xr\setminus\Sr$
and the new subset is accepted if the hypervolume indicator has improved.
Different schemes for replacing subsets of points were proposed in the literature (see~\cite{WFG2006, WFG2007,BDGLGecco16}). 
The last generic approach, GSEMO~\cite{Friedrich2014}, consists of a multiobjective evolutionary algorithm that
is expected to achieve a $(1-1/e)$-approximation to the $\hssp$ in $O(n^2(\log n + k))$ iterations.
In GSEMO, each individual in the population encodes a subset $\Sr$ of $\Xr$,
which represents a solution to the HSSP.
GSEMO considers the (maximization) of a bi-objective problem, where
the first objective is the hypervolume indicator of the encoded set $\Sr$ if $|\Sr|\leq k$ and is $-1$ otherwise,
while the second objective is the number of points left out of $\Sr$, \ie, $|\Xr\setminus\Sr|$.
In each iteration of GSEMO, a new individual (obtained through parent selection and variation operators)
is inserted in the population only if it is not weakly dominated.

In contrast to the above greedy approaches, 
Bringmann~\etal~\cite{Bringmann2017} recently proposed an efficient
polynomial-time approximation scheme (EPTAS)~{\myparref\label{HSSPg:EPTAS}}. The algorithm is based on
Dynamic Programming, and approximates the HSSP for any constant number of dimensions $d$,
in $O(n\epsilon^{-d}(\log n + k+ 2^{O(\epsilon^{-2}\log 1/\epsilon)^d)}))$ time
and has an approximation guarantee of $O(1-\epsilon)$ to the optimal solution.
For any constant $d$ and $\epsilon$, the time complexity is $O(n(k+\log n))$.

\subsubsection{Remarks}
\label{s:rmk:greedy}

\begin{table}[t]
    \small
    \center
  \caption[Greedy Algorithms for the HSSP]{Greedy Algorithms for the HSSP.}
\begin{tabular}{|r|r|c|c|c|c|c|}
    \hline
    Algorithm & Auxiliar Alg. & $d$ & $d'$ & Time complexity & Avail. \\
    \hline
    \multirow{6}{*}{gHSSD \ref{HSSPg:GD}} & UHVC2D \ref{HVC:UHVC2D} & $2$ & $2$ & $O((n-k)\log n)$ & Yes* \\
    & gHSSD3D (HVC3D-U) \ref{HVC:HVC3D-U} & $3$ & $3$ & $O((n-k)n + n\log n)$ & Yes \\
    & HVC4D \ref{HVC:HVC4D} & $4$ & $4$ & $O((n-k)n^2)$ & Yes \\
    & BF1 \ref{HVC:BF1} & $\geq 2$ & $\geq 5$ & $O((n-k)n^{d/2} \log n)$ & No \\
    & IWFG \ref{HVC:IWFG} & $\geq 2$ & $\geq 5$ & - & Yes \\
    \hline
    \multirow{6}{*}{gHSS \ref{HSSPg:GI}} & gHSSEx \ref{HSSPg:GI:gHSSEx} & $2$ & $2$ & $O(n\log n)$ & No \\
     & gHSS2D-u (gHSS2D)~\ref{HVC:gHSS2D-u} & $2$ & $2$ & $O(nk + n \log n)$ & Yes \\
     & gHSS3D-u (gHSS3D)~\ref{HVC:gHSS3D-u} & $3$ & $3$ & $O(nk + n \log n)$ & Yes \\
    & HV4D$^+$ \ref{HV:HV4D+} & $4$ & $4$ & $O(k^3n)$ & Yes* \\
    & WFG-c \ref{HV:WFG} & $\geq 2$ & $\geq 5$ & $\Omega(nk^{d/2+1} \log k)$ & Yes* \\
    & Chan \ref{HV:Chan} & $\geq 2$ & $\geq 4$ & $O(nk^{d/3+1}\text{polylog } k)$ & No \\
    \hline
\multirow{5}{*}{\shortstack{LS \ref{HSSPg:LS}/\\ GSEMO \ref{HSSPg:GSEMO}}} & FPL \ref{HV:FPL} & $2$ & $2$ & $O(tk\log k)$ & Yes* \\
     & HV3D$^+$ \ref{HV:HV3D+} & $3$ & $3$ & $O(tk\log k)$ & Yes* \\
    & HV4D$^+$ \ref{HV:HV4D+} & $4$ & $4$ & $O(tk^2)$ & Yes* \\
    & HBDA-NI~\ref{HV:HBDA} & $\geq 2$ & $5,6$ & $O(tk^{\lfloor\frac{d}{2}\rfloor})$ & Yes* \\
    & WFG \ref{HV:WFG} & $\geq 2$ & $\geq 7$ & $\Omega(tk^{d/2} \log k)$ & Yes* \\
    & QHV/QHV-II~\ref{HV:QHV} & $\geq 2$ & $\geq 7$ & $O(t2^{d(k-1)})$ & Yes* \\ 
    & Chan \ref{HV:Chan} & $\geq 2$ & $\geq 4$ & $O(tk^{d/3} \text{polylog }k)$ & No \\
    \hline
    EPTAS \ref{HSSPg:EPTAS} & - & $\geq 2$ & $\geq 2$ & $O(n(k+\log n))$ & No \\
    \hline
  \end{tabular}
  \label{res:HSSP:greedy}
\vspace{-0.1cm}
\end{table}

The most advantageous combination of each greedy approach with state-of-the-art hypervolume-based algorithms,
for each number of dimensions, are shown in Table~\ref{res:HSSP:greedy}.
This list includes the combinations that lead to algorithms with the best time complexity and includes
the currently fastest alternatives than can be constructed from implementations found online.
In the reported time complexity of LS and GSEMO, $t$ stands for the number of evaluated subsets, of at most $k$ points. In the case of GSEMO, $t$ must be $O(n^2(\log n+k))$ in order to achive a
$(1-1/e)$-approximation.
Hypervolume-based algorithms with available implementations that are not
available in combination with the greedy approach but should be easily integrated,
have \textit{Availability} marked as ``Yes*".
Note that in the case of $d=2$, there are very efficient algorithms to compute HSSP exactly
(see Section~\ref{ch:hvstoa:algs:hssp}) and thus, it is questionable whether a greedy algorithm is useful
in this case. Anyway, this case is also considered in this summary for completeness.

For the decremental greedy approach, the recommendation is to use the best algorithms to solve the
$\mincrp$ problem or, in the absence of one, the $\allcp$ problem: UHVC2D, HVC3D-U, and HVC4D 
for $d=2$, $d=3$, and $d=4$ cases, respectively.
The greedy versions based on the latter two are referred in~\cite{Guerreiro:PhD}
as gHSSD3D and gHSSD4D, respectively.
For the $d>4$ case,
BF1 could be used but since it is not available online, the alternative is to use IWFG.
For the incremental greedy approach for $d\leq 3$, the dedicated greedy algorithms are recommended:
gHSSEx and gHSS2D for the $d=2$ case, and gHSS3D for the $d=3$ case.
Note that, gHSSEx is asymptotically faster than gHSS2D only for $k > \log n$,
but at the cost of more complex data structures~\cite{HSSPsmallk}.
For the $d=4$ and $d>4$ cases, the recommendation is to iteratively solve $\onecp$ problems with
HV4D$^+$ and WFG-c (the version of WFG discussed in Section~\ref{HVC:rmks}), respectively.
For the remaining greedy approaches, LS and GSEMO, the recommended algorithms are the fastest ones for $\hvp$
problem with available implementations:
HV3D$ ^+$ for $d=3$, HV4D$^+$ for $d=4$, HBDA-NI for $d=5,6$, and QHV/QHV-II and WFG for $d\geq 7$.

Currently, only gHSS, gHSSD, and EPTAS provide approximation quality.
However,
the best one cannot be decided based on these guarantees as these only provide a lower bound
and do not imply which algorithm performs better in practice.

Finally, when comparing generic greedy algorithms such as gHSSD, gHSS, LS and GSEMO,
it is important to have in mind that the runtime required by them
relies on an algorithm for a hypervolume-based problem and the choice of such an algorithm
does not impact the approximation quality.
Thus, it is important that comparisons in terms of approximation quality
are (also) conducted independently of the
algorithms runtime as better algorithms for hypervolume-based problems may be
developed/implemented in the future, which
would lead to faster instances of such generic greedy algorithms.

\section{Concluding Remarks}\label{stoa:conc}

In this paper, the properties of the hypervolume indicator were reviewed,
namely, strict monotonicity,
scaling independence, optimal $\mu$-distributions, and submodularity.
These properties tell us what sets the indicator prefers, \ie, which sets the indicator is expected
to benefit (closer to the Pareto front and well spread, but depending on the slope of the front).
All of these properties reinforce the interest in hypervolume-based selection in EMOAs.
Computational cost has been pointed out as its main drawback, but hypervolume-based assessment and selection has become more affordable in the recent years,
particularly in low-dimensional cases and/or small problem sizes. Although hypervolume-based
selection is still limited for $d\geq 3$, the existence of (fast) greedy algorithms with known
approximation ratio are a good alternative.

Regarding algorithmic aspects, the dimension-sweep approach is at the core of faster algorithms tailored for specific low-dimensional cases,
but is also effectively used in more general-purpose algorithms when combined with techniques such as bounding. Approaches such as those based on spatial divide-and-conquer and local upper bounds so far have shown to be more competitive in higher-dimensional cases.
The exploitation of update algorithms and of the relation between hypervolume-based problems play an important role as well.
This is clear with greedy algorithms that greatly benefit from faster algorithms to compute/update
hypervolume contributions.

The hypervolume indicator continues to be widelly used, which makes the development of more efficient hypervolumen-based algorithms desirable.
By discussing the existing techniques,
how the hypervolume problems relate, and describing the core ideas of the existing algorithms, this paper attempts to help and motivate for further algorithmic development.
Moreover, for both benchmarking and faster optimization problem-solving purposes, it is important
to make use of the fastest algorithms available. This paper provides a rundown of the existing ones, the corresponding
links to source code, and recommends which algorithm to use in each case.
Hence, this paper attemps to be the starting point for the development of new algorithms
for hypervolume-related problems, and be a reference for
 hypervolume-based benchmarking and problem solving.

\begin{acks}
{\small
This work is financed by national funds through the \grantsponsor{fct}{FCT -- Foundation for Science and Technology, I.P.}{} within the scope of the project CISUC -- \grantnum{fct}{UID/CEC/00326/2020}, and was
partially carried out in the scope of project
MobiWise: From Mobile Sensing to Mobility Advising (\grantnum{poci,erdf,fct}{P2020 SAICTPAC/0011/2015}),
co-financed by \grantsponsor{comp}{COMPETE 2020}{}, \grantsponsor{poci}{Portugal 2020-POCI}{}, \grantsponsor{erdf}{European Regional Development
Fund of the European Union}, and \grantsponsor{fct}{FCT}{}. 
A. P. Guerreiro acknowledges \grantsponsor{fct}{FCT}{} for Ph.D. studentship
\grantnum{fct}{SFRH/BD/77725/2011}, co-funded by the \grantsponsor{esf}{European Social Fund}{} and by the \grantsponsor{sb}{State
Budget of the Portuguese Ministry of Education and Science}{} in the scope of
NSRF--HPOP--Type 4.1--Advanced Training, and financial support
by national funds through \grantsponsor{fct}{FCT}{} with reference \grantnum{fct}{PTDC/CCI-COM/31198/2017}.
}%
\end{acks}

\bibliographystyle{ACM-Reference-Format}
\bibliography{hv-stoa}

\appendix
\section{Links to source code}
\label{s:hvstoa:links}

Table~\ref{tab:links} summarizes the links to source code of implementations of the
algorithms described in Sections~\ref{ch:hvstoa:algs:hv} to~\ref{ch:hvstoa:algs:hssp}.

\begin{table}[h]
  \caption{List of links to source codes.}
  \begin{tabular}{|r|l|}
  \hline
  Algorithm & Link \\
  \hline
  HSO~\ref{HV:HSO}      & \url{ftp://ftp.tik.ee.ethz.ch/pub/people/zitzler/hypervol.c} \\
  FPL~\ref{HV:FPL}      & \url{http://lopez-ibanez.eu/hypervolume}\\
  HOY~\ref{HV:HOY}      & \url{ls11-www.cs.tu-dortmund.de/people/beume/publications/hoy.cpp}\\
  WFG~\ref{HV:WFG}      & \url{http://www.wfg.csse.uwa.edu.au/hypervolume/}\\
  QHV~\ref{HV:QHV}      & \url{http://web.tecnico.ulisboa.pt/luis.russo/QHV/\#down}\\
  QHV-II~\ref{HV:QHV}      & \url{https://sites.google.com/prod/view/qhv-ii/qhv-ii}\\
  HBDA~\ref{HV:HBDA}    & \url{https://github.com/renaudlr/hbda}\\
  HV3D~\ref{HV:HV3D}    & \url{http://lopez-ibanez.eu/hypervolume}\\
  HV3D$^+$~\ref{HV:HV3D+} & \url{https://github.com/apguerreiro/HVC}\\
  HV4D~\ref{HV:HV4D}    & \url{https://doi.org/10.5281/zenodo.285214}\\
  HV4D$^+$~\ref{HV:HV4D+} & \url{https://github.com/apguerreiro/HVC}\\
  \hline
  exQHV~\ref{HVC:exQHV} & \url{http://web.tecnico.ulisboa.pt/luis.russo/QHV/\#down}\\
  EF~\ref{HVC:EF}       & \url{http://liacs.leidenuniv.nl/~csnaco/index.php?page=code}\\
  HVC3D~\ref{HVC:HVC3D} & \url{https://github.com/apguerreiro/HVC}\\
  HVC4D~\ref{HVC:HVC4D} & \url{https://github.com/apguerreiro/HVC}\\
  IWFG~\ref{HVC:IWFG}   & \url{http://www.wfg.csse.uwa.edu.au/hypervolume/}\\
  \hline
  HypSSP~\ref{HSSP:HypSSP} & \url{http://hpi.de/friedrich/docs/code/ssp.zip}\\
  extreme2d~\ref{HSSP:extreme2d} & \url{https://gitlri.lri.fr/groz/hssp-hypervolume-contributions}\\
  $k$-LSP~\ref{HSSP:k-LSP} & \url{https://eden.dei.uc.pt/~paquete/HSSP/hypervolume-subset.zip}\\
  B\&B~\ref{HSSP:BnB}   & \url{https://github.com/rgoomes/hssp}\\
  GM-sc~\ref{HSSP:GM-sc} & \url{https://gitlri.lri.fr/groz/hssp-hypervolume-contributions} \\
  \hline
  gHSSD3D~\ref{HSSPg:GD} & \url{https://github.com/apguerreiro/HVC}\\
  gHSSD4D~\ref{HSSPg:GD} & \url{https://github.com/apguerreiro/HVC}\\
  gHSS2D~\ref{HSSPg:GI} & \url{https://doi.org/10.5281/zenodo.284559}\\
  gHSS3D~\ref{HSSPg:GI} & \url{https://doi.org/10.5281/zenodo.284559}\\
  \hline
  \end{tabular}
  \label{tab:links}
\end{table}

\end{document}